%% file: main.tex
\pdfoutput=1
\documentclass[twocolumn]{aastex631}

\usepackage{xspace}
\usepackage{amsmath}
\usepackage{mathrsfs}
\usepackage{catchfile}
\usepackage{pgffor}
\usepackage{graphicx}
\usepackage{graphbox}
\usepackage{ifthen}
\usepackage{subfloat}

\usepackage{booktabs}
\usepackage{afterpage}
\newenvironment{rotatepage}%
    {\clearpage\pagebreak[4]\global\pdfpageattr\expandafter{\the\pdfpageattr/Rotate 90}}%
    {\clearpage\pagebreak[4]\global\pdfpageattr\expandafter{\the\pdfpageattr/Rotate 0}}%
\newenvironment{ownpage}{}{\clearpage\pagebreak[4]}
\def\fnum@table{{\bf\tablename~\thetable}}
\def\tablecommentsmod#1{\vskip1pt
\parbox{\textwidth}{
{\small\vskip1sp\indent\vrule height 11pt depth 2pt
width 0pt\currtabletypesize{\sc Note}---{#1}\vskip1pt}}}
\def\tablecommentsmodhalf#1{\vskip1pt
\parbox{\columnwidth}{
{\small\vskip1sp\indent\vrule height 11pt depth 2pt
width 0pt\currtabletypesize{\sc Note}---{#1}\vskip1pt}}}
\def\tablerefsmod#1{\vskip1pt
\parbox{\textwidth}{
{\small\vskip1sp\indent\vrule height 11pt depth 2pt
width 0pt\currtabletypesize{\sc Note}---{#1}\vskip1pt}}}

\newcommand{\Kepler}{\textit{Kepler}\xspace}
\newcommand{\TESS}{TESS\xspace}
\newcommand{\Gaia}{\textit{Gaia}\xspace}

\newcommand{\Exofast}{\texttt{EXOFASTv2}\space}

\newcommand{\Rstar}{\ensuremath{R_{\star}}\xspace} 
\newcommand{\Mstar}{\ensuremath{M_{\star}}\xspace}
\newcommand{\Rjup}{\ensuremath{R_\mathrm{J}}\xspace} 
\newcommand{\Mjup}{\ensuremath{M_\mathrm{J}}\xspace}
\newcommand{\Rearth}{\ensuremath{R_\oplus}\xspace} 

\newcommand{\Rsun}{\ensuremath{R_\odot}\xspace} 
\newcommand{\Msun}{\ensuremath{M_\odot}\xspace}
\newcommand{\Rp}{\ensuremath{R_p}\xspace}

\newcommand{\Teff}{\ensuremath{T_\mathrm{eff}}\xspace}
\newcommand{\logg}{\ensuremath{\log{g}}\xspace}
\newcommand{\feh}{\ensuremath{\mathrm{[Fe/H]}}\xspace}
\newcommand{\vsini}{\ensuremath{v\sin{i}}\xspace}
\newcommand{\vmac}{\ensuremath{v_\mathrm{mac}}\xspace}

\newcommand{\ms}{\ensuremath{\mathrm{m}\,\mathrm{s}^{-1}}\xspace}
\newcommand{\kms}{\ensuremath{\mathrm{km}\,\mathrm{s}^{-1}}\xspace}
\newcommand{\gcc}{\ensuremath{\mathrm{g}\,\mathrm{cm}^{-3}}\xspace}

\begin{document}
\title{The TESS Grand Unified Hot Jupiter Survey.\ II. Twenty New Giant Planets.
\protect{\footnote{This paper includes data gathered with the 6.5 meter Magellan Telescopes located at Las Campanas Observatory, Chile}}}
\shorttitle{Twenty New TESS Giant Planets}

\author[0000-0001-7961-3907]{Samuel W.\ Yee}
\author[0000-0002-4265-047X]{Joshua N.\ Winn}
\author[0000-0001-8732-6166]{Joel D.\ Hartman}
\affiliation{Department of Astrophysical Sciences, Princeton University, 4 Ivy Lane, Princeton, NJ 08544, USA}
\author[0000-0002-0514-5538]{Luke G. Bouma}
\altaffiliation{51 Pegasi b Fellow}
\affiliation{Department of Astronomy, California Institute of Technology, Pasadena, CA 91125, USA}
\author[0000-0002-4891-3517]{George Zhou}
\affiliation{University of Southern Queensland, Centre for Astrophysics, West Street, Toowoomba, QLD 4350, Australia}
\author[0000-0002-8964-8377]{Samuel~N.~Quinn}
\affiliation{Center for Astrophysics \textbar \ Harvard \& Smithsonian, 60 Garden St, Cambridge, MA 02138, USA}
\author[0000-0001-9911-7388]{David~W.~Latham}
\affiliation{Center for Astrophysics \textbar \ Harvard \& Smithsonian, 60 Garden St, Cambridge, MA 02138, USA}
\author[0000-0001-6637-5401]{Allyson~Bieryla}
\affiliation{Center for Astrophysics \textbar \ Harvard \& Smithsonian, 60 Garden St, Cambridge, MA 02138, USA}
\author[0000-0001-8812-0565]{Joseph~E.~Rodriguez}
\affiliation{Center for Data Intensive and Time Domain Astronomy, Department of Physics and Astronomy, Michigan State University, East Lansing, MI 48824, USA}
\author[0000-0001-6588-9574]{Karen~A.~Collins}
\affiliation{Center for Astrophysics \textbar \ Harvard \& Smithsonian, 60 Garden St, Cambridge, MA 02138, USA}
\input{authors_alpha}

\begin{abstract}
NASA's Transiting Exoplanet Survey Satellite (\TESS) mission promises to improve our understanding of hot Jupiters by providing an all-sky, magnitude-limited sample of transiting hot Jupiters suitable for population studies.
Assembling such a sample requires confirming hundreds of planet candidates with additional follow-up observations.
Here, we present twenty hot Jupiters that were detected using \TESS data and confirmed to be planets through photometric, spectroscopic, and imaging observations coordinated by the \TESS Follow-up Observing Program (TFOP).
These twenty planets have orbital periods shorter than 7 days and orbit relatively bright FGK stars ($10.9 < G < 13.0$).
Most of the planets are comparable in mass to Jupiter, although there
are four planets with masses less than that of Saturn.
TOI-3976\,b, the longest period planet in our sample ($P = 6.6$~days), may be on a moderately eccentric orbit ($e = 0.18\pm0.06$), while observations of the other targets are consistent with them being on circular orbits.
We measured the projected stellar obliquity of TOI-1937A\,b, a hot Jupiter on a 22.4 hour orbit with the Rossiter-McLaughlin effect, finding the planet's orbit to be well-aligned with the stellar spin axis ($|\lambda| = 4.0\pm3.5^\circ$).
We also investigated the possibility that TOI-1937 is a member of the NGC\,2516 open cluster, but ultimately found the evidence for cluster membership to be ambiguous.
These objects are part of a larger effort to build a complete sample of hot Jupiters to be used for future demographic and detailed characterization work.
\end{abstract}

\section{Introduction}

Hot Jupiters were among the first extrasolar planets to be discovered, initiating the longest-running mystery in exoplanet science -- how did these gas giant planets come to occupy such small orbits?
The core-accretion theory for giant planet formation, which was devised before the discovery of any exoplanets, held that giant planets could only form at distances of a few AU, beyond the ice line (see, e.g., \citealt{Lissauer1993}).
\citet{Dawson2018} and \citet{Fortney2021} reviewed the three main categories of theory that have emerged to explain the existence of hot Jupiters -- \textit{in situ} formation, inward migration of initially wide-orbiting planets via gravitational interactions with the protoplanetary gas disk, and eccentricity excitation followed by tidal dissipation.

Even though we now know of hundreds of hot Jupiters, it is not yet clear which of these processes, if any, is primarily responsible for the production of hot Jupiters.
One line of evidence that may help us shed light on this mystery is the study of the demographics of the hot Jupiter population -- the distribution of planet properties, the joint distribution of planetary and stellar properties, and the dependence of all these distributions on other aspects of planetary systems such as wide-orbiting companions.
However, hot Jupiter demographics have been difficult to determine, in part because hot Jupiters are rare, occurring around 0.5--1\% of Sun-like stars \citep[e.g.,][]{Mayor2011,Wright2012,Fressin2013}, and also because most of the currently known hot Jupiters were discovered by a heterogeneous collection of surveys with poorly characterized selection biases.

NASA's Transiting Exoplanet Survey Satellite (\TESS) mission provides an opportunity to clarify hot Jupiter demographics.
As an all-sky survey with the photometric precision to be nearly complete to transiting hot Jupiters around relatively bright stars \citep{Zhou2019a}, \TESS will not only discover hundreds of new hot Jupiters, but will also allow us to unify the previously discovered hot Jupiters into a homogeneous sample with a well-characterized selection function.
A magnitude-limited sample of stars brighter than a \Gaia $G$ magnitude of 12.5 will host $\sim 400$ transiting hot Jupiters \citep{Yee2021b}, an order-of-magnitude larger than the previous best statistical samples (from the \Kepler mission and radial-velocity surveys).
Of these 400 planets, $\approx 40\%$ are already known from the previous ground-based transit surveys.
In fact, \TESS has already begun to provide some insights into the demographics of hot Jupiters.
\citet{Zhou2019a} and \citet{Beleznay2022} investigated the dependence of hot Jupiter occurrence on stellar mass based on 18 and 97 \TESS planet candidates respectively, finding tentative evidence for an anticorrelation between the occurrence rate and stellar mass for AFG stellar hosts.

However, while the \TESS mission has announced hundreds of new hot Jupiter planet candidates as \TESS Objects of Interest (TOIs) \citep{TESS_TOIs_Guerrero2021,TESS_Faint_Kunimoto2021b},
follow-up observations are key to separating the true planets from the false positives in this sample, as well as characterizing the planets and their host stars.
A major factor in our ability to confirm hundreds of planets
is the successful operation of the \TESS Follow-Up Observing Program (TFOP; \citealt{TFOP_Collins2018,ExoFoPTESS})\footnote{\url{https://heasarc.gsfc.nasa.gov/docs/tess/followup.html}}\textsuperscript{,}\footnote{\url{https://exofop.ipac.caltech.edu/tess/}\label{footnote:exofop_url}}, in which any interested astronomer is invited to participate.  The TFOP helps to organize and coordinate follow-up observations between members of the community, thereby maximizing observing efficiency.

We began the \TESS Grand Unified Hot Jupiter Survey (\citealt{Yee2022a}, hereafter Paper I) to accelerate the process of building up a magnitude-complete sample of hot Jupiters ($P < 10$~days, $8\,\Rearth \leq \Rp \leq 24 \Rearth$), by coordinating between follow-up groups, performing new observations, and characterizing each planet candidate.
The first 10 planets found in our survey were described in Paper I.
In this paper, we present 20 new planets discovered by \TESS and confirmed by ground-based follow-up observations.
These planets orbit FGK stars brighter than $G = 13$~mag, have orbital periods $P < 7$~days, and have masses between $0.18$~\Mjup and $2.3$~\Mjup, as determined by high-precision radial-velocity (RV) observations.
A summary of the new planets is provided in Table \ref{tab:target_summary}.

Section \ref{sec:obs} of this paper describes the photometric, imaging, and spectroscopic observations; Section \ref{sec:stellar_char} presents our characterization of the planet host stars; while Section \ref{sec:planet_char} describes our global modelling of the planetary systems with \Exofast.
Because our data collection and analysis procedures are similar to those described in Paper I, we describe them more concisely in this paper.
We discuss our results and place the new systems in the context of the broader hot Jupiter sample in Section \ref{sec:discussion}.

\begin{deluxetable*}{cccccccc}
\tablecaption{Summary of New Planetary Systems \label{tab:target_summary}}
\tablehead{
    \colhead{TOI} & \colhead{TIC} & \colhead{$G$} & \colhead{Stellar \Teff} & \colhead{Stellar Radius} & \colhead{Orbital Period} & \colhead{Planet Radius} & \colhead{Planet Mass} \\
    & & \colhead{(mag)} & \colhead{(K)} & \colhead{(\Rsun)} & \colhead{(days)} & \colhead{(\Rjup)} & \colhead{(\Mjup)}
}
\startdata
\input{target_summary}
\enddata
\tablecomments{We summarize the key stellar and planetary properties for the twenty new hot Jupiter systems described in this paper, as derived from our global fits (\S\ref{sec:planet_char}).}
\end{deluxetable*}

\section{Observations and Data} \label{sec:obs}

\subsection{TESS Photometry} \label{ssec:tess}

The twenty planets described in this paper were first identified as transiting planet candidates in the \TESS photometry.
None of them orbit stars that had been preselected for 2-minute observations during the \TESS Prime Mission (Sectors 1 -- 26).
Instead, the \TESS photometry for these targets during the Prime Mission comes from the full-frame images (FFIs), which were combined and downloaded at 30-minute cadence.
Following the conclusion of its Prime Mission in July 2020, \TESS re-observed most of the sky as part of the first Extended Mission (EM1).
Six of our targets (TOI-1937\,b, -2583\,b, -3807\,b, -3819\,b, -3912\,b, and -4087\,b) were identified as planet candidates based on Prime Mission data and selected
for 2-minute observations during EM1.
The remaining objects continued to be observed as part of the FFIs, which are available with a 10-minute cadence in EM1.
Table \ref{tab:tess_summary} summarizes the \TESS observations for each target.

The short-cadence data were reduced by the \TESS Science Processing Operations Center (SPOC) pipeline \citep{TESS_SPOC_Jenkins2016}, with light-curves extracted from the ``postage stamp'' images around each selected target.
The SPOC pipeline computes optimal apertures to extract light curves from each target, and estimates the contamination from stars within the same aperture to correct the flux levels.
Meanwhile, the FFI data were calibrated with the \texttt{tica} software \citep{TESS_QLP_Fausnaugh2020} and light-curves were extracted with the MIT Quick-Look Pipeline (QLP; \citealt{TESS_QLP_Huang2020a,TESS_QLP_Huang2020b,TESS_QLP_Kunimoto2021}).
The QLP produces light-curves for each target using difference imaging, subtracting each frame from a reference frame generated from the median of 40 frames with minimal scattered light.
This approach automatically accounts for any contamination from other stars within the same aperture.
The \TESS SPOC pipeline has also recently begun extracting light-curves for a subset of the targets in the FFIs \citep{TESS_SPOC_Caldwell2020}, and we use these light-curves when available.

The transit signals of the majority of our targets were first found through a box-least squares (BLS) transit search \citep{BLS_Kovacs2002,VARTOOLS_Hartman2016} of the QLP light-curves.
These transit events were then triaged and vetted by the \TESS Science Office \citep{AstroNetTriage_Yu2019,TESS_Faint_Kunimoto2021b} before being announced to the community as \TESS Objects of Interest (TOIs; \citealt{TESS_TOIs_Guerrero2021}).

Eight of the targets described here were first identified by other investigators and announced as ``community TOIs'' (cTOIs).
TOI-1937A\,b and TOI-4145\,b were found by the Cluster Difference Imaging Photometric Survey (CDIPS; \citealt{CDIPS_Bouma2019}), which extracted light-curves of potential members of stellar clusters from the FFI data and searched the data for transiting planet candidates.
We discuss the potential cluster membership of these two targets in Section \ref{ssec:cluster}.
TOI-2364\,b and TOI-2796\,b were identified and vetted as planet candidates by \citet{Montalto2020}, who used the DIAmante difference imaging pipeline to extract light-curves from the first year of \TESS FFIs and performed an independent BLS transit search.
\citet{Olmschenk2021} flagged TOI-2583\,b, TOI-2818\,b, TOI-2842\,b, TOI-4087\,b, and TOI-4145\,b in their search of the FFI light-curves extracted by the \texttt{eleanor} pipeline \citep{Eleanor_Feinstein2019}, using the Quasiperiodic Automatic Transit Search (QATS; \citealt{QATS_Carter2013,QATS_Kruse2019}) and Discovery and Vetting of Exoplanets (DAVE; \citealt{DAVE_Kostov2019}) pipelines to identify and vet the transit candidates.
All six targets underwent further vetting by the \TESS Science Office and were subsequently also flagged as TOIs \citep{TESS_cTOIs_Mireles2021}.

We used the \texttt{lightkurve} Python package \citep{Lightkurve18} to download all available \TESS photometry from the Mikulski Archive for Space Telescopes (MAST).
When available, we used the Presearch Data Conditioning (PDC; \citealt{TESS_PDC_Stumpe2012,TESS_PDC_Smith2012,TESS_PDC_Stumpe2014}) light-curves produced by the SPOC pipeline, which have been corrected for instrumental systematics.
When SPOC light-curves were unavailable, we used those produced by the QLP, except for the case of TOI-1937A.
For the Sector 7 and 9 observations of TOI-1937A, we used the CDIPS light-curves based on the smallest choice of aperture (\texttt{IRM1}).

While inspecting the light curves, we noticed that the TESS Sector 9 QLP light curve for TOI-2977 showed a different transit depth from the QLP light curves from Sectors 36 and 37.
We determined that this was due to a 0.6 mag difference in the TESS magnitudes between versions 7 and 8 of the TESS Input Catalog (TIC; \citealt{TIC_Stassun2018,TIC_Stassun2019}).
These magnitudes are used by the QLP to convert the difference fluxes into absolute fluxes, with TICv7 used for light curves from the TESS Prime Mission, while TICv8 was used for TESS EM1 light curves.
We therefore used the updated TICv8 magnitude to correct the Sector 9 light curve for TOI-2977, bringing the resulting transit depth into agreement with those from the later sectors and ground-based photometry.

Before using the TESS data to help determine the system properties (\S\ref{sec:planet_char}), we ``flattened'' the SPOC light-curves using the \texttt{Keplerspline}\footnote{\url{https://github.com/avanderburg/keplersplinev2}} routine \citep{Keplerspline_Vanderburg2014,Keplerspline_Shallue2018}, which attempts to eliminate variations in the light-curve due to stellar variability and residual instrumental effects.
In this step, we also normalized the light curves such that the
data obtained outside of transits has a mean flux of unity.
We note that the QLP data were already detrended using this procedure during the light-curve production process.
For our analysis, we used only the segments of the light-curves that are centered on transits and span 3 transit durations, excluding the remaining out-of-transit data.
All the raw \TESS data used in this manuscript can be found on MAST, while the flattened and normalized TESS photometry data are provided as online supplementary material.

\begin{deluxetable}{cccr}
\tablecolumns{4}
\tablecaption{Summary of TESS Observations \label{tab:tess_summary}}
\tablehead{
    \colhead{Target} & \colhead{Sector} & \colhead{Source$^{\mathrm{a}}$} & \colhead{Cadence (s)}
}
\startdata
\input{tess_summary} 
\enddata
\tablenotetext{a}{The source column indicates the High-Level Science Product (HLSP) source of the TESS light-curves used in the analysis.}
\tablecomments{The raw TESS data are available on MAST, while the flattened \& normalized TESS photometry used in our analysis are provided as online supplementary material (Data behind the Figure for Figure Set \ref{fig:all_multiplots}).}
\end{deluxetable}


\subsection{WASP Photometry} \label{ssec:wasp_photometry}

\begin{figure}
\includegraphics[width=0.95\linewidth]{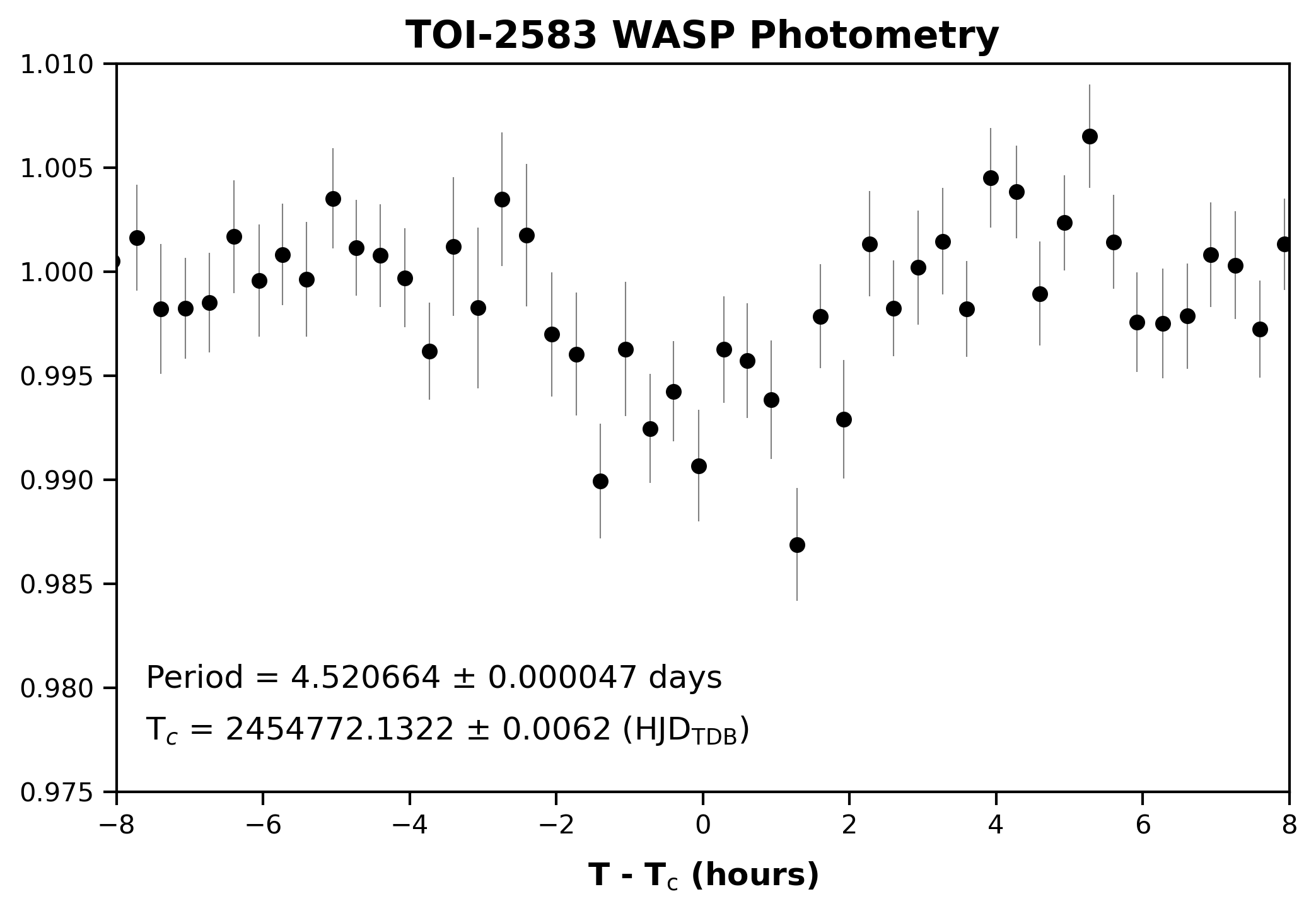}
\includegraphics[width=0.95\linewidth]{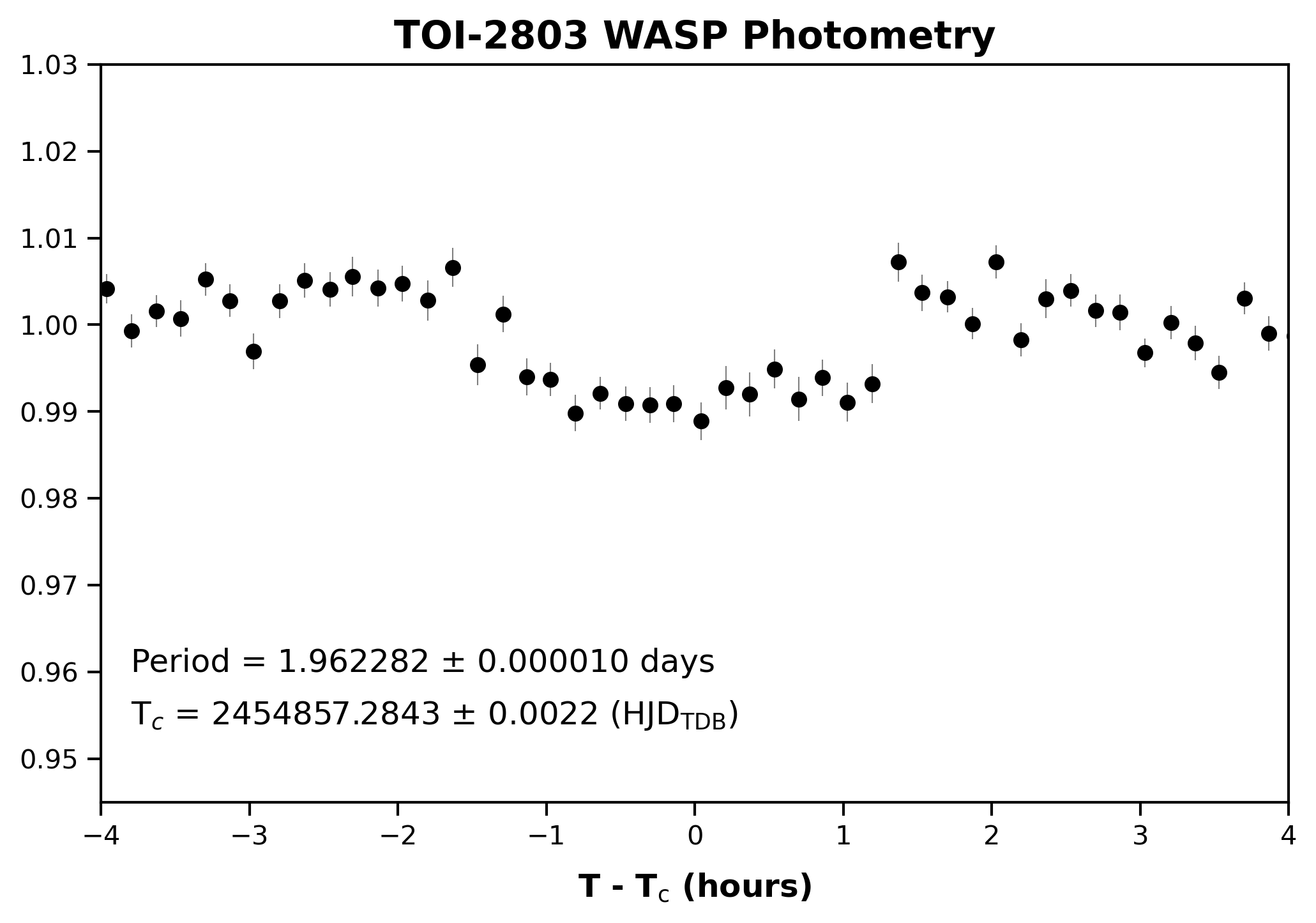}
\caption{WASP photometry for TOI-2583 (top) and TOI-2803 (bottom), binned and phase-folded onto the ephemeris found by the WASP transit-finding algorithm.
\label{fig:wasp_lightcurves}}
\end{figure}

Two of our targets, TOI-2583\,b and TOI-2803\,b, were also detected as planet candidates by the WASP transit search \citep{WASP_Pollacco2006}.
The WASP survey comprises two wide-field camera arrays at the Observatorio del Roque de los Muchachos on La Palma and the Sutherland Station of the South African Astronomical Observatory (SAAO).
TOI-2583 was observed by WASP between 2004 and 2010, while TOI-2803 was observed between 2006 and 2012, with the transit events detected at the same period as those found by TESS.
Figure \ref{fig:wasp_lightcurves} shows the phase-folded WASP photometry for these two objects.
To incorporate the power from the long baseline of the WASP photometry in our fits, we use the times of conjunction found by the WASP transit search algorithm ($T_c = 2454772.1322 \pm 0.0062$ HJD$_\mathrm{TDB}$ for TOI-2583\,b; $T_c = 2454857.2843 \pm 0.0022$ HJD$_\mathrm{TDB}$ for TOI-2803\,b) as a prior for our global fits (\S\ref{sec:planet_char}).

In addition to TOI-2583 and TOI-2803, archival WASP photometry was also available for TOI-2587, TOI-3364, TOI-3819, TOI-3912, and TOI-3976.
We searched the WASP photometry of all targets to check for rotational modulation.
In all cases, no significant modulations were detected, with 95\% upper limits between 1--3~mmag.

\subsection{Follow-up Ground-Based Photometry} \label{ssec:sg1}

Apart from the \TESS and archival WASP photometry, we obtained additional light-curves from a wide range of ground-based facilities, organized by the TFOP Seeing-limited Photometry Sub-Group 1 (SG1).
The superior angular resolution of the ground-based photometry (typically a few arcseconds) compared with that of \TESS helped to rule out the possibility that the
transit-like fading events are actually due to a nearby eclipsing binary rather than the intended target star.
In some cases, the target was observed in multiple photometric bands, and the lack of chromatic variation in transit depths was evidence that the transit events are due to a planetary companion rather than a
luminous companion.
The ground-based observations also extended the timespan over which transits have been detected, thereby allowing the orbital period to be determined with greater precision.

We summarize all the ground-based follow-up photometry in Table \ref{tab:sg1_lcs}.
The observations from the Perth Exoplanet Survey Telescope (PEST) were reduced by a custom software package\footnote{\url{http://pestobservatory.com/the-pest-pipeline/}}, while the TRAPPIST-North \citep{TRAPPIST_Gillon2011,TRAPPIST-North_Barkaoui2019} observations were reduced with the procedures described by \citet{TRAPPIST-North_Gillon2013}.
The five light-curves of TOI-1937 obtained from the Las Cumbres Observatory Global Telescope (LCOGT; \citealt{LCOGT_Brown2013}) were extracted from the calibrated science images produced by the LCOGT network using aperture photometry routines from the {\sc FITSH} package \citep{FITSH_Pal2012}.
For all remaining observations, data reduction and aperture photometry was performed using the \texttt{AstroImageJ} software \citep{AstroImageJ_Collins17}, with scheduling assisted by the \texttt{TAPIR} software \citep{TAPIR_Jensen2013}.

We included most of the ground-based time-series photometry in our global fits (Section \ref{sec:planet_char}), simultaneously fitting a transit model while detrending against the columns listed in Table \ref{tab:sg1_lcs}. 
The TRAPPIST-North light-curve of TOI-2364, the MLO light-curve of TOI-3688, and the 2021 Nov 19 GMU light-curve of TOI-3819 were excluded from the fits because no transits were detected in those light-curves.
The non-detections are all consistent with the ephemerides derived from the rest of the data.
For TOI-2583, we also excluded the OAUV light-curve, for which the data during the transit were strongly affected by variable sky conditions.
All of the ground-based time-series photometry data are available through ExoFoP\textsuperscript{\ref{footnote:exofop_url}}, and as supplementary material accompanying this article.

\begin{rotatepage}
\movetabledown=0.25in
\begin{longrotatetable}
\begin{deluxetable*}{cccccrcrc}
\tablecolumns{9}
\tablecaption{Summary of Ground-Based Photometric Follow-Up Observations \label{tab:sg1_lcs}}
\tablehead{
    \colhead{Target} & \colhead{Facility/Instrument} & \colhead{Aperture} & \colhead{Filter} &
    \colhead{Date} & \colhead{Cadence} & \colhead{Used in Fit} & \colhead{Precision$^\mathrm{a}$} & \colhead{Detrending Vectors} \\
    & & \colhead{(m)} & & \colhead{(UT)} & \colhead{(s)} & & \colhead{(mmag)} &
}
\startdata
\input{sg1_summary}
\enddata
\tablenotetext{a}{Precision is computed as the rms of the residuals when the observed data points are subtracted from the transit and detrending model.}
\tablecomments{
The ground-based follow-up ohotometry are publicly avilable via ExoFoP\textsuperscript{\ref{footnote:exofop_url}} and are also provided as online supplementary material (Data behind the Figure for Figure Set \ref{fig:all_multiplots}).
\\The following facilities were used for ground-based photometric observations: 0.4m and 1.0m telescopes of the Las Cumbres Observatory Global Telescope (LCOGT; \citealt{LCOGT_Brown2013}) using sites at Siding Spring Observatory (SSO), Cerro Tololo Inter-American Observation (CTIO), and the South African Astronomical Observatory (SAAO); the 0.36m and 0.51m telescopes at the El Sauce Observatory; TRAPPIST-North at the Oukaimeden Observatory \citep{TRAPPIST_Jehin2011,TRAPPIST_Gillon2011,TRAPPIST-North_Barkaoui2019}; the Hazelwood Observatory; the Observatori Astronòmic de la Universitat de València (OAUV) T50 0.5m telescope; KeplerCam on the Fred Lawrence Whipple Observatory (FLWO) 1.2m telescope; the University of Louisville Manner Telescope (ULMT) at Mt. Lemmon; the Caucasian Mountain Observatory (CMO); the Brierfield Observatory; the Perth Exoplanet Survey Telescope (PEST); the Maury Lewin Astronomical Observatory (MLO); the 0.8m telescope at George Mason University (GMU) with automation described in \citet{GMU_Reefe2022}; the Acton Sky Portal; the Silesian University of Technology Observatories (SUTO) OTIVAR 0.3m telescope; the Villa '39 observatory; the Private observatory of the Mount at Saint-Pierre-du-Mont, France (OPM); and the Wellesley College Whitin Observatory.}
\end{deluxetable*}
\end{longrotatetable}
\end{rotatepage}


\subsection{High Angular Resolution Imaging} \label{ssec:imaging}

\begin{figure}
\centering
\includegraphics[width=0.95\linewidth]{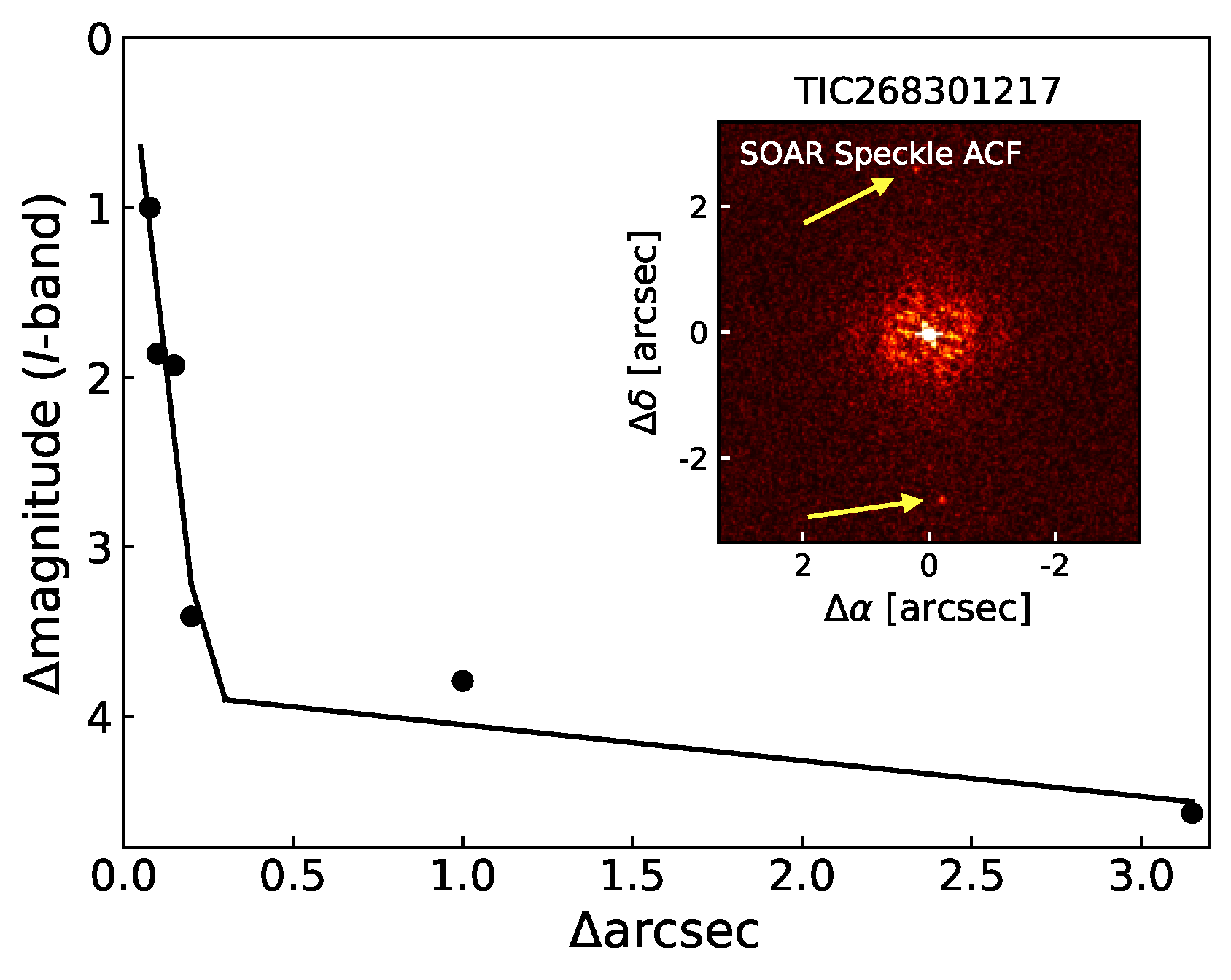}
\includegraphics[width=0.95\linewidth]{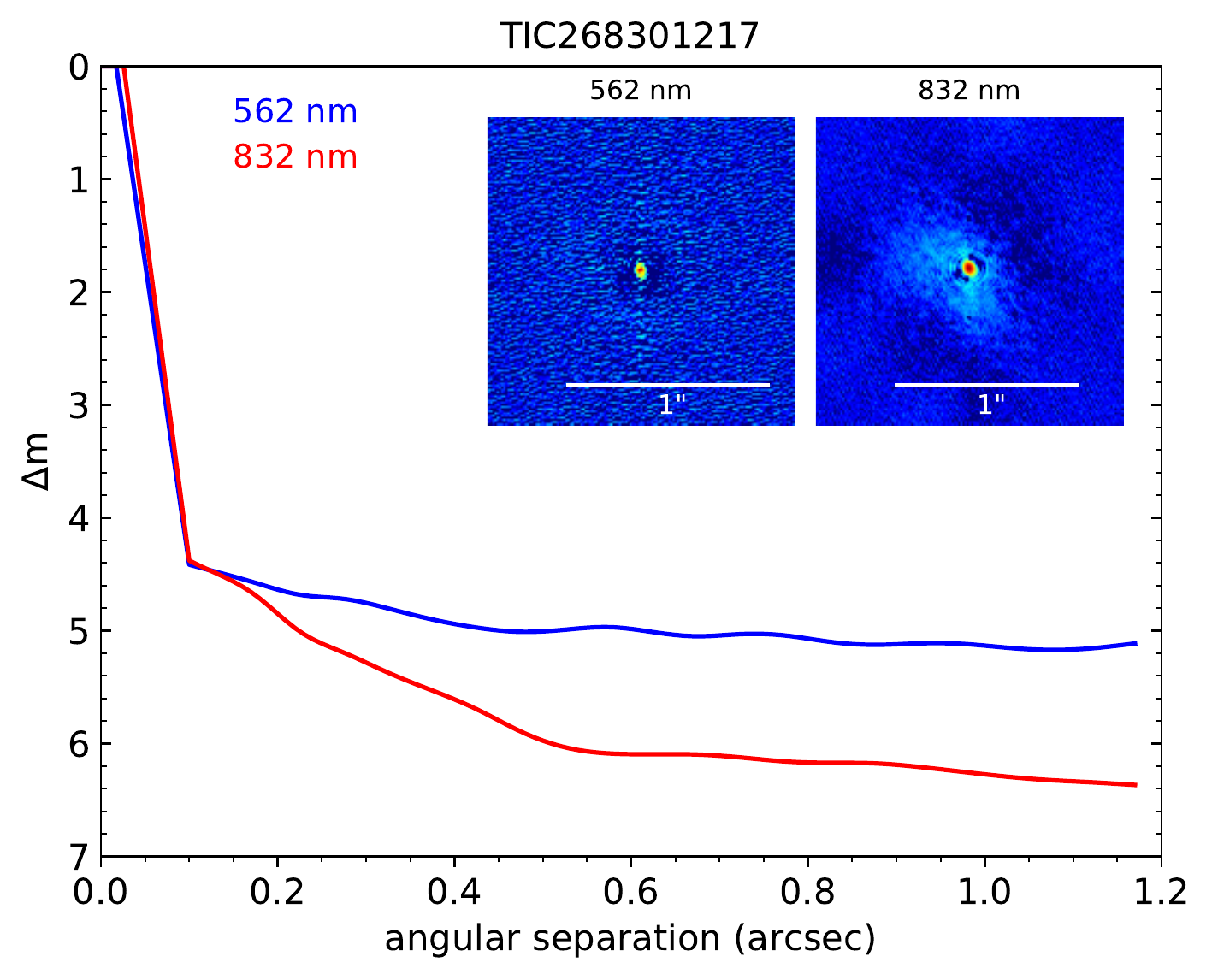}
\caption{High-resolution imaging data for TOI-1937.
\textbf{Top}: SOAR HRCam speckle sensitivity curve (solid line) and auto-correlation function (ACF, inset image).
The detected companion appears as two dots $\approx 2\farcs5$ to the north and south of the primary (marked by yellow arrows).
\textbf{Bottom}: Gemini-South Zorro 5-$\sigma$ sensitivity curve and reconstructed image (inset), taken at 562~nm and 832~nm.
The field of view for these images is smaller than the separation of the $2\farcs5$ companion, and no further stars were detected down to instrumental detection limits.
\label{fig:toi1937_imaging}}
\end{figure}

\begin{figure}
\centering
\includegraphics[width=0.95\linewidth]{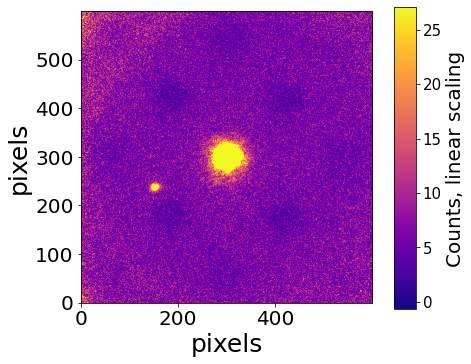}
\includegraphics[width=0.95\linewidth]{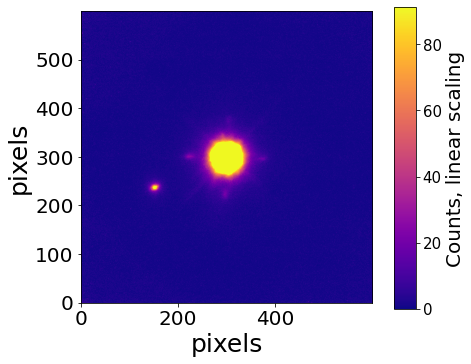}
\caption{ShARCS AO imaging of TOI-2583, taken in $J$-band (top) and $K_s$-band (bottom), revealing a nearby companion at $\approx 5"$.
\label{fig:toi2583_imaging}}
\end{figure}

\begin{figure}
\centering
\includegraphics[width=0.95\linewidth]{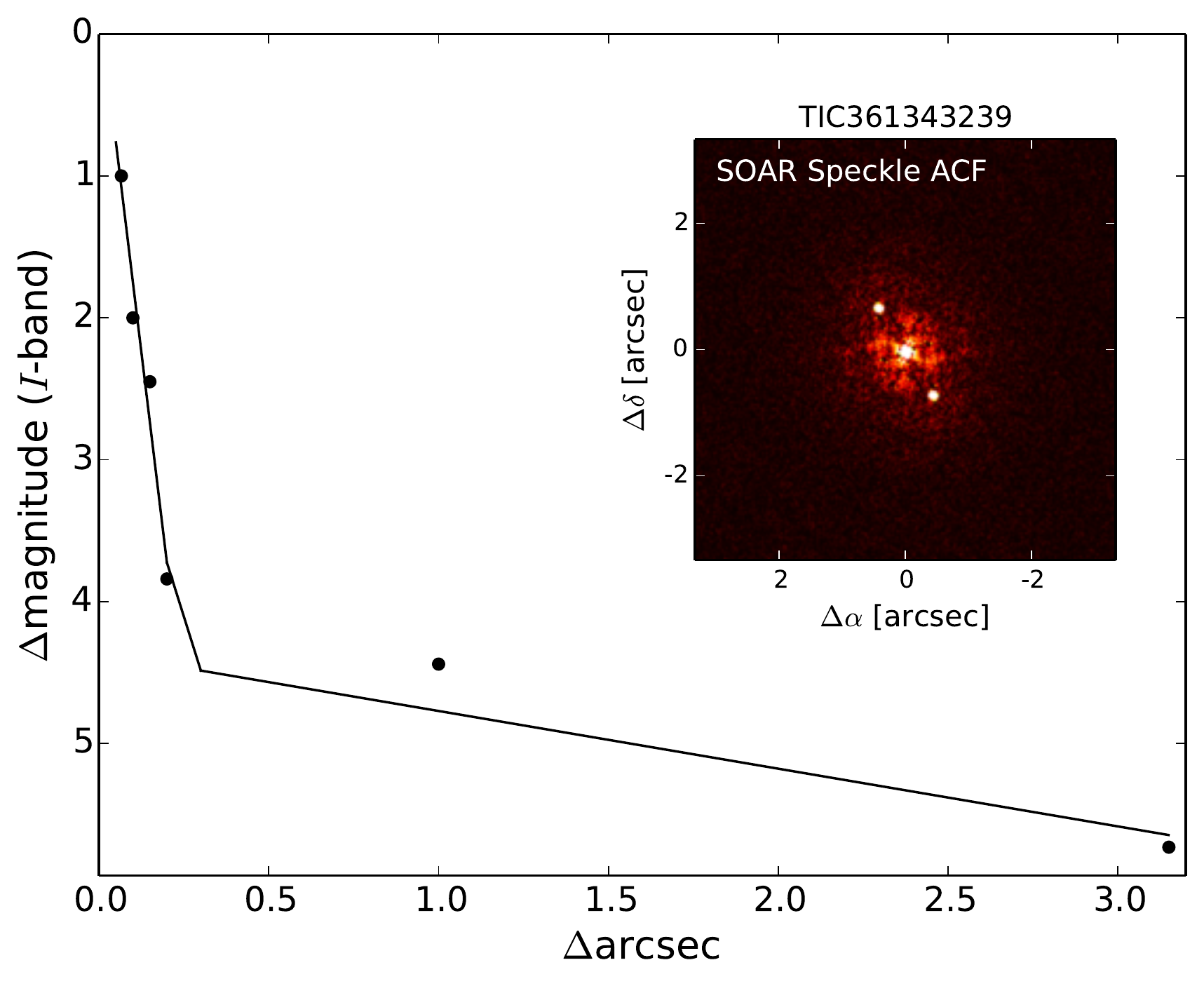}
\caption{Speckle sensitivity curve (solid line) and auto-correlation function (ACF, inset image) from the SOAR HRCam observations of TOI-2977.
In the ACF image, the two dots to the northwest and southeast of the center represent the detection of a stellar companion at $0\farcs77$ from the primary.
\label{fig:toi2977_imaging}
}
\end{figure}

As part of follow-up observations coordinated by the TFOP High-Resolution Imaging Sub-Group 3 (SG3), we obtained high angular-resolution imaging of all the targets described here.
Observations were made using the 'Alopeke and Zorro speckle cameras on the Gemini-North and Gemini-South telescopes respectively \citep{Gemini_Zorro_Alopeke_Scott2021} and reduced according to the procedures in \citet{Speckle_Reduction_Howell2011}; the High-Resolution Camera (HRCam; \citealt{SOAR_Tokovinin2008}) speckle imaging instrument on the Southern Astrophysical Research (SOAR) 4.1m telescope; the ShARCS camera using the adaptive optics system on the Shane 3m telescope at Lick Observatory \citep{ShaneAO_Kupke2012,ShaneAO_Gavel2014,ShaneAO_McGurk2014}; the NN-explore Exoplanet Stellar Speckle Imager (NESSI; \citealt{NESSI_Scott2018}) on the WIYN 3.5m telescope at Kitt Peak National Observatory (KPNO); the speckle polarimeter on the 2.5m telescope at the Caucasian Mountain Observatory (CMO) of Sternberg Astronomical Institute (SAI) of Lomonosov Moscow State University \citep{SAI_Safonov2017}; and the Palomar High Angular Resolution Observer (PHARO; \citealt{PHARO_Hayward2001}) on the 200-in Hale telescope at Palomar Observatory.
The observation strategy and data reduction procedures for the SOAR observations are described in \citet{SOAR_Tokovinin2018,SOAR_TESS_Ziegler2019} and \citet{SOAR_TESS_Ziegler2021}, while the ShARCS observations were reduced with the publicly available \texttt{SImMER} pipeline \citep{SIMmER_Savel2020}.\footnote{\url{https://github.com/arjunsavel/SImMER}}.
These imaging observations are summarized in Table \ref{tab:imaging_obs}.

Nearby companions were detected in these high-resolution imaging observations only for the targets TOI-1937 (Fig. \ref{fig:toi1937_imaging}), TOI-2583 (Fig. \ref{fig:toi2583_imaging}) and TOI-2977 (Fig. \ref{fig:toi2977_imaging}).
The SOAR speckle imaging of TOI-1937 detected a companion at an angular separation of $2\farcs5$ from the primary, which has $\Delta I = 4.3$~mag.
The ShARCS imaging of TOI-2583 detected a companion at an angular separation of $5\farcs4$ from the primary, with $\Delta J = 4.4$~mag and $\Delta K_s = 4.0$~mag.
Finally, for the target TOI-2977, a fainter companion ($\Delta\,I_c$ = 1.7~mag) at an angular separation of $0\farcs77$ was revealed by the SOAR observations (Figure \ref{fig:toi2977_imaging}).
We discuss the treatment of these companions in Section \ref{ssec:companions}.
No other companions were detected in the high angular resolution imaging down to the detection limits for the remaining targets, and we show these data in Figure Set \ref{fig:high_res_imaging_nocomp}.

\begin{deluxetable*}{ccccccr}
\tablecolumns{7}
\tablecaption{Summary of High-Resolution Imaging Observations \label{tab:imaging_obs}}
\tablehead{
    \colhead{Target} & \colhead{Telescope} & \colhead{Instrument} & \colhead{Filter} &
    \colhead{Date} & \colhead{Image Type} & \colhead{Contrast} 
}
\startdata
\input{sg3_summary}
\enddata
\end{deluxetable*}


\begin{subfigures}
\label{fig:high_res_imaging_nocomp}
\makeatletter\onecolumngrid@push\makeatother
\clearpage
\begin{figure*}
\centering
\includegraphics[align=c,width=0.29\linewidth]{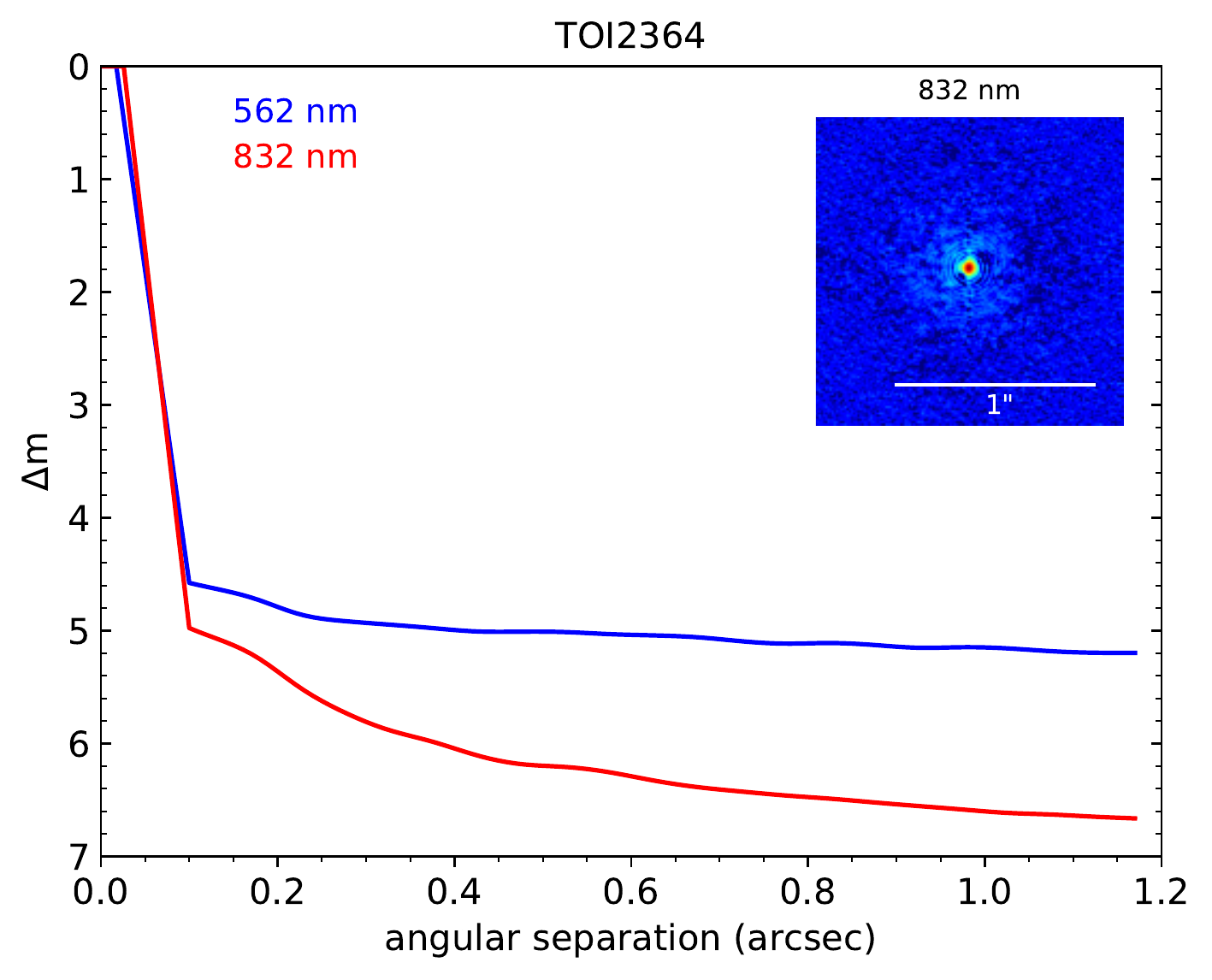}
\includegraphics[align=c,width=0.39\linewidth]{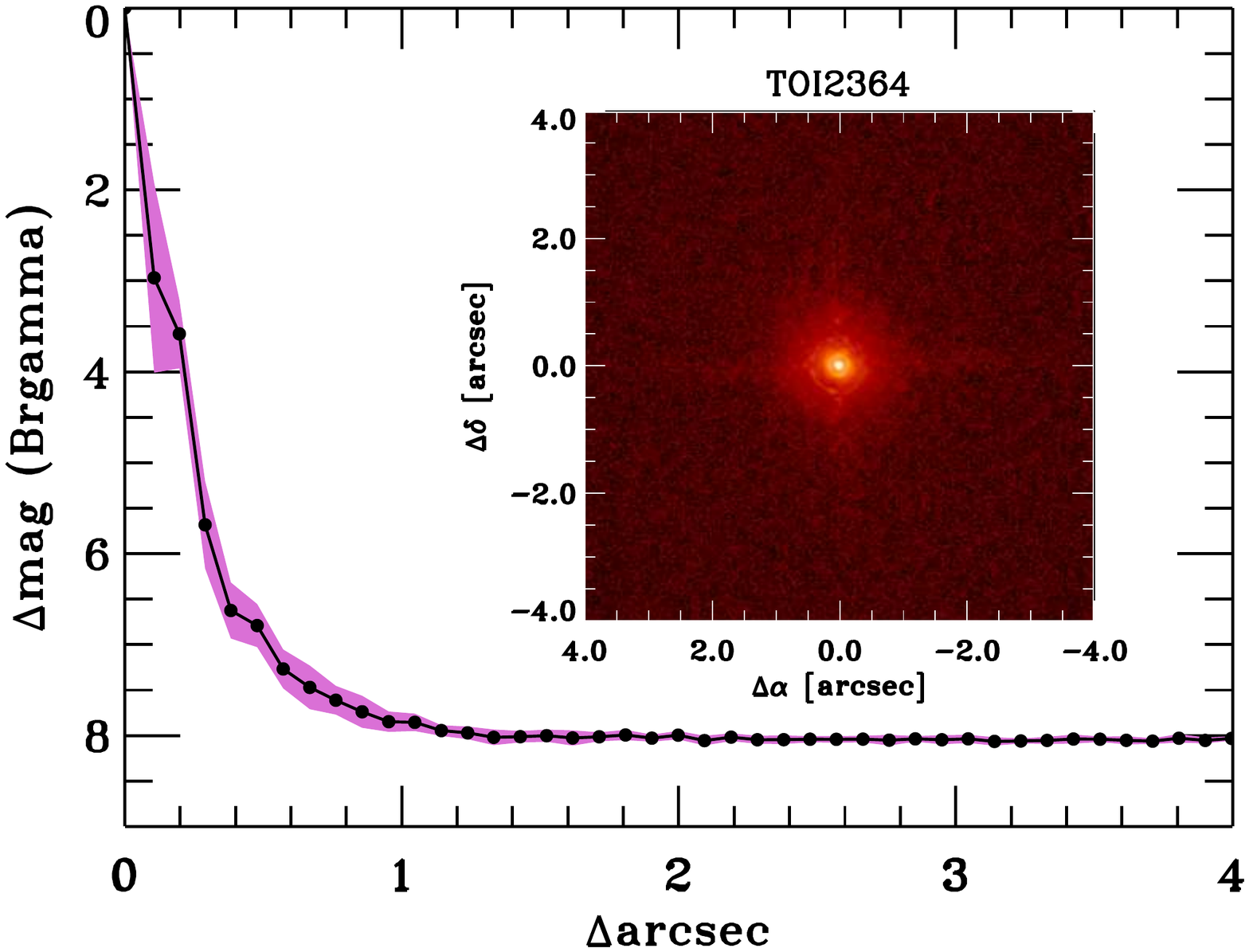}
\includegraphics[align=c,width=0.29\linewidth]{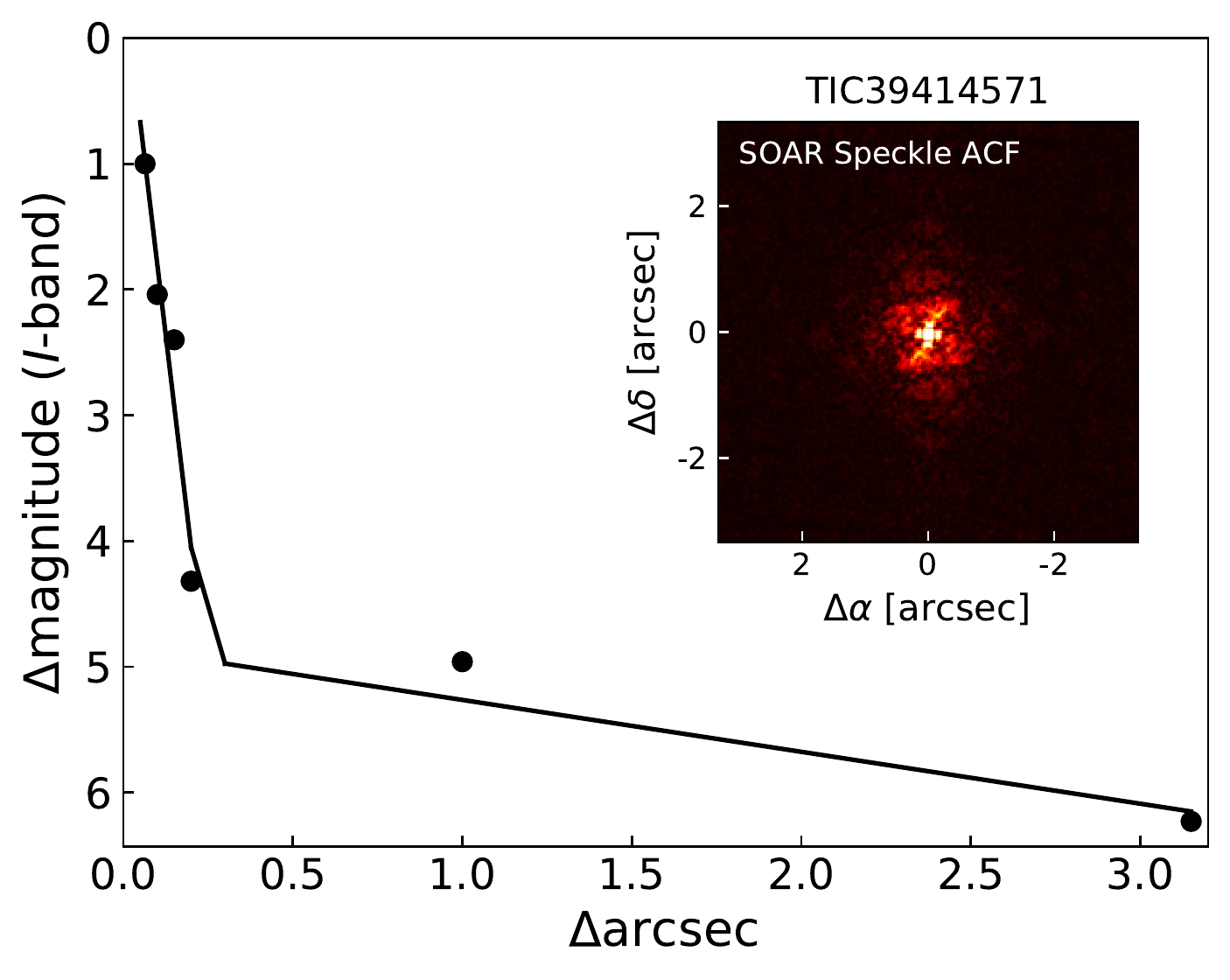}\\
\includegraphics[width=0.32\linewidth]{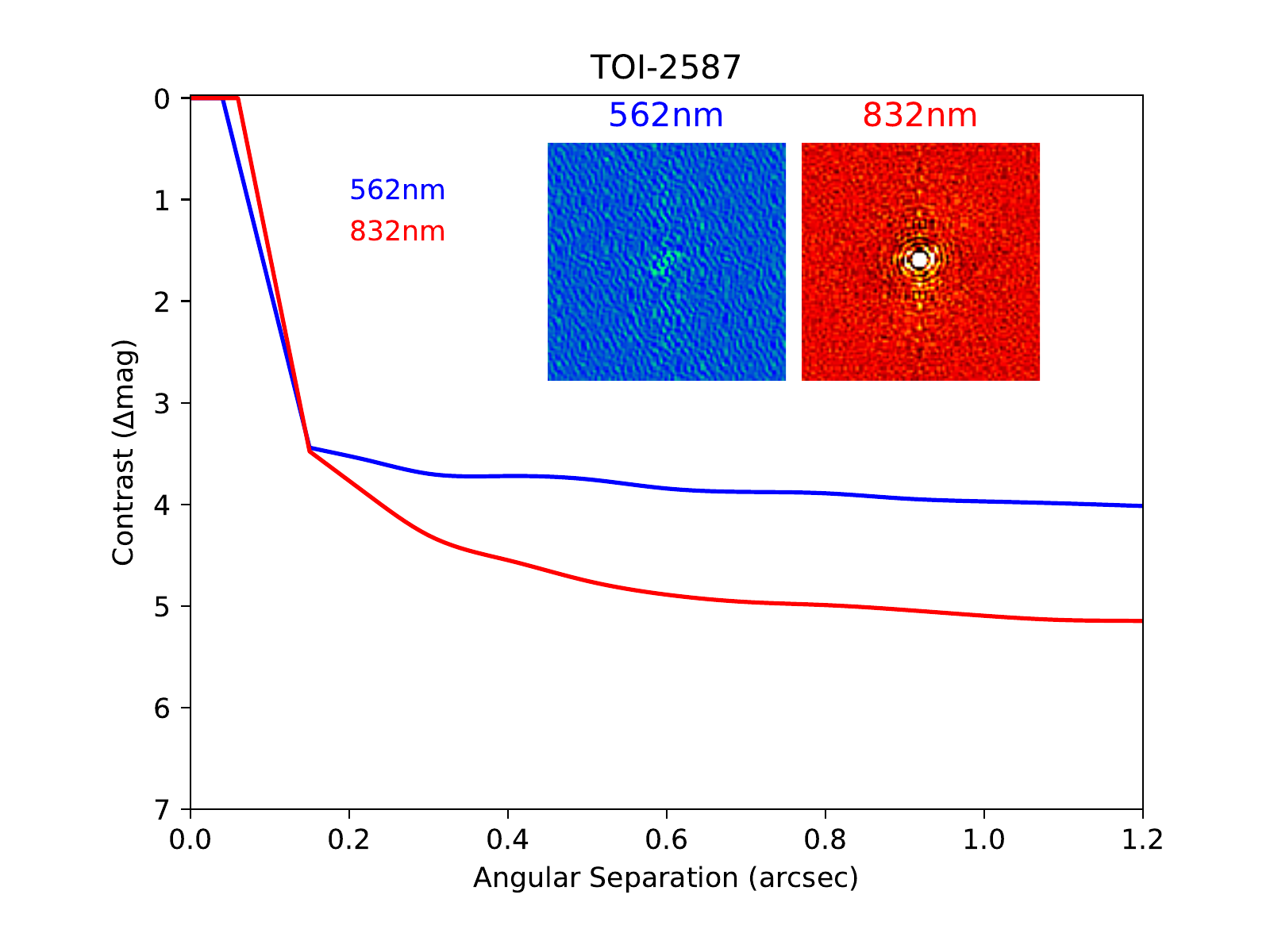}
\includegraphics[width=0.32\linewidth]{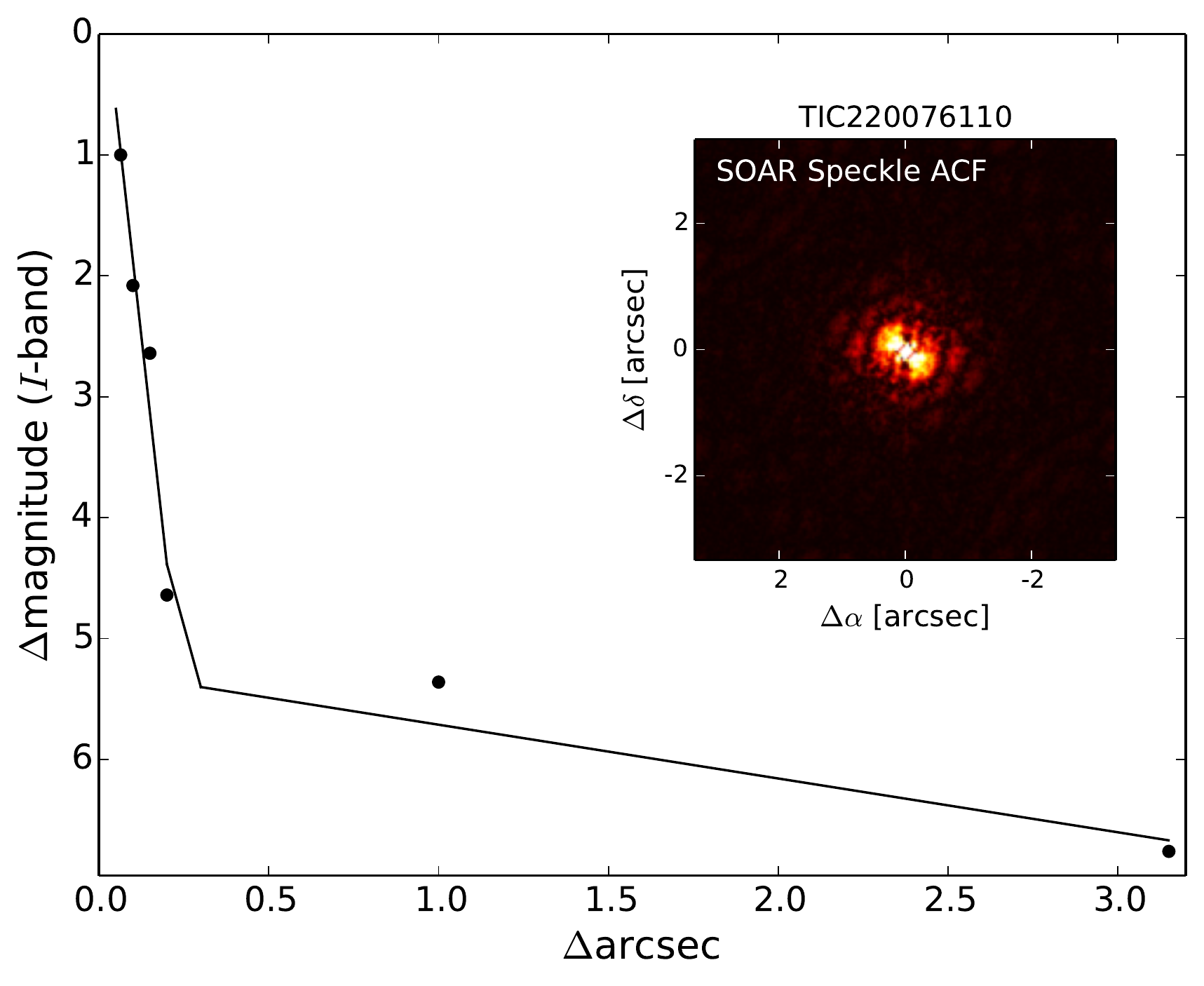}
\includegraphics[width=0.32\linewidth]{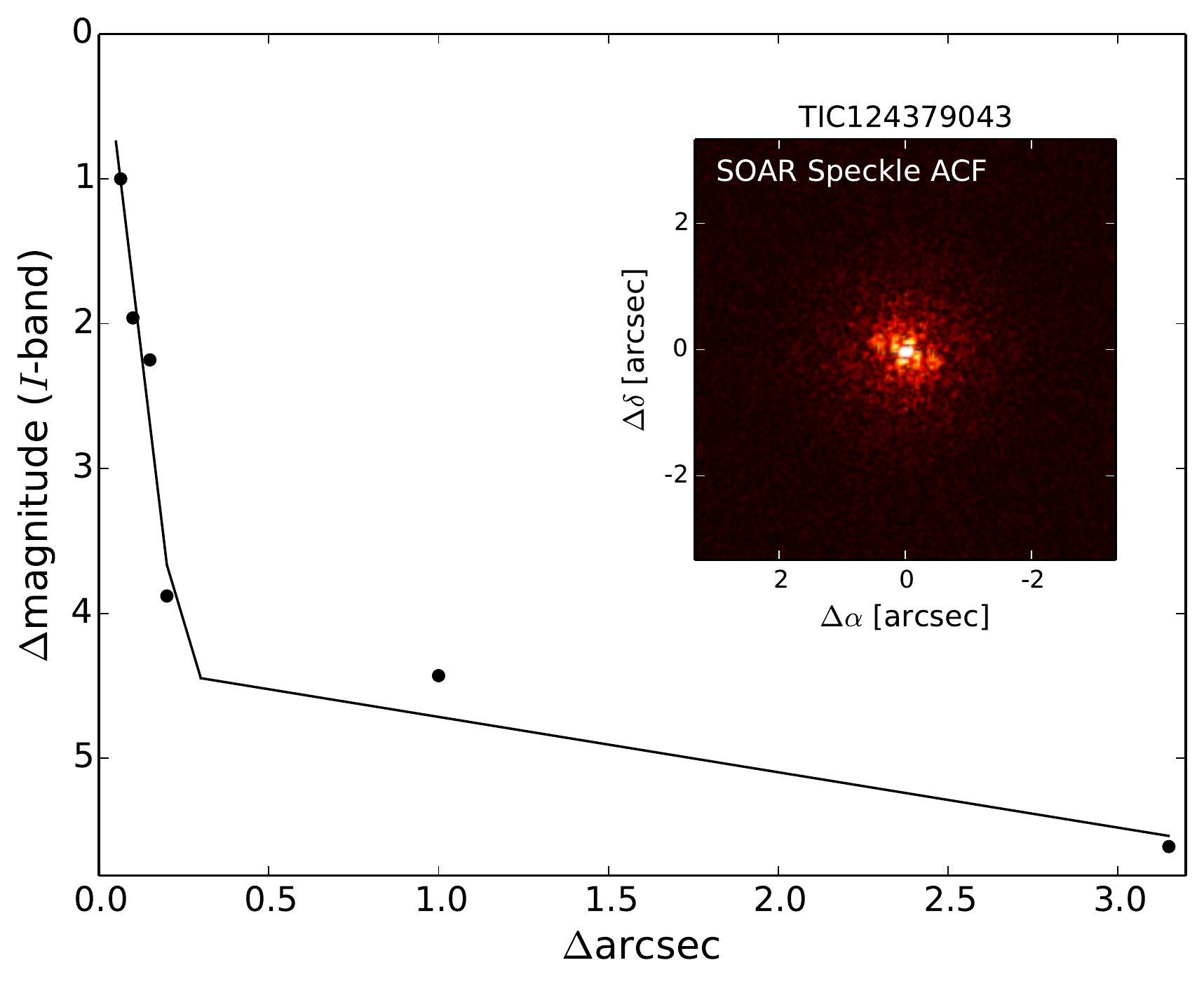} \\
\includegraphics[width=0.32\linewidth]{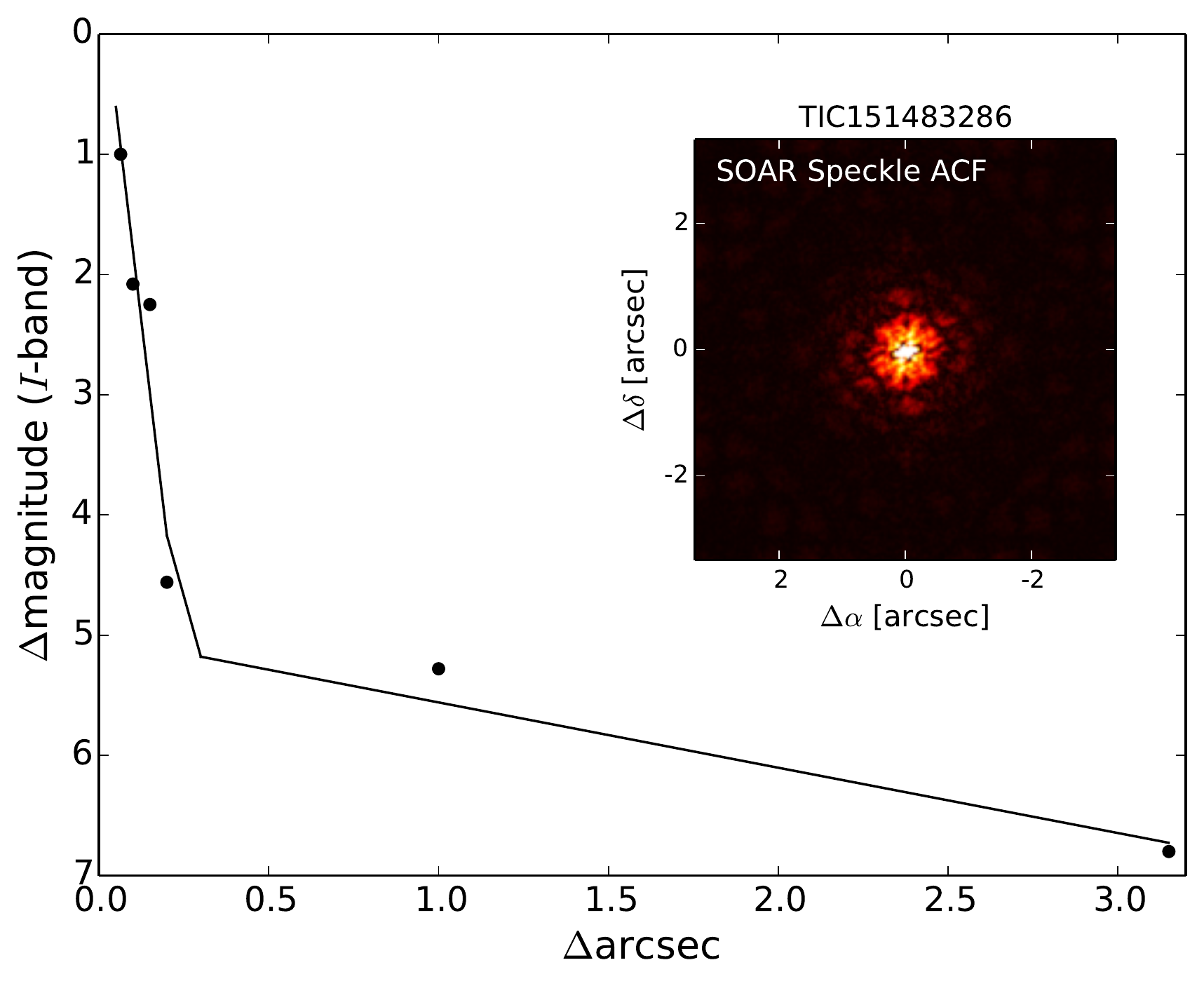}
\includegraphics[width=0.32\linewidth]{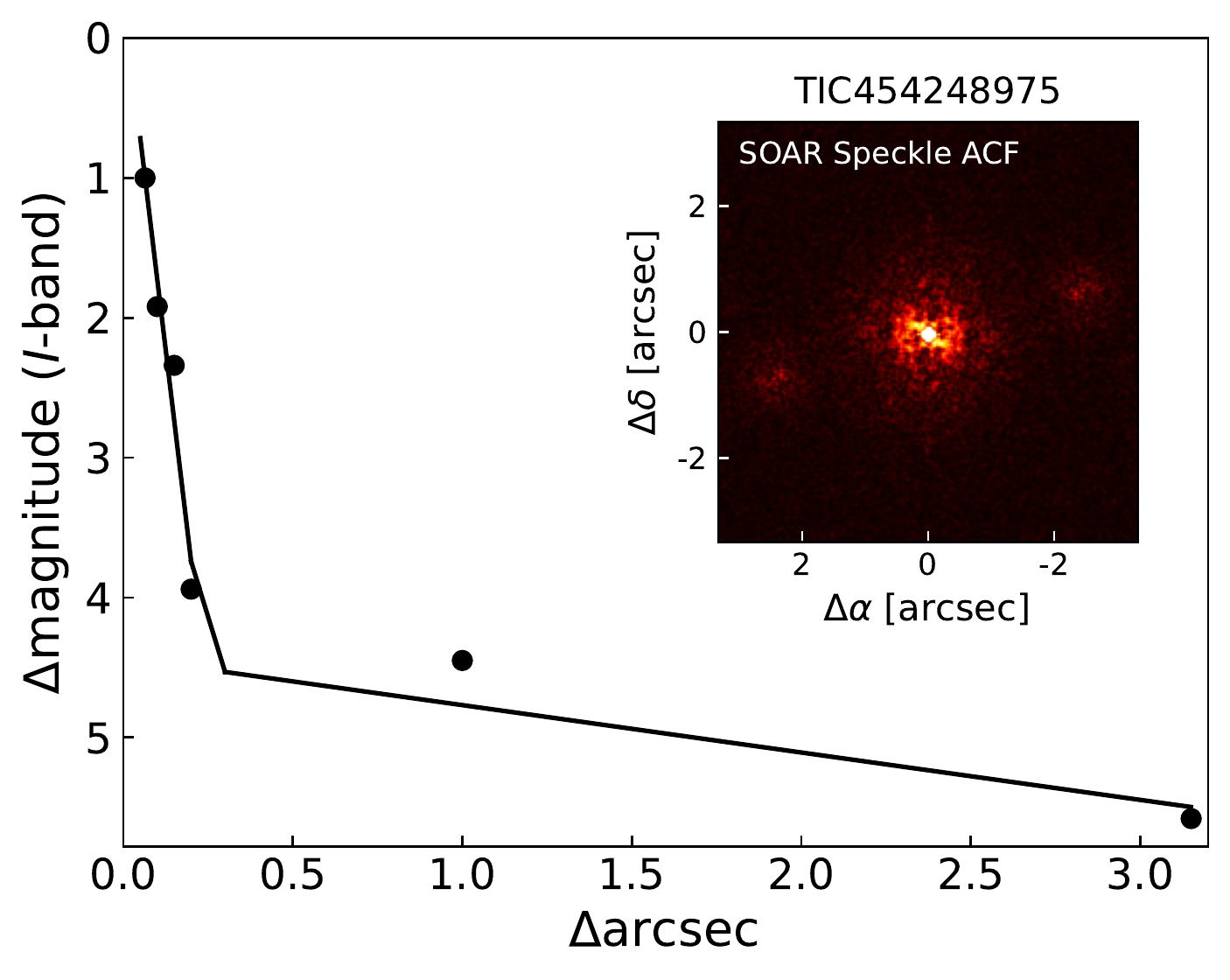}
\includegraphics[width=0.32\linewidth]{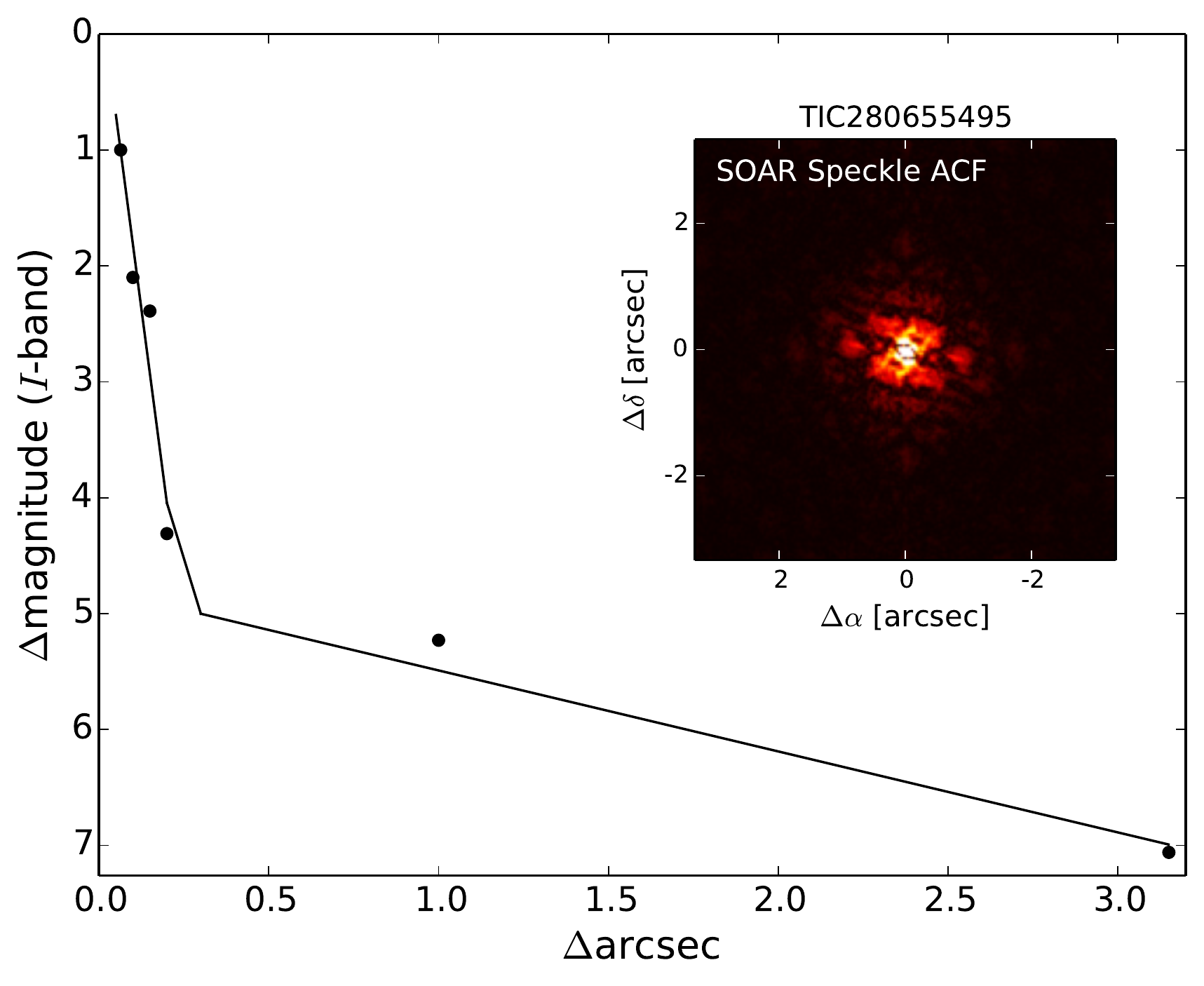} \\
\includegraphics[width=0.32\linewidth]{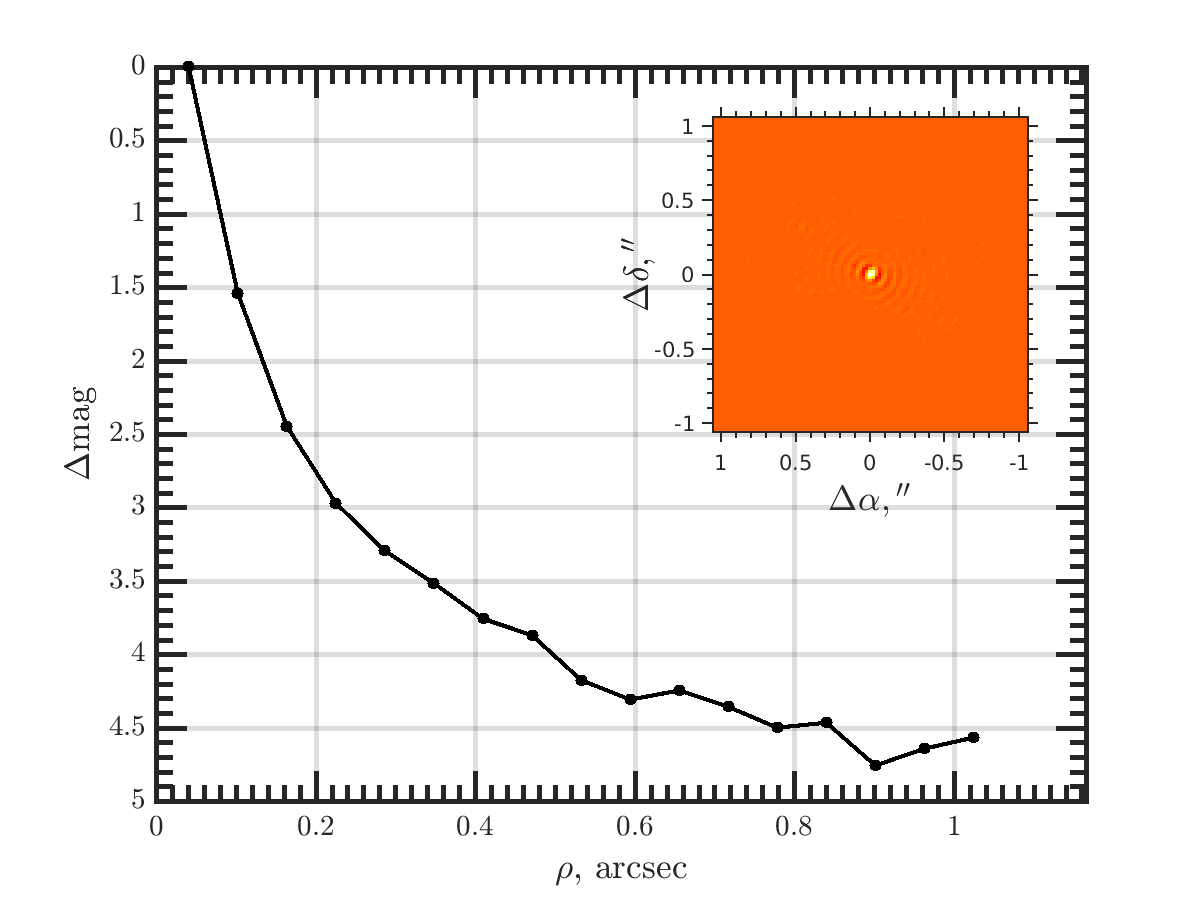}
\includegraphics[width=0.32\linewidth]{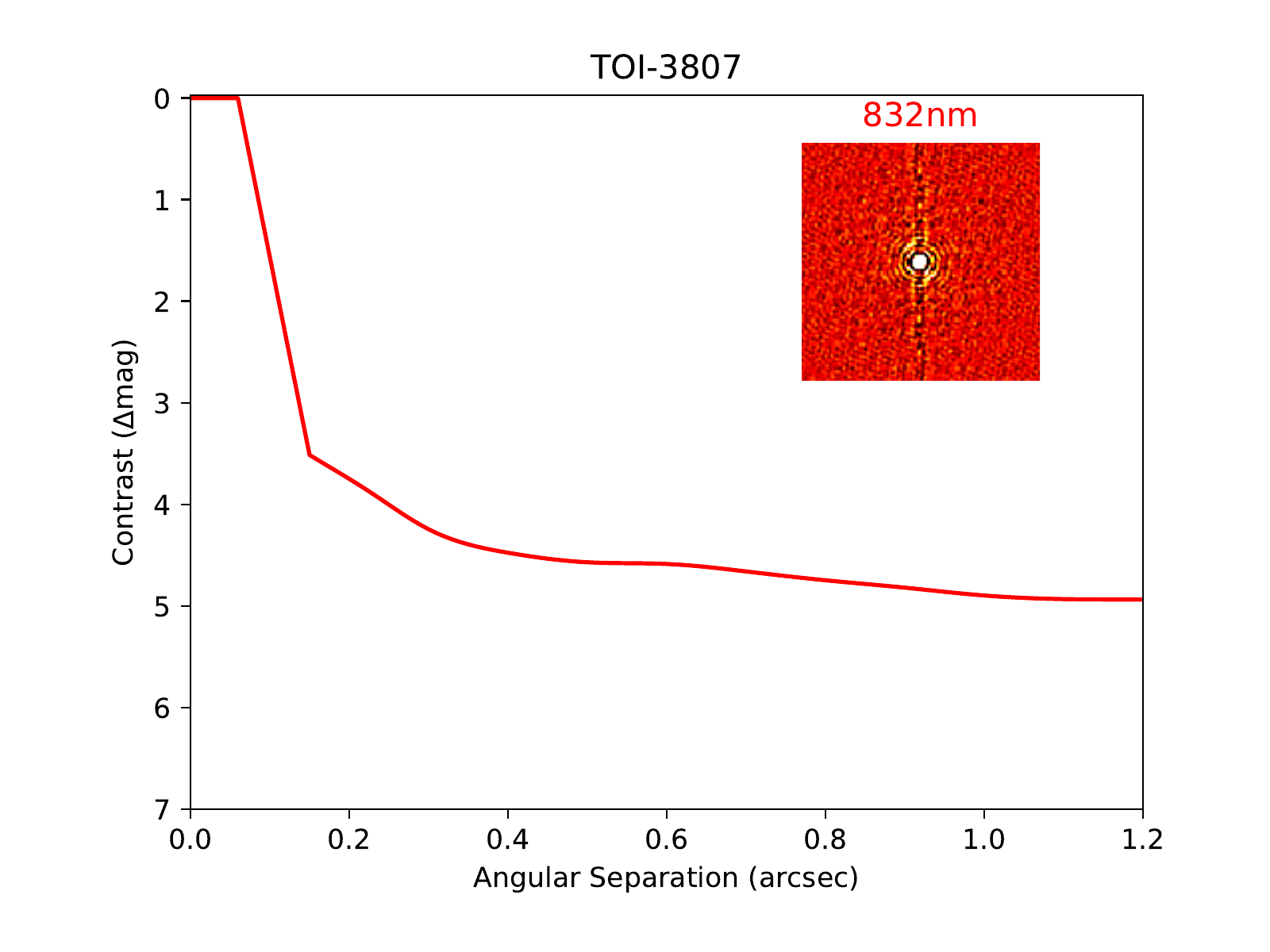}
\includegraphics[width=0.32\linewidth]{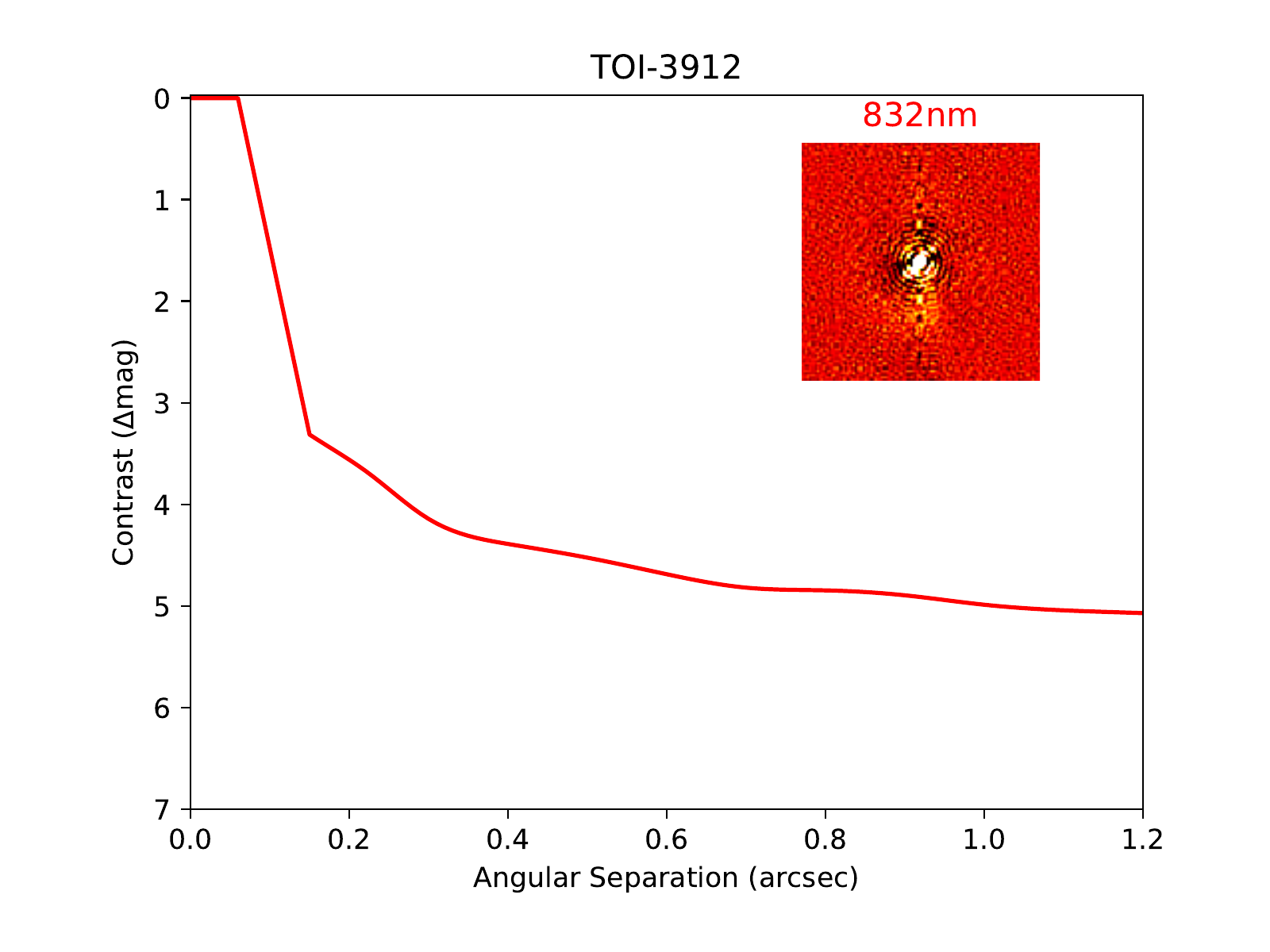}
\caption{High-Resolution imaging of hot Jupiter hosts described in this paper.
From top to bottom, left to right:
\textbf{Row 1:}
Gemini-South Zorro, Palomar PHARO, and SOAR HRCam observations of TOI-2364;
\textbf{Row 2:}
NESSI observations of TOI-2587; SOAR HRCam observations of TOI-2796 and TOI-2803;
\textbf{Row 3:}
SOAR HRCam observations of TOI-2818, TOI-3023 and TOI-3364;
\textbf{Row 4:}
SAI-2.5m Speckle Polarimeter observations of TOI-3688; NESSI observations of TOI-3807 and TOI-3912. \\
\textit{Note:} The inset reconstructed images from NESSI show a field $2"\times2"$ centered on the target.}
\end{figure*}
\makeatletter\onecolumngrid@pop\makeatother
\clearpage

\makeatletter\onecolumngrid@push\makeatother
\clearpage
\begin{figure*}
\centering
\includegraphics[width=0.32\linewidth]{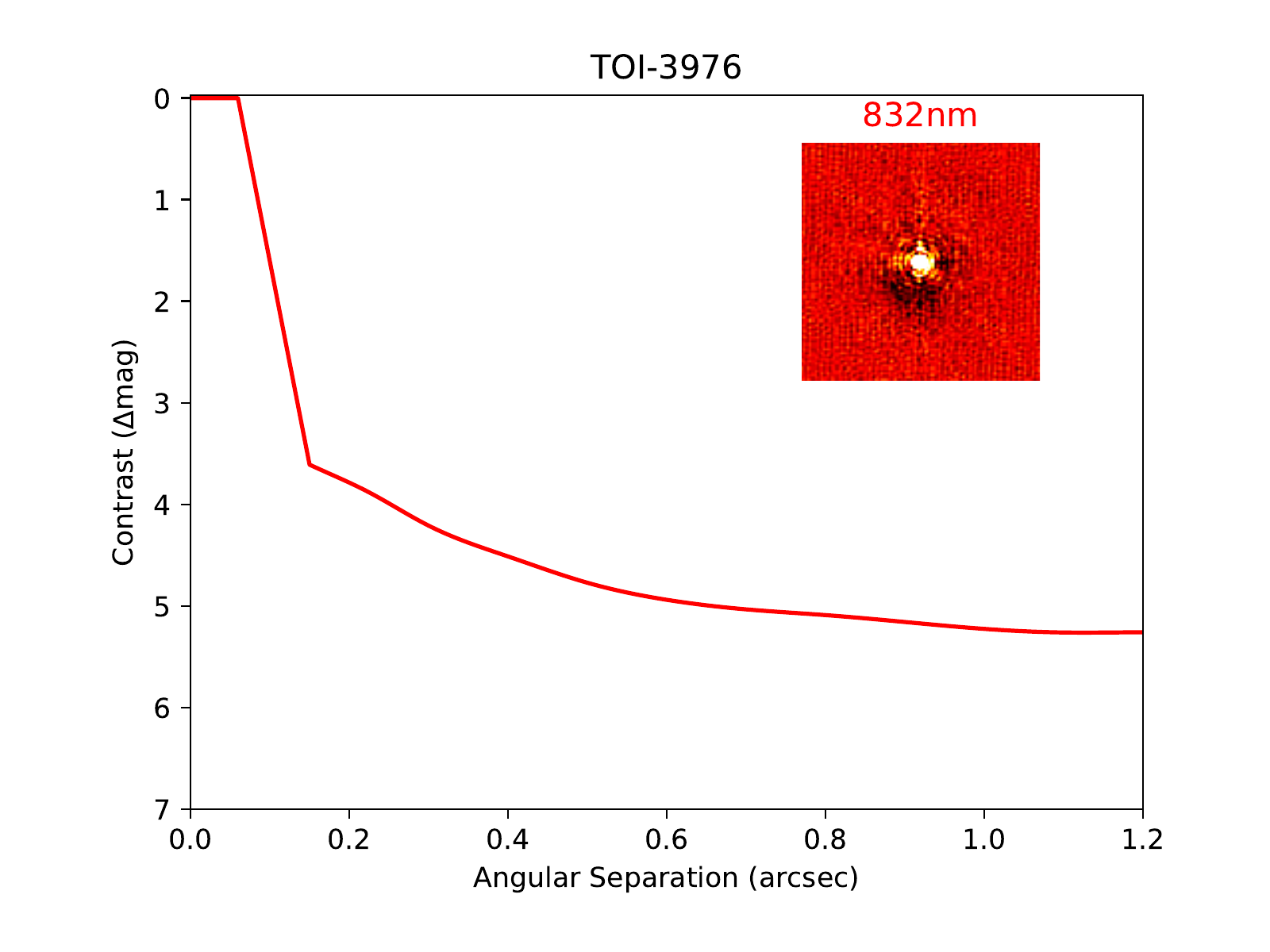}
\includegraphics[width=0.32\linewidth]{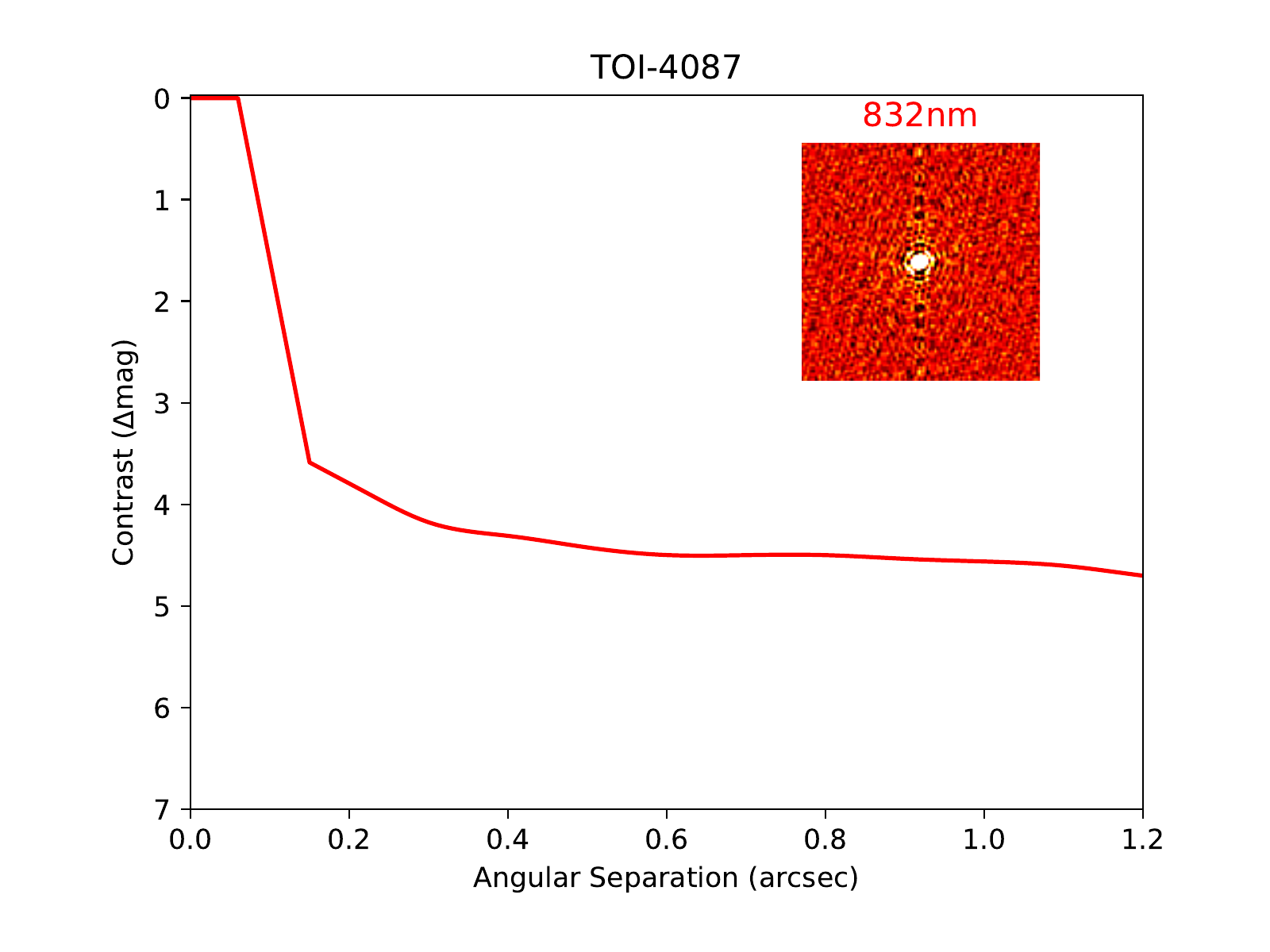}
\includegraphics[width=0.32\linewidth]{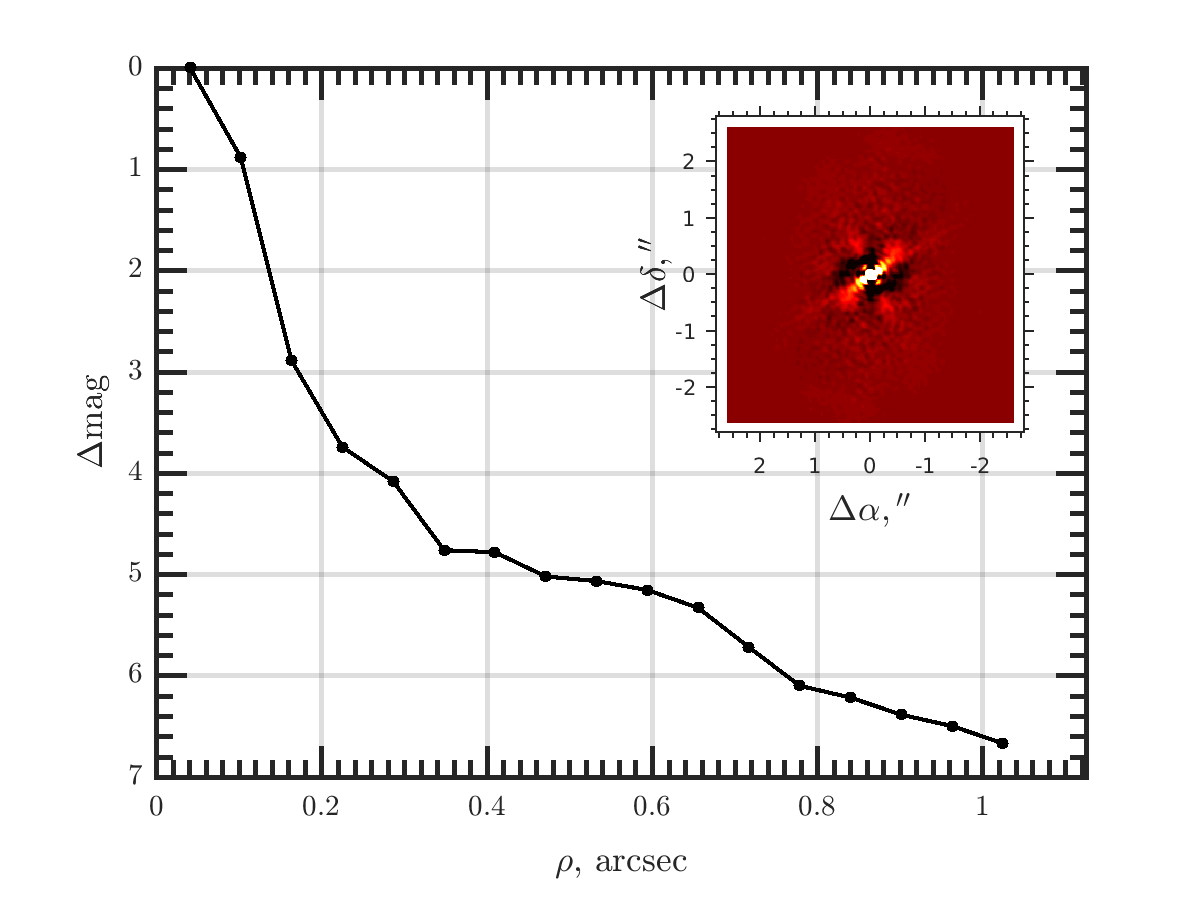} \\
\includegraphics[width=0.32\linewidth]{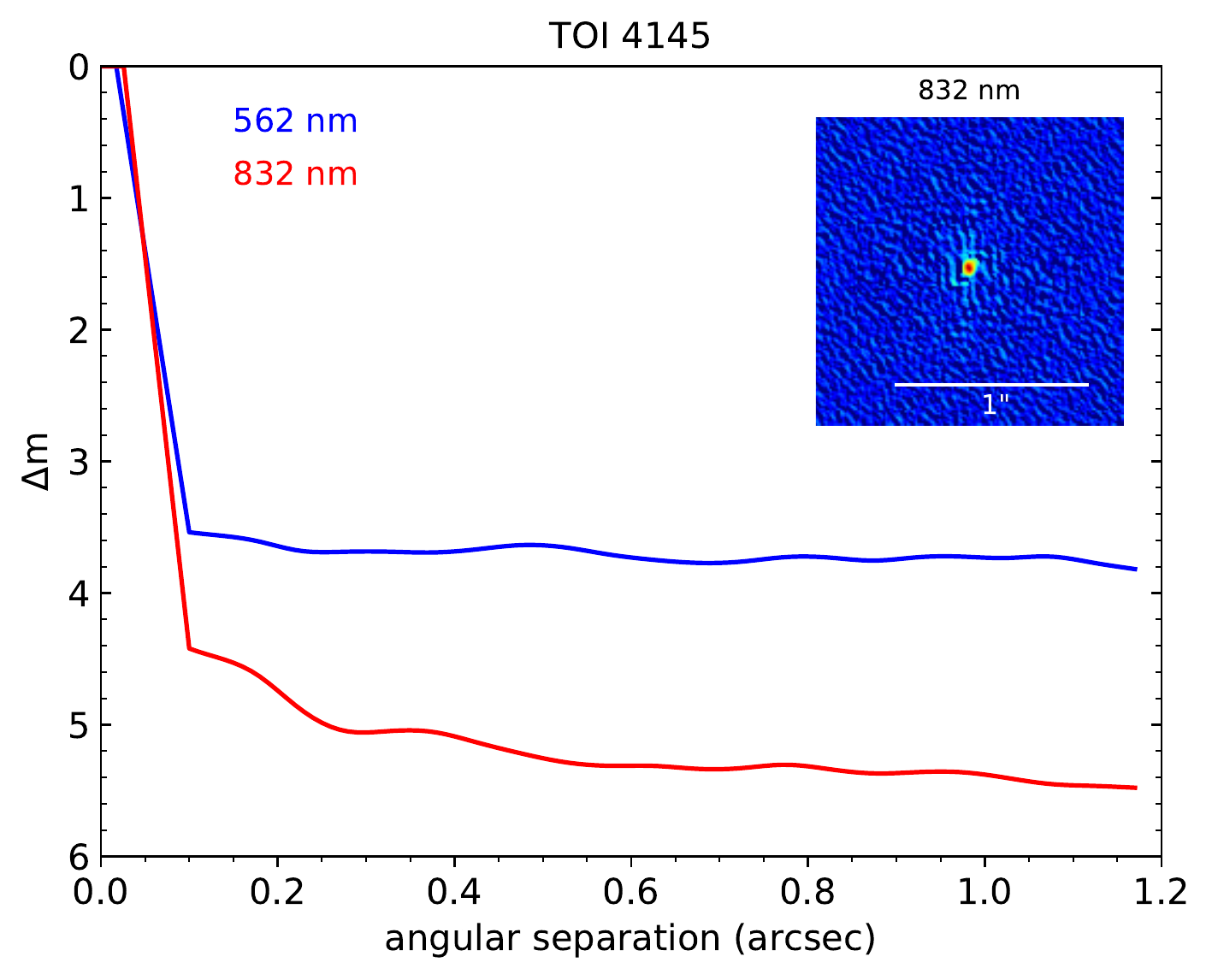}
\includegraphics[width=0.32\linewidth]{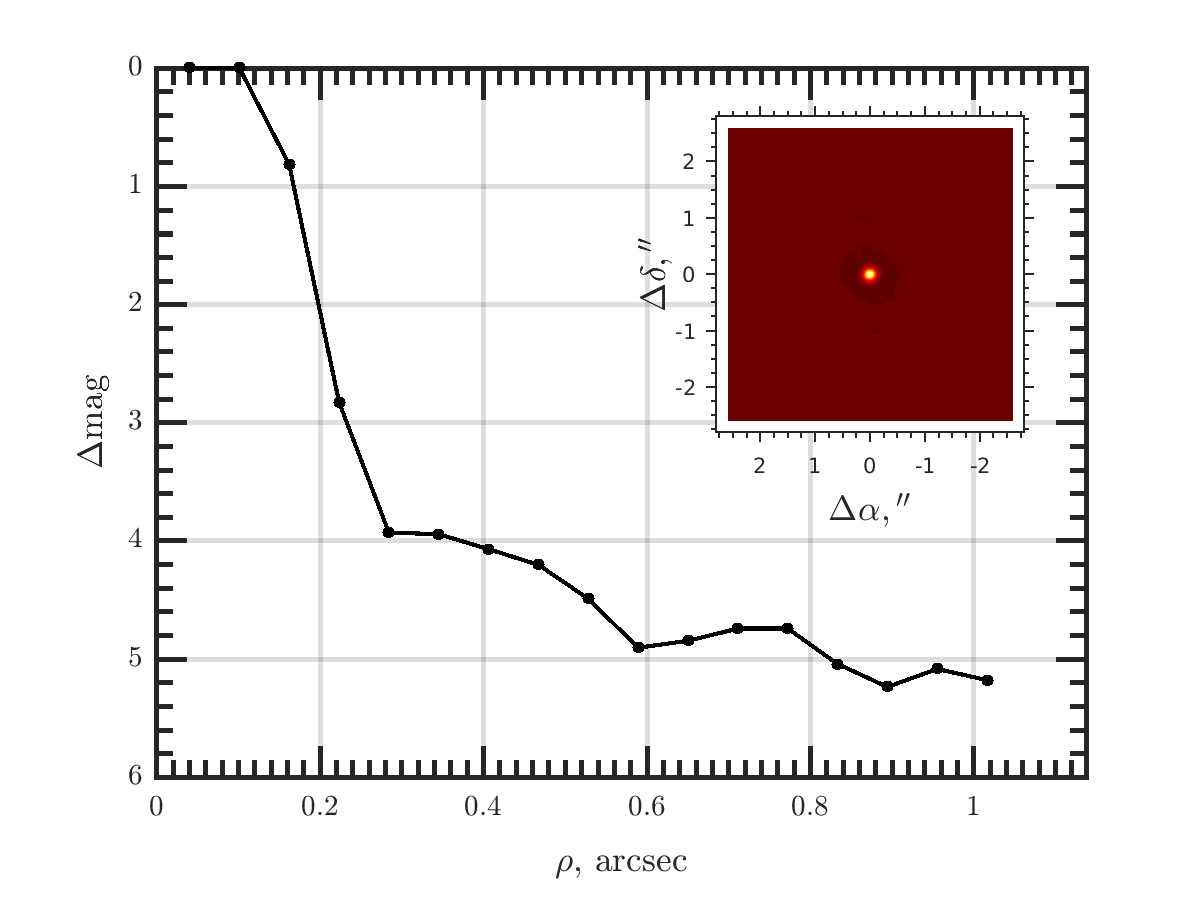} \\
\includegraphics[width=0.29\linewidth]{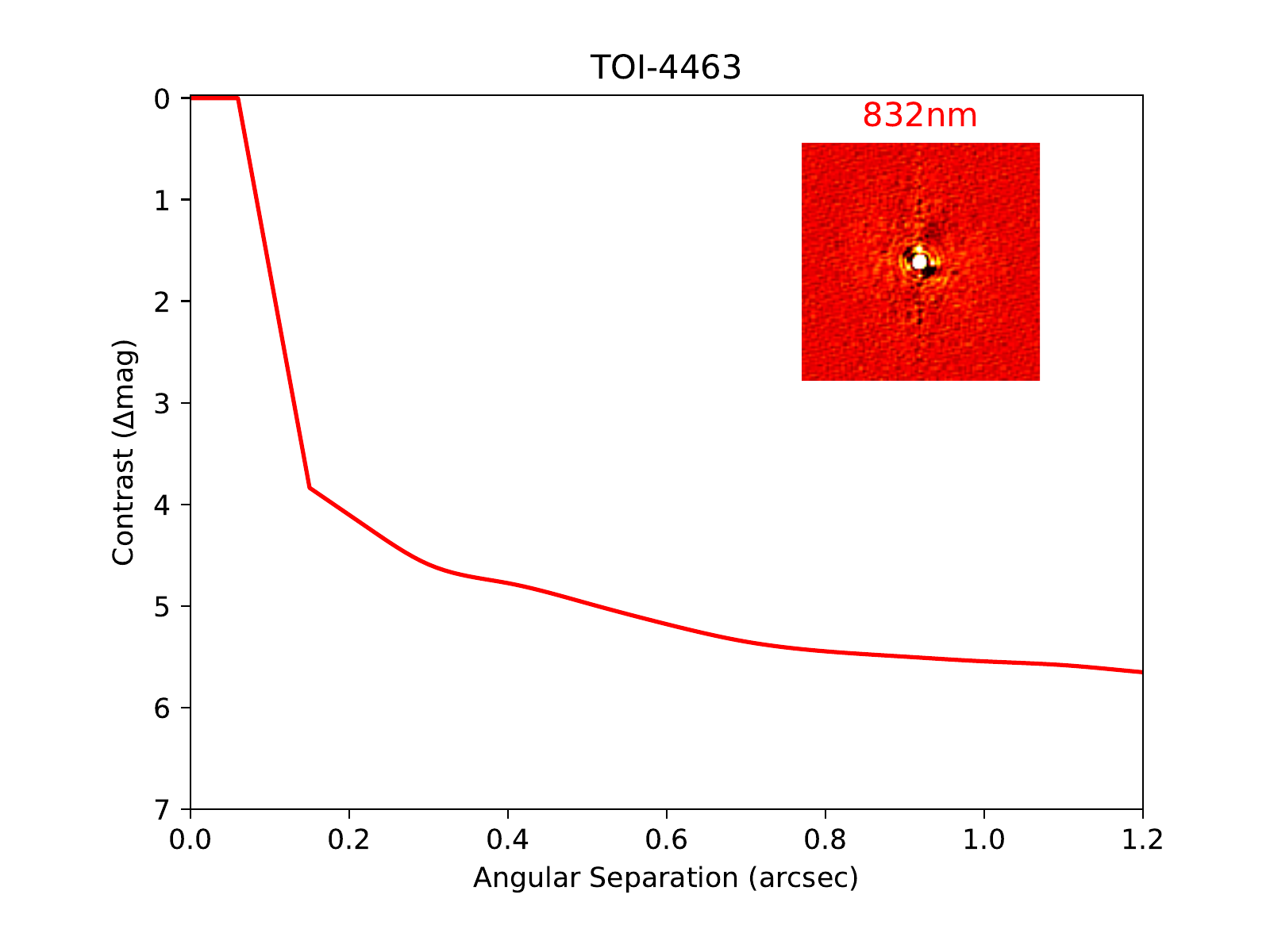}
\includegraphics[width=0.34\linewidth]{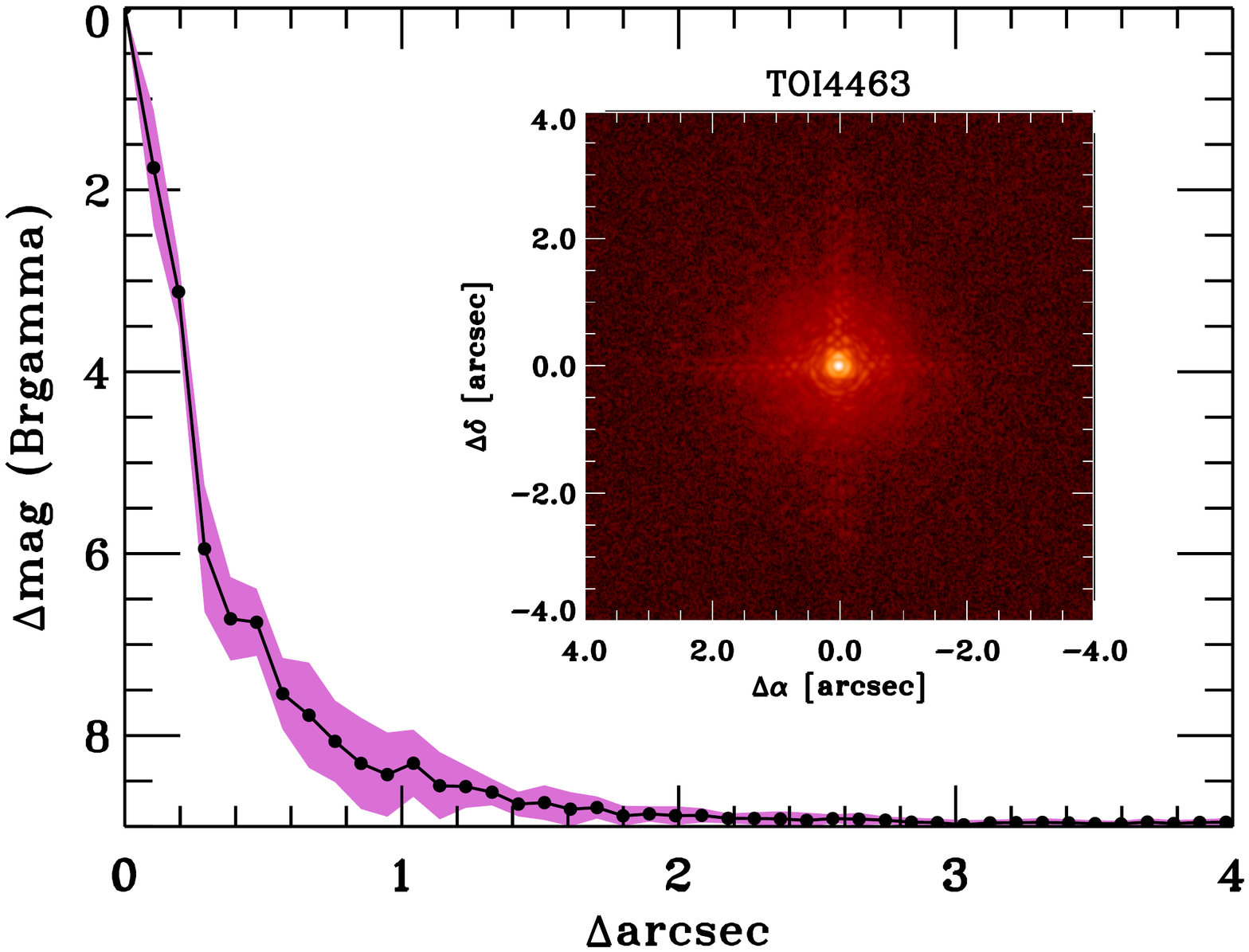}
\includegraphics[width=0.34\linewidth]{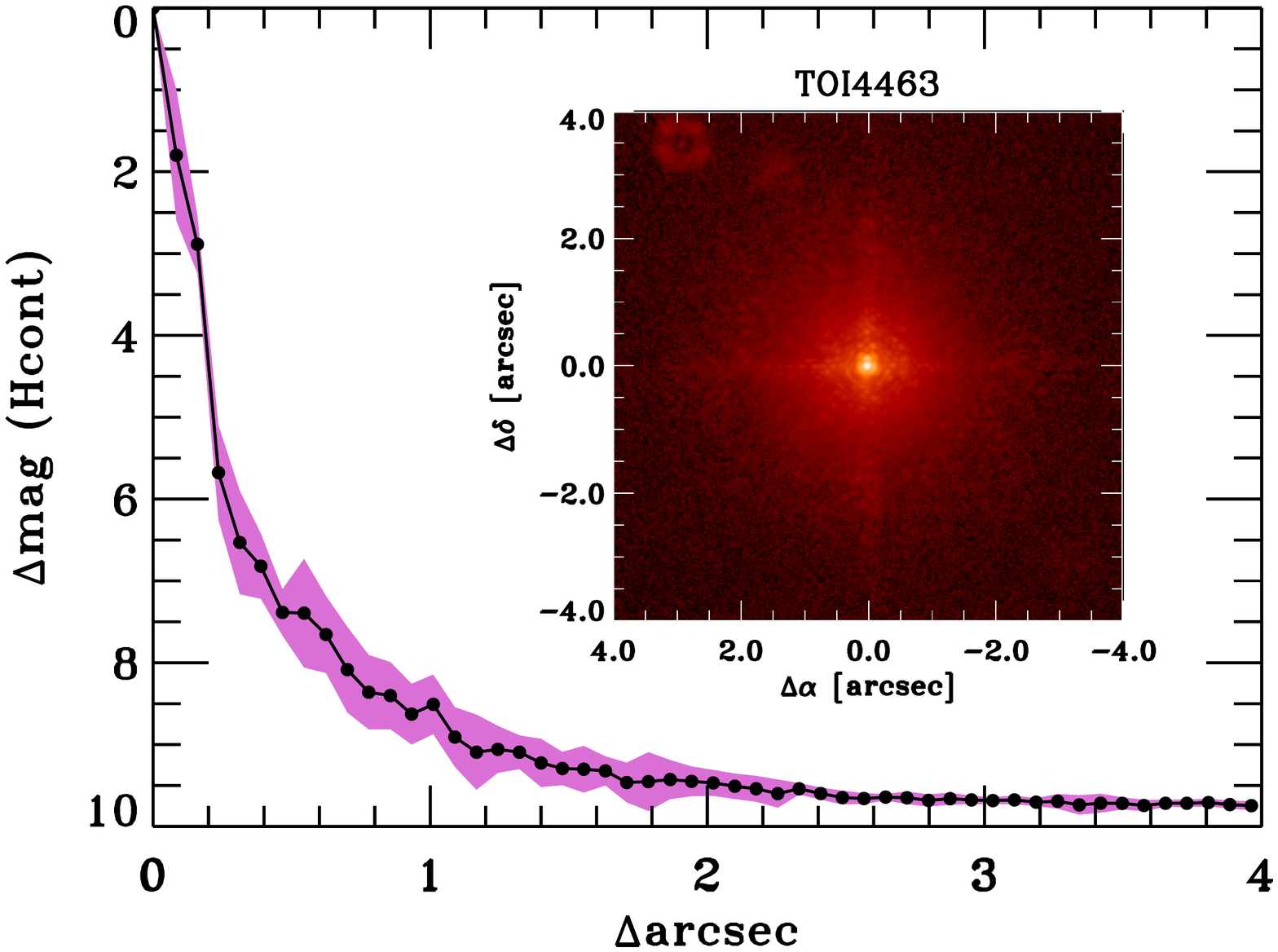} \\
\includegraphics[width=0.32\linewidth]{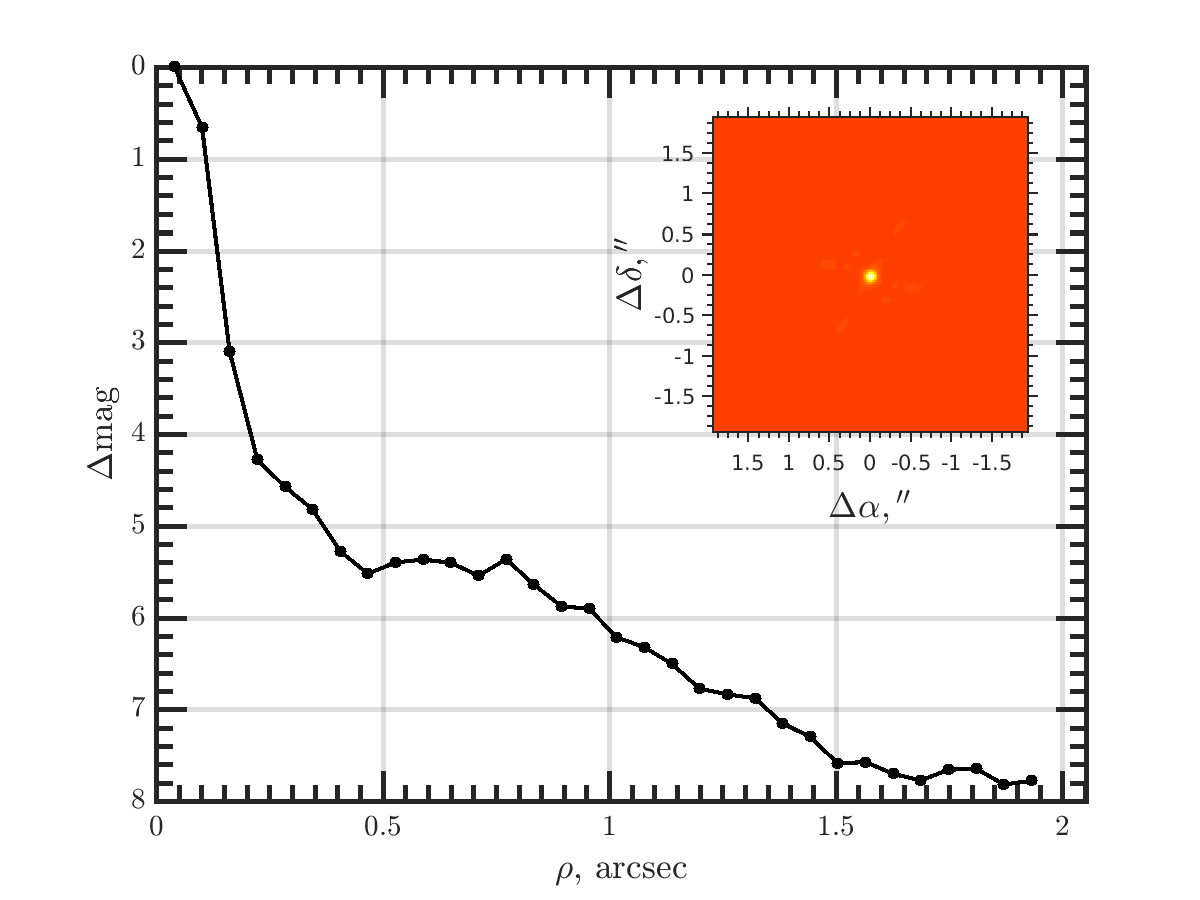}
\includegraphics[width=0.32\linewidth]{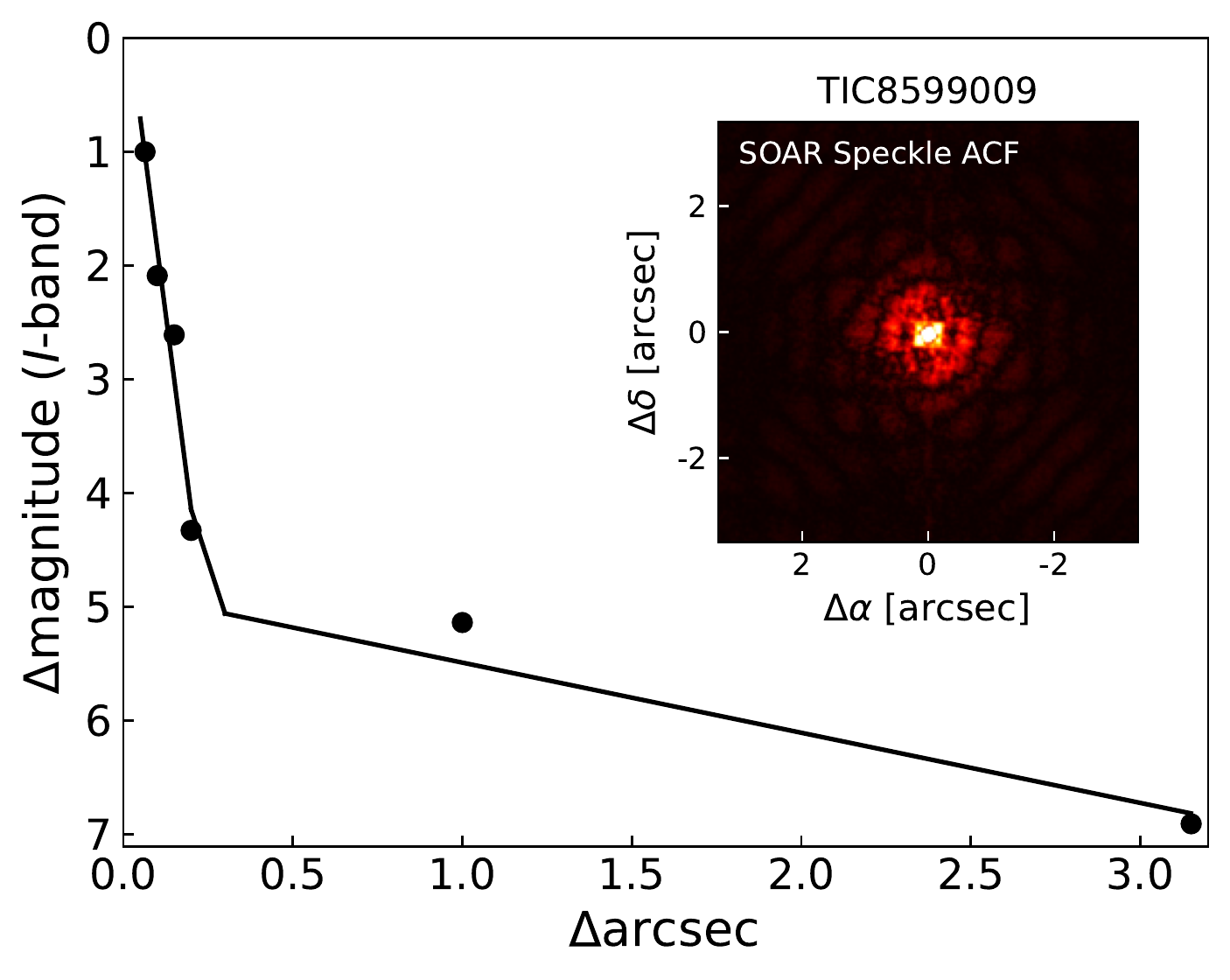} \\
\caption{High-Resolution imaging of hot Jupiter hosts described in this paper (continued).
From top to bottom, left to right:
\textbf{Row 1:}
NESSI observation of TOI-3976; NESSI and SAI Speckle Polarimeter observations of TOI-4087;
\textbf{Row 2:}
Gemini-North 'Alopeke and SAI Speckle Polarimeter observations of TOI-4145;
\textbf{Row 3:}
NESSI, Palomar PHARO Br$\gamma$ and H\textit{cont} observations of TOI-4463;
\textbf{Row 4:}
SAI Speckle Polarimeter and SOAR HRCam observations of TOI-4463. \\
\textit{Note:} All imaging data used in this paper are available via ExoFoP\textsuperscript{\ref{footnote:exofop_url}.}
\label{fig:high_res_imaging_1}}
\end{figure*}
\makeatletter\onecolumngrid@pop\makeatother
\clearpage
\end{subfigures}

\subsection{High-Resolution Spectroscopy} \label{ssec:spec}

In order to confirm each planetary candidate and measure its mass, we obtained high-resolution spectroscopy of their host stars for the purpose of measuring precise relative radial velocities (RVs).
We obtained 5--13 observations per target, scheduled primarily at orbital quadrature when possible, with the goal of obtaining at least a $\approx4\sigma$ mass measurement of the planetary companion.
These observations are summarized in Table \ref{tab:rvs}, with the full table of extracted RVs provided in machine-readable form as supplementary material to this article.

The instruments used for these observations were the Planet Finder Spectrograph (PFS) on the Magellan II Clay 6.5m telescope \citep{PFS_Crane2006,PFS_Crane2008,PFS_Crane2010}; the High Resolution Echelle Spectrometer (HIRES; \citealt{HIRES_Vogt94}) on the Keck-I 10m telescope; the NEID spectrograph \citep{NEID_Schwab2016,NEID_Halverson2016} on the WIYN 3.5m telescope at Kitt Peak National Observatory (KPNO); the CTIO High Resolution Spectrometer (CHIRON; \citealt{CHIRON_Tokovinin2013,CHIRON_Paredes2021}) on the CTIO 1.5m telescope; and the Tillinghast Reflector Echelle Spectrograph (TRES; \citealt{TRES_Furesz2008}) on the FLWO 1.5m Tillinghast Reflector.
Our observing strategy and data reduction procedures are detailed in Paper I, but we also describe them briefly here.

The HIRES spectra were observed through the queue organized by the California Planet Search (CPS; \citep{CPS_Howard2010,CPS_Howard2016}), using the standard CPS observing setup and data reduction procedures. We used the matched template technique from \citet{CPS_Dalba2020a} to extract the RVs without the need for an expensive high S/N template, instead using an archival HIRES template spectrum matched to a low S/N reconnaissance observation of the target. This procedure increases the RV scatter by $\approx 4.7$~\ms, which we add in quadrature to the instrumental uncertainties.

The NEID spectra were observed through the NEID queue in NEID's high-resolution (HR) mode.
The data were reduced using v1.1.2--v1.1.4 of the standard NEID Data Reduction Pipeline (NEID-DRP)\footnote{\url{https://neid.ipac.caltech.edu/docs/NEID-DRP}}, which extracts RVs via cross-correlation with a stellar line mask following procedures developed by \citet{Baranne1996,Pepe2002}.
We note that the NEID data from the nights of May 9th and May 10th were affected by an RV drift due to the failure of the Fabry-Perot etalon laser used to calibrate nightly drifts.
Using observations of standard stars taken on those nights, we estimated the offsets to be $35.8\pm1.2$~\ms and $31.5\pm1.3$~\ms respectively, and used these to correct one observation each of TOI-3807 and TOI-4087 taken on those nights.

The PFS observations were made in 3x3 binning mode with an iodine cell. An additional high S/N iodine-free template observation was obtained and used in the pipeline from \citet{PFS_Butler1996} to extract RVs.

The CHIRON spectra were taken using the echelle spectrograph with an image slicer, and bracketed with calibration observations of a ThAr lamp. These spectra were reduced using the standard CHIRON pipeline, and RVs extracted via least-squares deconvolution \citep{CHIRON_Donati1997,CHIRON_Zhou2020}.
The TRES observations were reduced and RVs extracted using the pipeline described by \citet{TRES_Buchhave2010} and \citet{TRES_Quinn2012}.

For the PFS, HIRES, NEID, and CHIRON spectra, we also used the procedures from \citet{Hartman2019} to measure bisector inverse slopes (BIS) from the iodine-free orders. These measurements were used to check for spurious RV variations that are actually due to variations in the spectral line profiles, rather than orbital motion.
For all the systems presented here, we did not observe any significant correlations between the RVs and BIS measurements.

\makeatletter\onecolumngrid@push\makeatother
\clearpage
\begin{deluxetable*}{ccrrcc}
\tablecolumns{6}
\tablecaption{Summary of Radial-Velocity Measurements \label{tab:rvs}}
\tablehead{
    \colhead{Target} & \colhead{Instrument} & \colhead{N$_\mathrm{obs}$} & \colhead{Median $\sigma_\mathrm{RV}$ } & \colhead{First Observation Date} & \colhead{Last Observation Date} \\
    &  &  & \colhead{(m/s)$^{a}$} & \colhead{(UT)} & \colhead{(UT)}
}
\startdata
\input{rv_summary}
\enddata
\tablenotetext{a}{Median instrumental RV uncertainty for each target and instrument.}
\tablecomments{The complete table of RV measurements is available in machine-readable form as Data behind the Figure for Figure Set \ref{fig:all_multiplots}, provided as relative RVs with an arbitrary target- and instrument-specific offset subtracted.}
\end{deluxetable*}  
\makeatletter\onecolumngrid@pop\makeatother
\clearpage


\subsection{Rossiter-McLaughlin Effect of TOI-1937}

We observed a transit of TOI-1937A\,b on the night of 2020 Dec 29 both
spectroscopically and photometrically, in order to measure the projected
stellar obliquity through the Rossiter-McLaughlin (RM; \citealt{Rossiter1924,McLaughlin1924}) effect.
The photometric observations
were acquired from El Sauce, and the spectroscopy was acquired using
Magellan/PFS.
The photometric results are shown in
Figure~\ref{fig:toi1937_multiplot}, with the transit occurring at the
expected time.
The PFS observations were made in the same manner as described in \S\ref{ssec:spec},
with thirteen exposures of twelve minutes each, covering the full transit duration.

\begin{figure}[t]
    \begin{center}
        \leavevmode
            \includegraphics[width=0.49\textwidth]{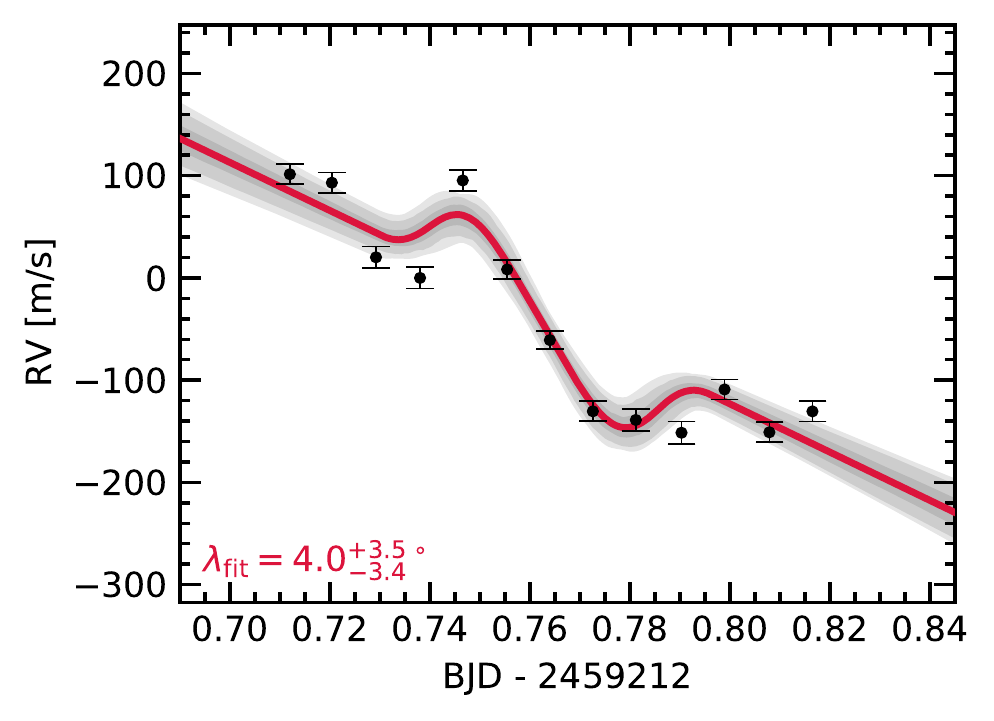}
    \end{center}
    \vspace{-0.6cm}
    \caption{
    Radial velocities of TOI-1937A from the night of 2020 Dec 29.
    The best-fit model of the Rossiter-McLaughlin effect is shown in
    red, with the corresponding 1-,2-, and 3-$\sigma$ uncertainties in
    gray.
    \label{fig:toi1937_rm}
    }
\end{figure}

The resulting velocities are shown as a function of time in
Figure~\ref{fig:toi1937_rm}.
The expected RM anomaly has an amplitude $\Delta v_{\rm RM} \approx \delta \cdot v\sin
i \cdot \sqrt{1-b^2} \approx 52\,{\rm m\,s}^{-1}$, which agrees
visually with the data.  We therefore fitted the RVs using the
\citet{hirano_2010,hirano_2011} models for the RM effect.  We assumed
a Gaussian prior on $v\sin i$ and $a/R_\star$ from Table~\ref{tab:fitted_props},
and also allowed for a white-noise jitter term to be added in
quadrature to the measurement uncertainties.  We fixed the
limb darkening using the $V$-band tabulation from
\citet{Claret2011}.  We then varied the sky-projected
obliquity, the projected stellar equatorial velocity, and the Gaussian
dispersion of the spectral lines, along with a linear trend accounting
for the out-of-transit variability.
Given that the \TESS light-curves (Fig. \ref{fig:toi1937_tess_lc}) show
photometric variability only at the level of $\sim1\%$, with no evidence
of spot-crossing events in the simultaneous transit photometry, we did not
account for starspot-induced variability in the RM model \citep[e.g.,][]{Oshagh2013,Oshagh2018}.

The model shown in
Figure~\ref{fig:toi1937_rm} suggests a good fit to the data for a model
that has a projected stellar obliquity that is well-aligned with the orbit
of TOI-1937A\,b: $\lambda = 4.0 \pm 3.5^\circ$.

\subsection{Catalog Photometry and Astrometry}

To ensure we have a complete view of each planetary system, we used the \texttt{astroquery} package \citep{Astroquery_Ginsburg2019} to gather information about each target from the literature.
We obtained photometric and astrometric observations from the \TESS Input Catalog (TIC; \citealp{TIC_Stassun2018,TIC_Stassun2019}), \Gaia DR3 \citep{GaiaEDR3_Brown2021,GaiaEDR3_Riello2021,GaiaEDR3_Lindegren2021}, \textit{2MASS} \citep{TMASS_Cutri2003,TMASS_Skrutskie2006}, \textit{WISE} \citep{WISE_Cutri2012}, and Tycho-2 \citep{Tycho2_Hog2000} catalogs.
We corrected the \Gaia photometry for the known parallax zero-point error as described in  \citet{GaiaEDR3_Lindegren2021}\footnote{\url{https://gitlab.com/icc-ub/public/gaiadr3_zeropoint}}.
These data are shown in Table \ref{tab:stellar_props}. 

\begin{rotatepage}
\movetabledown=1.5in
\begin{rotatetable}
\begin{deluxetable*}{lcccccc}
\tablecaption{Catalog Photometry and Astrometry of Planet Host Stars \label{tab:stellar_props}}
\tabletypesize{\small}
\input{target_props_0}
\tablenotetext{a}{The 2MASS and WISE photometry of TOI-1937 are a blend of both the primary and secondary components (\S\ref{ssec:companions}).}
\tablenotetext{b}{The WISE photometry of TOI-2583 is a blend of both primary and secondary components.}
\end{deluxetable*}
\end{rotatetable}
\addtocounter{table}{-1}

\movetabledown=1.5in
\begin{rotatetable}
\begin{deluxetable*}{lcccccc}
\tablecaption{\textit{(Continued)}}
\tabletypesize{\small}
\input{target_props_1}
\tablenotetext{c}{All catalog photometry of TOI-2977 are a blend of both primary and secondary components.}
\end{deluxetable*}
\end{rotatetable}
\addtocounter{table}{-1}

\movetabledown=1.5in
\begin{rotatetable}
\begin{deluxetable*}{lcccccc}
\tablecaption{\textit{(Continued)}}
\tabletypesize{\small}
\input{target_props_2}
\end{deluxetable*}
\end{rotatetable}
\addtocounter{table}{-1}

\movetabledown=1.5in
\begin{rotatetable}
\begin{deluxetable*}{lcccccc}
\tablecaption{\textit{(Continued)}}
\tabletypesize{\small}
\input{target_props_3}
\tablerefsmod{\textbf{Sources:} (1) - \Gaia DR3 \citep{GaiaEDR3_Brown2021};
(2) - \TESS Input Catalog \citep{TIC_Stassun2019};
(3) - Tycho-2 \citep{Tycho2_Hog2000};
(4) - 2MASS \citep{TMASS_Cutri2003,TMASS_Skrutskie2006};
(5) - WISE \citep{WISE_Cutri2012}.}
\tablecommentsmod{The catalog photometry presented here has not been corrected for contamination by nearby stellar companions (\S\ref{ssec:companions}).\\
The data in this table is available in machine-readable form.}
\end{deluxetable*}
\end{rotatetable}
\end{rotatepage}

\pdfpageattr{}
\section{Stellar Characterization} \label{sec:stellar_char}

\subsection{Spectroscopic Parameters} \label{ssec:spec_char}

We used the high-resolution spectra of each target to perform an initial characterization of the stellar properties of these planet hosts.

For stars with observations from TRES, spectroscopic atmospheric properties were derived using the Stellar Parameter Classification code (SPC; \citealt{SPC_Buchhave2012}), which cross-correlates the observed spectrum against a grid of synthetic spectra from \citet{Kurucz1993}.
We classified each observation of the same target separately, adopting the median value in each parameter and using the scatter in the results as the uncertainty, with an error floor of 50\,K in \Teff, 0.1~dex in \logg, 0.08~dex in \feh, and 0.5~\kms in \vsini.
The CHIRON spectra were analyzed by matching against a library of $\sim$10,000 observed spectra that were previously classified using SPC, as described by \citet{CHIRON_Zhou2020}.
As we did for the TRES observations, we adopted the median value and scatter of the results derived from the spectra of the same target, with error floors of 50\,K in \Teff, 0.1~dex in \logg, 0.1~dex in \feh, and 0.5~\kms in \vsini.

Because not all of our targets have spectra from TRES or CHIRON, we used the \texttt{SpecMatch-Emp} code \citep{SpecMatchEmp_Yee2017}\footnote{\url{https://github.com/samuelyeewl/specmatch-emp}} to derive spectroscopic parameters for all our targets, using the highest S/N iodine-free spectrum from PFS, HIRES, NEID, or CHIRON per object.
This code matches the target spectrum to a library of observed Keck/HIRES spectra from stars with empirically well-determined stellar properties, yielding \Teff, \Rstar, and \feh for the target with median uncertainties of $\sigma(\Teff) = 110$~K, $\sigma(\Delta \Rstar/\Rstar) = 15\%$ and $\sigma(\feh) = 0.09$~dex, as derived from a cross-validation test.
To derive \vsini, we used \texttt{SpecMatch-Synth}\footnote{\url{https://github.com/petigura/specmatch-syn}} \citep{SpecMatchSynth_Petigura2015}, which performs a similar matching procedure using the \cite{Kurucz1993} synthetic spectral library.
We combined the \vsini and macroturbulent velocity $v_{\mathrm{mac}}$ (assumed using the relationship with stellar \Teff from \citealt{Valenti2005}) from this code together with \Teff, \Rstar, and \feh from \texttt{SpecMatch-Emp}, reporting these in Table \ref{tab:spec_props}.

We decided to adopt the \texttt{SpecMatch} spectroscopic parameters for the rest of our analysis, using them as priors for our global \Exofast fits, since we could apply the same code homogeneously to all our targets.
We compared the \texttt{SpecMatch}-derived parameters with those from SPC when available, and found agreement in almost all cases to better than 2-$\sigma$.

\subsubsection{Iterative Solution for TOI-2803} \label{ssec:toi2803_spec}
The case with the largest discrepancy was for TOI-2803, where \texttt{SpecMatch-Emp} reported $\feh = -0.43 \pm 0.09$~dex based on the PFS template spectrum, whereas SPC reported a metallicity of $-0.11 \pm 0.10$~dex based on two TRES spectra.
Furthermore, the spectroscopic stellar $\Teff$, $\logg$, and $\Rstar$ also differed from the results derived from the global \Exofast fits (\S\ref{sec:planet_char}).
The global fits provide better constraints on the stellar surface gravity (\logg) from the measured transit duration \citep{Seager2003}, as well as on the stellar $\Teff$ based on the \Gaia parallaxes and measured Spectral Energy Distribution (SED).

To resolve the discrepancy and arrive at a self-consistent solution, an iterative approach was required.
First, we fixed $\logg = 4.3$~dex (based on the fits to the transit light-curve) in the SPC analysis of the TRES spectrum.
We also performed a new analysis of the PFS spectrum using the Zonal Atmospheric Stellar Parameters Estimator (\texttt{ZASPE}; \citealt{ZASPE_Brahm2017}) code, also holding $\logg$ fixed.
\texttt{ZASPE} derives spectroscopic parameters using only the most sensitive spectral regions to compare the target with a grid of spectra computed with the \texttt{SPECTRUM} spectral synthesis program \citep{SPECTRUM_Gray1999} and the ATLAS model atmospheres \citep{ATLAS_Castelli2003}.
The results from these analyses are included in Table \ref{tab:spec_props}.
We then used only the spectroscopic metallicities as priors in two separate global \Exofast fits.

The global fit returned $\Teff = 6280\pm100$~K, $\logg = 4.30\pm0.01$~dex, $\feh = -0.11\pm0.07$~dex when the SPC TRES metallicity prior was used ($\feh = -0.11\pm0.08$~dex).
Meanwhile, when the \texttt{ZASPE} PFS metallicity prior was used ($\feh = -0.35\pm0.07$~dex), the global \Exofast fit required a hotter star ($\Teff = 6436 \pm 80$~K) than found by the spectroscopic analysis ($\Teff = 6341 \pm 122$~K).
Given that the results were most self-consistent when using the SPC TRES metallicity, we chose to adopt that set of stellar properties.


\subsection{Stars with Nearby Companions} \label{ssec:companions}

\subsubsection{\Gaia-detected Companions}

We investigated the stellar multiplicity of the twenty planet host stars by querying the \Gaia DR3 catalog \citep{GaiaDR3_Vallenari2022} for nearby stars within $30"$ with similar proper motions to the planet host.
Eight of our targets were found to have such companions, with angular separations between $1\farcs7$ and $20"$ away, and are listed in Table \ref{tab:gaia_comp_props}.
We also cross-matched our sample with the catalog of wide binaries from \citet{GaiaEDR3_Binaries_El-Badry2021}, who computed probabilities that each potential binary pair is due to a chance alignment, given their parallaxes, projected separation, and requiring that their proper motions are consistent with a Keplerian orbit.
All eight binary pairs we identified were also listed in the \citet{GaiaEDR3_Binaries_El-Badry2021} catalog, with chance alignment probabilities $< 3\times10^{-3}$, suggesting that they are bound.
In each case, the planet host star was the brighter of the two; as such, we refer to them subsequently as the ``A'' component.

For these eight cases, the secondary component was also identified in the TIC; hence any dilution of the TESS light curve due to their light would have already been accounted for.
In the cases of TOI-1937 and TOI-4145, the secondaries were located $2\farcs48$ and $1\farcs74$ from the primary respectively, and would have contaminated the ground-based light curves.
We therefore computed a dilution factor for each band using the spectral energy distribution (SED)-fitting procedures described in \S\ref{ssec:ao_companions} to correct the light curves for these targets.

\begin{longdeluxetable}{lcc}
\tablecaption{Properties of Stellar Companions \label{tab:gaia_comp_props}}
\tabletypesize{\footnotesize}
\tablehead{
    & \colhead{Primary} & \colhead{Secondary}
}
\startdata
\input{gaia_comp_props}
\enddata
\tablecommentsmodhalf{The angular and projected separation between the two components are from the catalog of \citet{GaiaEDR3_Binaries_El-Badry2021}. $T$ magnitude is from the TESS Input Catalog \citep{TIC_Stassun2018,TIC_Stassun2019}, while all remaining parameters, including systemic RVs, are drawn from \Gaia DR3 \citep{GaiaDR3_Vallenari2022,GaiaDR3_RVs_Katz2022}.}
\end{longdeluxetable}

\subsubsection{Companions from High-Resolution Imaging \label{ssec:ao_companions}}

For the majority of targets described in this paper, high-resolution imaging did not detect any stellar companions (Figure Set \ref{fig:high_res_imaging_nocomp}).

The SOAR HRCam speckle imaging of the target TOI-2977 (\S\ref{ssec:imaging}, Fig. \ref{fig:toi2977_imaging}) revealed a nearby star 1.7 magnitudes fainter in the $I_c$ band located $0\farcs77$ away from the primary.
This star was not identified in the \Gaia DR3 catalog \citep{GaiaDR3_Vallenari2022}.\footnote{\Gaia DR3 did identify a faint star about $2\farcs4$ away and $\approx 6.25$~mag fainter than the primary in the \Gaia $G$ band, which we neglect in the remainder of our analysis due to its faintness.} Although we did not detect any lines of this nearby star in our spectroscopic observations, we need to correct the broad-band catalog and time-series photometry for contamination by this nearby star.

We used the same procedures as described in Paper I to perform this correction.
Briefly, we fitted the catalog photometry from \Gaia, 2MASS and WISE, as well as the $\Delta$mag measured by SOAR HRCam, with a blended two-component model of the spectral energy distribution (SED) using the \texttt{isochrones} package \citep{Isochrones_Morton2015,MIST0_Dotter2016,MISTI_Choi2016}.
We made the assumption that the two stars are physically associated and have the same parallax as measured for the primary by \Gaia, as is the case for most stars separated by $< 1$" \citep{Horch2014,Matson2018}.
We determined an upper limit on the line-of-sight extinction from the dust maps of \citet{Schlegel1998} and \citet{Schlafly2011}.
Priors were placed on the spectroscopic properties of the primary from those derived by the \texttt{SpecMatch-Emp} analysis (\S\ref{ssec:spec_char}).
We imposed an error floor of 0.02~mag for the \Gaia and 2MASS photometry, and 0.03~mag for the WISE photometry.

\begin{figure}
\epsscale{.98}
\plotone{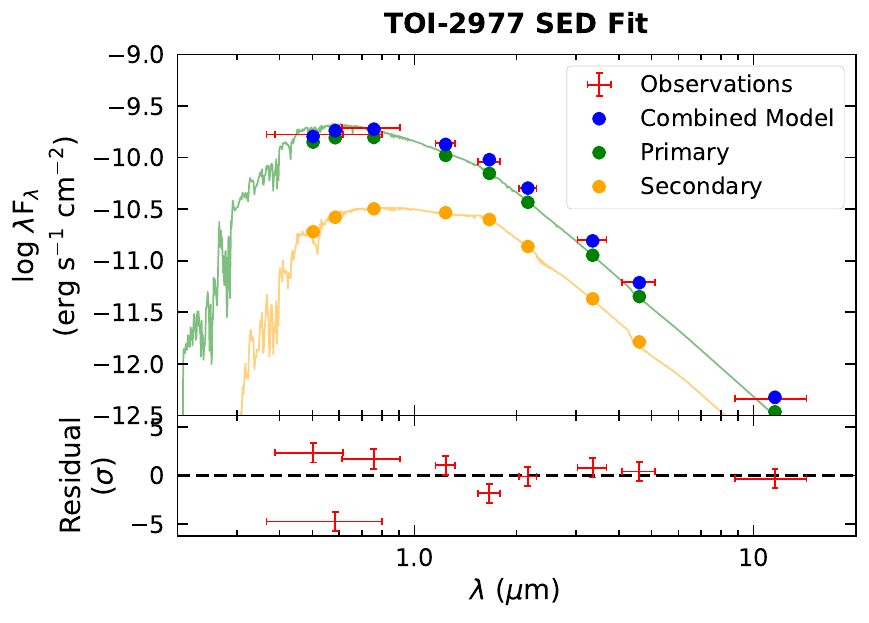}
\plotone{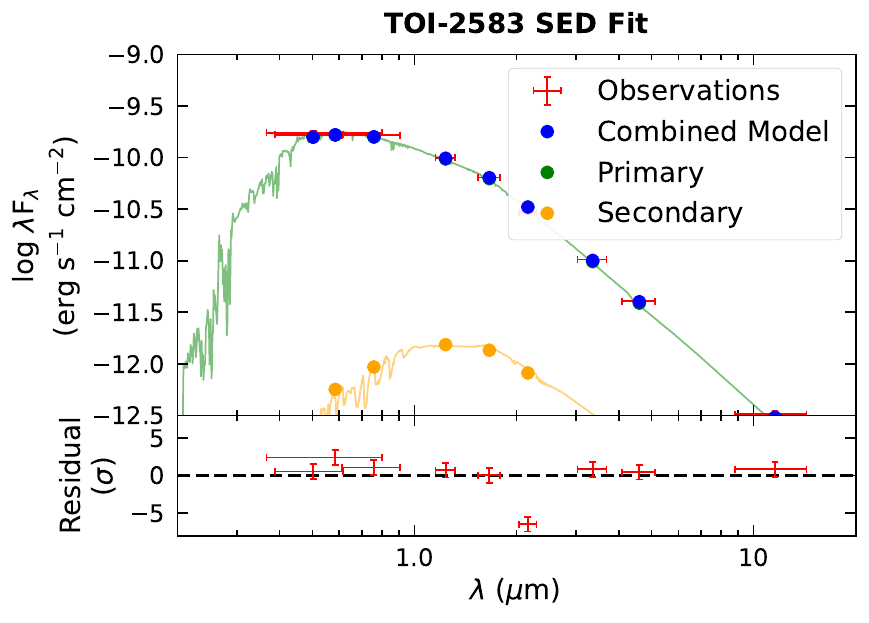}
\plotone{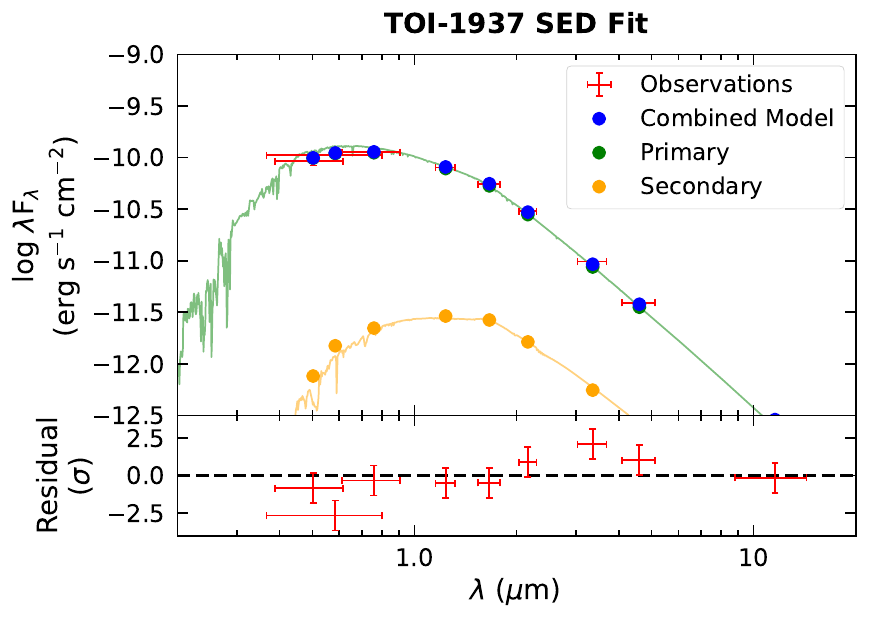}
\caption{Multi-component SED fit for the TOI-2977 system (top), TOI-2583 system (middle), and TOI-1937 system (bottom).
The red points with vertical error bars show the observed catalog fluxes and uncertainties from \Gaia, 2MASS, and WISE, while the horizontal error bars illustrate the width of the photometric band.
The green and orange points show the model fluxes of the primary and secondary respectively, while the blue points show the combined flux of both stars.
For illustrative purposes, we underplot extinction-corrected \citet{Kurucz1993} atmospheric models for the two stellar components, although these models were not used directly in the fit, which was based on MIST bolometric correction tables.
For TOI-2583, the 6-$\sigma$ residual between the model and observed flux in the $K_s$ band reflects a large discrepancy between the difference as measured by 2MASS ($\Delta$(mag) = 1.7) and by the ShARCS AO imaging ($\Delta$(mag) = 4.0).
We excluded the 2MASS $K_s$-band flux from the SED fit, relying on the AO measurements instead, but show the residual here.
\label{fig:multi_sed_plot}
}
\end{figure}

We plot the best-fit SED model in the top panel of Figure \ref{fig:multi_sed_plot}, and present the stellar properties of the primary and secondary components in Table \ref{tab:toi2977_sec_properties}.
If the two components are indeed bound, TOI-2977B is a mid-K dwarf with a sky-projected separation of $\approx 260$~AU from the primary.
We subtracted the model fluxes of the secondary from the observed catalog fluxes, and also used them to compute dilution factors for the \TESS and ground-based time-series photometry.
These corrected fluxes and dilution factors were then included in our global modelling of the TOI-2977 system (\S\ref{sec:planet_char}).

Apart from the TOI-2977 system, the high-resolution imaging observations detected companions for TOI-1937 and TOI-2583, which were also identified by \Gaia (Table \ref{tab:gaia_comp_props}).
For TOI-1937, SOAR speckle imaging detected the nearby star at a separation of $2\farcs5$ and $\Delta I = 4.3$~mag (Fig. \ref{fig:toi1937_imaging}).
Similarly for TOI-2583, the ShARCS AO imaging detected the companion at $5\farcs3$ which was also resolved by \Gaia\,(Fig. \ref{fig:toi2583_imaging}).
We performed a multi-component SED fit (lower panels of Figure \ref{fig:multi_sed_plot}) for these two targets similarly to TOI-2977.
The broadband photometry are compatible with an M0.5V stellar companion to TOI-1937A and an M2.5V stellar companion to TOI-2583A.
We only included dilution corrections to the ground-based photometry for TOI-1937A, but not for TOI-2583A, where the ground-based photometry resolved both stellar components.

In the case of TOI-1937A and TOI-2583A, our spectroscopic observations resolved the primary, thus the spectroscopic properties and radial-velocity variations are measured for the primary star.
The TESS SPOC Data Validation reports\footnote{\dataset[10.17909/t9-2tc5-a751]{http://dx.doi.org/10.17909/t9-2tc5-a751}, \dataset[10.17909/t9-yjj5-4t42]{http://dx.doi.org/10.17909/t9-yjj5-4t42}} for these two targets also showed that the difference image centroid offset compared with the TIC positions were $0\farcs89\pm2\farcs54$ for TOI-1937A and $1\farcs75 \pm 2\farcs5$ for TOI-2583A, indicating that the transits did occur on the primary star.
For TOI-2977A, while the spectroscopic observations did not resolve the $0\farcs77$ companion, we did not detect its line in the measured spectrum, nor were there large BIS variations, suggesting that the measured radial-velocity variations are also due to motion of the primary star.

\begin{deluxetable}{lcc}
\tablecaption{TOI-2977 Stellar Properties from SED Fit \label{tab:toi2977_sec_properties}}
\tablehead{
    & \colhead{Primary} & \colhead{Secondary}
}
\startdata
\input{comp_props}
\enddata
\end{deluxetable}

\subsection{Potential Cluster Membership} \label{ssec:cluster}

\subsubsection{TOI-1937: ambiguous member of NGC 2516}
\label{ssec:toi1937}

TOI-1937 was reported by \citet{kounkelcovey2019} to be a member of
the southern open cluster NGC\,2516 ($d\approx 400\,pc$, $t\approx
150\,{\rm Myr}$, $M_{\rm tot}\approx 1400\,M_\odot$;
\citealt{bouma2021,meingast2021}).  Very few if any hot Jupiters have
well-measured ages below a few hundred million years.  If TOI-1937
were a member of NGC\,2516, it would therefore likely represent a new
upper limit for the arrival time of hot Jupiters on their close-in
orbits, which might in turn help to constrain the mechanism by which
these planets migrate.  We investigated the possible cluster membership
of TOI-1937 by considering the available six-dimensional positions and
kinematics of the star relative to the cluster, the stellar rotation
period, the photospheric lithium content, and the stellar metallicity.
Our analysis drew heavily from the rotation and lithium analyses
performed by \citet{bouma2021}.  We ultimately found that current
evidence for the cluster membership of TOI-1937 is inconclusive,
as laid out in the following paragraphs. 
We do note however that multiple age indicators would need to be
compromised by the presence of the hot Jupiter if the star {\it were}
a member of NGC\,2516.  Implications for whether TOI-1937 is a young
hot Jupiter, or whether it is simply a tidally spun-up field star, are
discussed in Section~\ref{disc:toi1937}.

\paragraph{Positions, Kinematics, and False Positive Rates}

Analyses of the positions and kinematics of NGC\,2516 from Gaia DR2
yielded the discovery of tidal tails that lead and lag up to $\pm
250\,{\rm pc}$ from the central core of the cluster, relative to its
orbit in the Galaxy \citep{kounkelcovey2019,meingast2021}.  A
visualization is available
online\footnote{\url{https://homepage.univie.ac.at/stefan.meingast/coronae.html},
made by \citet{meingast2021}, last accessed 30 July 2022.}.  TOI-1937
was included as a member of the trailing tail by
\citet{kounkelcovey2019}.  The star was not reported as a cluster
member by \citet{meingast2021}, due to a more stringent cut on the
maximum tangential velocities out to which a star could be considered
as a candidate cluster member.  The reality of the tidal tails is
supported by inspection of the lowest-mass M-dwarfs, which are more
luminous than their field counterparts, as one would expect for
$\approx$150\,Myr pre-main-sequence stars.  The stellar rotation
periods also show a gyrochronal locus consistent with other open
clusters \citep{bouma2021}.  However, the expected field star
contaminant rate from the clustering algorithms used by both
\citet{kounkelcovey2019} and \citet{meingast2021} is expected to
increase with separation from the core of the cluster;
\cite{bouma2021} estimated that up to $\approx 40\%$ of the candidate members in the outermost regions of the cluster's tidal tails are in fact field stars.

\begin{figure}
\includegraphics[width=\linewidth]{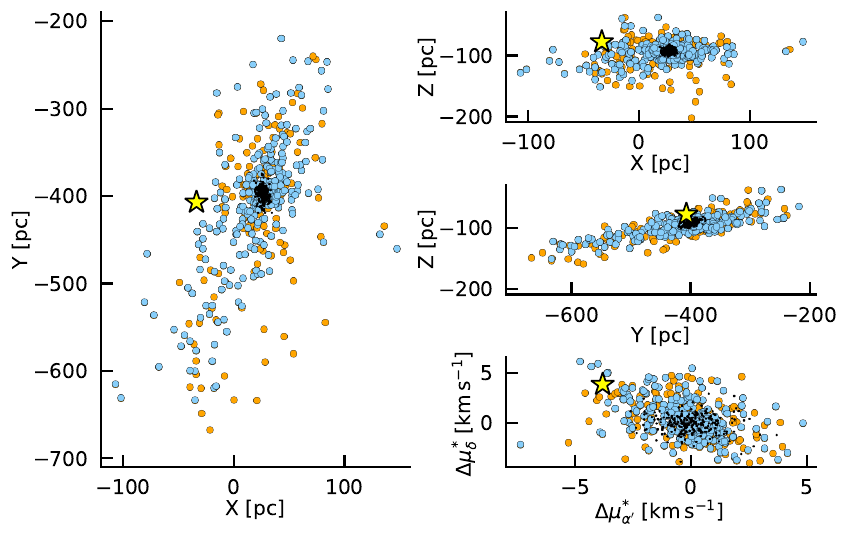}
\caption{XYZ galactic coordinates and on-sky velocities of TOI-1937 (yellow star)
in relation to the NGC\,2516 cluster. The galactic center is in the direction of $+\hat{X}$, and galactic rotation is in the direction of $+\hat{Y}$. The black dots represent stars in the core of the cluster; blue circles are stars that are rotationally confirmed to be in the tidal tails; orange circles are reported candidate members of the tidal tails for which stellar rotation was expected but not detected.  The latter are likely field interlopers.  In the vicinity of TOI-1937, the contamination rate from such stars is about one in three.
\label{fig:toi1937_kinematics}
}
\end{figure}

TOI-1937 is in this wide-separation regime (Figure \ref{fig:toi1937_kinematics}): 
it is $\approx$70\,pc outside the core of the cluster,
with a tangential on-sky velocity that is $\approx$5.4\,{\rm km}\,{\rm s}$^{-1}$
separate from that of the cluster core,
although its systemic velocity of $RV_{\rm sys} = 22.4\pm0.1\,{\rm
km}\,{\rm s}^{-1}$ is consistent with that of NGC\,2516
($23.8\pm1.1$\,{\rm km}\,{\rm s}$^{-1}$;
\citealt{healy_stellar_2020}). The spatial and kinematic evidence,
coupled with the fraction of nearby stars that are likely field
interlopers, leaves the cluster membership of TOI-1937 ambiguous.

\begin{figure}
\centering
\includegraphics[width=0.95\linewidth]{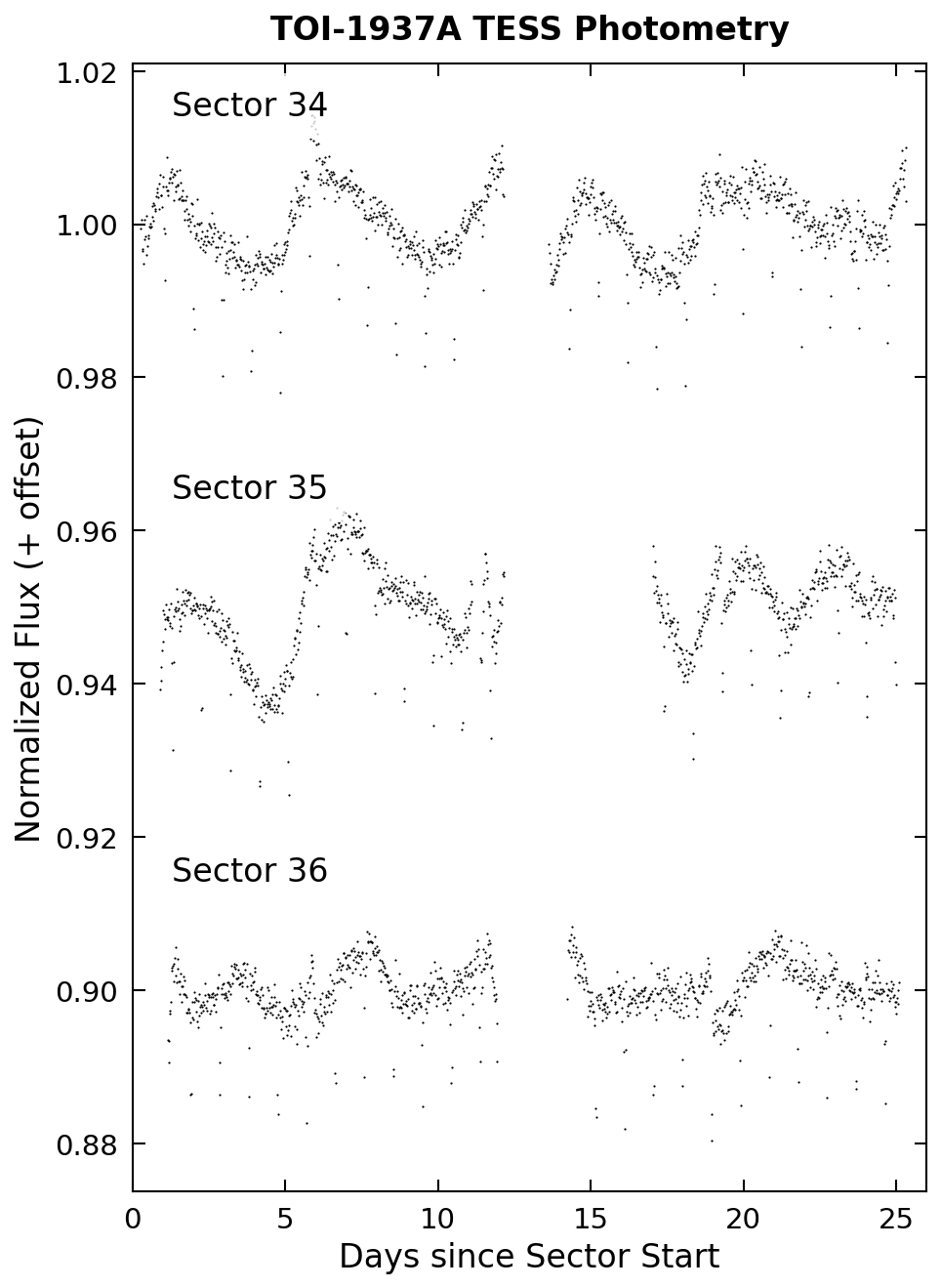}
\caption{TESS SPOC light-curves from Sectors 34-36 for TOI-1937, binned to 30-minute cadence.
The planetary transits as well as the 6.6~day rotational modulation are clearly visible in the light-curves.
\label{fig:toi1937_tess_lc}
}
\end{figure}

\paragraph{Rotation and Gyrochronology}
A Lomb-Scargle analysis of the TESS Sector 34, 35, and 36 light curves
yields a rotation period for TOI-1937 of $6.6 \pm 0.2\,$days (Figure \ref{fig:toi1937_tess_lc}.
At $(G_{\rm BP}-G_{\rm RP})_0 = 0.87$, comparable stars on the slow
sequence of NGC\,2516 have rotation periods between 4.0 and 5.5 days.
Similar-aged clusters such as the Pleiades have an indistinguishable
distribution of G-dwarf rotation periods.
\citep{fritzewski_rotation_2020}.  The rotation period of TOI-1937A
would therefore imply a gyrochronal age of $\approx$300-400\,Myr,
based on the gyrochronal loci of M\,48 and NGC\,3532
\citep{barnes_2015,fritzewski_2021}.  The rotation period is well
below that of the Praesepe slow sequence ($\approx$700\,Myr;
\citealt{douglas_poking_2017}).

The main assumption of gyrochronology -- that spin-down is dominated
by magnetic braking -- may not be applicable to TOI-1937 due to the
presence of the hot Jupiter. 
There is a significant amount of
population-level evidence that hot Jupiters can tidally spin up their
host stars \citep{maxted_2015,colliercameron_2018,penev_2018,tejada_2021}.
Given the short orbital period ($P < 1$~day) of the hot Jupiter TOI-1937A\,b,
this may be the case for TOI-1937A, which would imply that the gyrochronal age
of TOI-1937A is unreliable and the system may in fact be older than inferred.

\paragraph{Lithium}
Photospheric lithium can be used as an age diagnostic for FGKM stars
\citep{soderblom_ages_2010}.  One well-studied feature is the \ion{Li}{1}
6708\,\AA\ doublet, and at least 90\% of 150\,Myr old Sun-like stars
show it in absorption \citep[{\it e.g.},][]{soderblom_1993,bouma2021}.
Given its stellar mass, TOI-1937A would be expected to display an
equivalent width for this doublet of 110 to 140\,m\AA\ if it were a
bonafide member of NGC\,2516 \citep{bouma2021}.  Field stars of the
same mass typically show an equivalent width between 0 and 70\,m\AA\ 
\citep{berger_2018}.  Using the highest S/N PFS spectrum available,
from 11 November 2020, we find ${\rm EW}_{\rm Li} < 25.1$\,m\AA\ at
3-$\sigma$, after correcting for the neighboring \ion{Fe}{1} blend
at 6707.44\,\AA.  The non-detection of lithium would be quite
anomalous for a 150\,Myr Solar-mass star. While as with
gyrochronology, one could imagine scenarios in which the hot Jupiter
affects the stellar lithium abundance
\citep{israelian_enhanced_2009,figueira_2014,delgado_mena_2015}, 
this would require special justification.
The lack of lithium is the strongest
argument against the membership of TOI-1937 in NGC\,2516.

\paragraph{Metallicity}
The most up-to-date analysis of the metallicity of NGC\,2516 appears to
be that of \citet{baratella_2020} (but see \citealt{bailey_2018}).
The study by \citet{baratella_2020} suggested that for young stars,
using only Fe lines to estimate the microturbulence parameter could
yield over-estimates, which can propagate to yield under-estimated
cluster abundances.  Based on 10 spectra of Sun-like stars in NGC\,2516
acquired with VLT/FLAMES, they report a mean cluster metallicity of
$\feh = 0.08 \pm 0.01$.  Our application of \texttt{SpecMatch-Emp} to
the PFS spectra of TOI-1937 yielded $\feh = 0.22 \pm 0.09$, which while slightly
higher than the cluster metallicity, is approximately consistent
within the uncertainties.  

\subsubsection{TOI-4145: not a member of Platais-3}

TOI-4145 was reported, also by \citet{kounkelcovey2019}, to be a
candidate member of Platais-3 based on its 3D positions and 2D
sky velocities from \Gaia DR2.  The mean distance of the cluster is $d
\approx 180\,$pc, and its age could be between 210 and 440 Myr
\citep{bossini_2019,kounkelcovey2019}.  Similar to TOI-1937, the star
is in the positional and kinematic outskirts of the cluster -- it
would need to be in its ``corona'', or ``tidal tail'', but the false
positive rates in the membership lists in this region of Platais-3 are
much less well-quantified than they are for NGC\,2516.  Regardless, the
cluster's age range, combined with the stellar $T_{\rm eff} \approx
5400\,K$, leads to some immediate predictions that can confirm or
refute membership.  First, the rotation period should be between 7 and
9 days, with an amplitude of 0.2\% to 2\% \citep{fritzewski_2021}.
Second, the ${\rm EW}_{\rm Li}$ should be between 50 and 120\,m\AA. 

The TESS light curves of TOI-4145 do not show any evidence for a
rotation signal.  The photometric precision achieved is $\approx
400$\,ppm\,hr$^{1/2}$, well in excess of what would be necessary to
detect the relevant rotation signal.  The NEID spectra similarly yield
a non-detection with ${\rm EW}_{\rm Li}<33.5$\,m\AA at 3-$\sigma$,
after excluding the Fe blend.  There is therefore no corroborating
evidence for TOI-4145 being a bonafide member of Platais-3.

\begin{deluxetable*}{cccccccccc}
\tablecaption{Spectroscopic Stellar Properties \label{tab:spec_props}}
\tablecolumns{9}
\tablehead{
    \colhead{Target} & \colhead{Code} & \colhead{Instrument} &
    \colhead{\Teff~(K)} & \colhead{\Rstar~(\Rsun)} & \colhead{\logg~(dex)} & \colhead{\feh~(dex)} &
    \colhead{\vsini~(\kms)} & \colhead{\vmac~(\kms)} &
    \colhead{Adopted}
}
\startdata
\input{spec_props}
\enddata
\end{deluxetable*}

\section{Planetary System Characterization} \label{sec:planet_char}

We used the \Exofast code \citep{ExoFAST_Eastman2013,ExoFASTv2_Eastman19} to fit all the available data for each target and thereby characterize each planetary system.
\Exofast fits the broad-band photometry, transit light-curves, and RV time-series with a self-consistent stellar and planetary model.
For example, the constraint on the stellar density from the transit model is forced to be consistent with the constraint on the stellar properties from fitting the broad-band photometry.
The uncertainties on each parameter are measured through a differential evolution Markov Chain Monte Carlo (DE-MCMC) exploration of the posterior distribution.

We follow the same fitting strategy as used in Paper I, but reiterate the key points here.
We use the \texttt{SpecMatch-Emp}-derived spectroscopic parameters and uncertainties as Gaussian priors on the stellar \Teff, \Rstar, and \feh.
A Gaussian prior is also imposed on the stellar parallax, with the mean and standard deviation derived from the zero-point-corrected \Gaia DR3 parallax measurement \citep{GaiaEDR3_Lindegren2021}.
We incorporate the \Gaia and 2MASS photometry with an error floor of 0.02~mag, and the WISE photometry with an error floor of 0.03~mag.
An upper limit on the line-of-sight extinction $A_V$ is imposed based on the sky coordinates of each target and the dust maps from \cite{Schlegel1998} and \cite{Schlafly2011}.

We fit the radial velocities with independent offset and jitter terms for each instrument and target, but did not allow for any linear or quadratic RV trends, because the data did not suggest such terms were necessary.
For the systems with only two TRES observations, we did not include the TRES RVs in the fit, since doing so would require adding two more free parameters (a per-instrument offset and jitter term) for the same number of new data points.
However, we include those TRES data points in Figure Set \ref{fig:all_multiplots} for illustrative purposes, with the relative RV offset computed by a simple $\chi^2$-minimization to the best-fitting RV model.

The transit light-curves are fit by a quadratic limb-darkened transit model from \citet{Mandel2002,Agol2020}, with the limb-darkening coefficients constrained by the tables from \citet{Claret2011,Claret2017} during the \Exofast fit.
When fitting the long-cadence 30-minute and 10-minute data from \TESS, this model is computed at 2-minute intervals and integrated over the length of the exposure, to account for distortions in the transit shape arising from the finite integration time of the observations \citep[e.g.,][]{Kipping2010}.
Although the \TESS light-curves are in theory already corrected for dilution from neighboring stars in the same pixel, this assumes that all stars are accounted for with correct magnitudes in the \TESS Input Catalog (TIC; \citealt{TIC_Stassun2018,TIC_Stassun2019}).
To account for possible errors in this correction, we allowed for an additional dilution factor to be fit for the \TESS light-curves, imposing a Gaussian prior centered at zero and with a width equal to 10\% of size of the dilution factor already corrected for in the TIC.
We also fitted for a separate flux baseline $F_0$ and added variance $\sigma^2$ for each sector of \TESS data as well as each ground-based light-curve.
The ground-based light-curves are simultaneously detrended against the variables listed in Table \ref{tab:sg1_lcs}, assuming an additive detrending model.

We ran the \Exofast fit until the DE-MCMC algorithm converged under the criteria recommended by \citet{ExoFASTv2_Eastman19}: Gelman-Rubin statistic \citep{GelmanRubin} $\mathrm{GR} < 1.01$ and more than 1000 independent MCMC draws.
As in Paper I, we ran two fits per system -- one in which the eccentricity was fixed to zero, and one in which the eccentricity was allowed to float.
We adopted the zero-eccentricity set of parameters as our fiducial results, but also state the 1-$\sigma$ upper limits on the eccentricity derived from the latter fit.

Two of the objects: TOI-2796\,b and TOI-3807\,b, were found to be on orbits that lead to grazing transits.
In these cases, fitting the light curve leads to a strong covariance between the impact parameter $b$ and planet-to-star radius ratio $\Rp/\Rstar$.
The \Exofast MCMC exploration of the posterior distributions for these parameters have long tails out to unphysical values for the planet radius.
For these objects, instead of reporting the median of the posterior distribution for the planet radius \Rp and impact parameter $b$, we report the posterior modes, as well as the 95\% lower limits on these parameters (and corresponding lower or upper limits for those parameters which depend directly on \Rp and $b$).

We present the results and 1-$\sigma$ uncertainties from our MCMC fitting procedures in Table \ref{tab:fitted_props}, and plot the light-curve, radial-velocity and stellar flux data alongside the best-fit models in Figure Set \ref{fig:all_multiplots}.
The full set of fitted parameters are provided in Table \ref{tab:additional_fit_params} of the Appendix, as well as in a machine-readable table accompanying the electronic version of this article.

\section{Discussion} \label{sec:discussion}

\begin{figure}
\epsscale{1.2}
\plotone{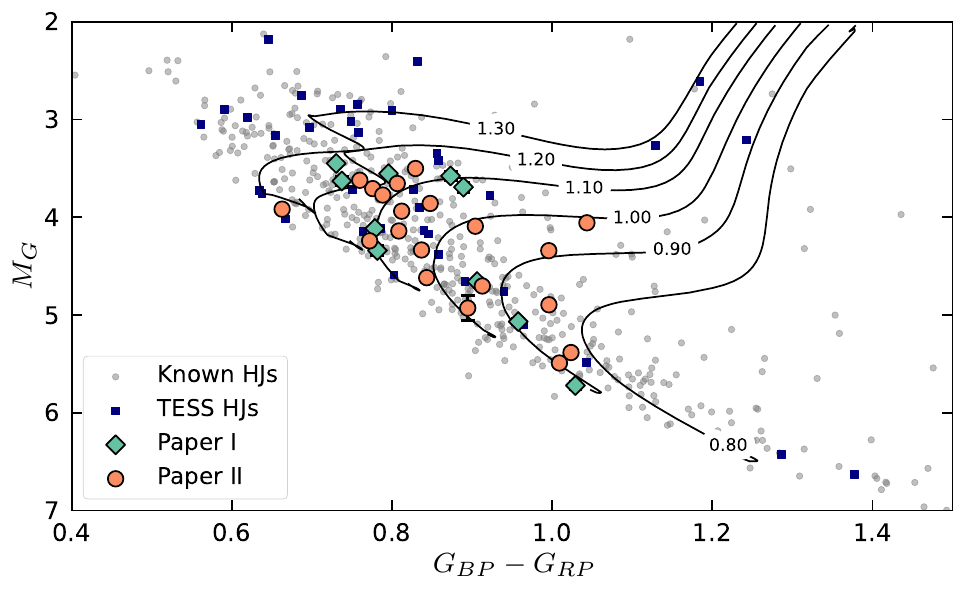}
\caption{\Gaia $G_\mathrm{BP} - G_\mathrm{RP}$ vs $G$ color-magnitude diagram for known hot Jupiter hosts in the NASA Exoplanet Archive (gray points).
The orange circles show the newly confirmed planets described in this paper, while the green diamonds show the planets from Paper I.
Navy blue squares represent the other confirmed hot Jupiters discovered by the \TESS mission.
We also plot the MIST evolutionary tracks for a metallicity of $\feh = +0.20$~dex, equal to the median metallicity of the stellar hosts in our sample, labelled with the corresponding stellar mass between 0.8 and 1.3~\Msun.
The symbols in this plot are consistent with those used in the following figures.
\label{fig:hj_pop_cmd}}
\end{figure}

\begin{figure}
\epsscale{1.2}
\plotone{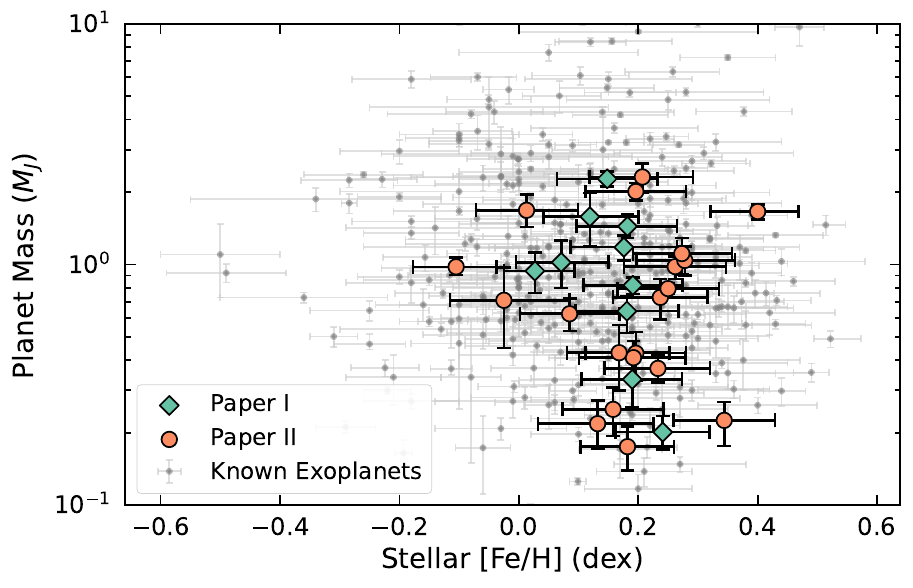}
\caption{The distribution of hot Jupiters in the planet mass-stellar metallicity plane.
Most gas giants orbit metal-rich stars, and our sample of planets follows this tendency, with a median metallicity of $\feh = +0.20$~dex.
\label{fig:hj_pop_met}}
\end{figure}

The twenty newly-confirmed hot Jupiters in this paper bring us closer to a complete, magnitude-limited sample of transiting giant planets that may reveal new insights into their formation and subsequent evolution.
In this section, we place them in context by comparing them with the previously known hot Jupiter sample from the NASA Exoplanet Archive \citep{ExoplanetArchive_PSCompPars}\footnote{\url{https://exoplanetarchive.ipac.caltech.edu/}, accessed 16 Aug 2022}.

Figure \ref{fig:hj_pop_cmd} shows the \Gaia color-magnitude diagram of all currently known hot Jupiters.
Our twenty new planets all orbit FGK dwarfs, which was one of our survey's selection criteria for follow-up observations of planet candidates.
All but one of them orbit host stars consistent with super-solar metallicities, with the median \feh of our sample being $+0.2$~dex (Figure \ref{fig:hj_pop_met}), as expected based on the steep dependence of the occurrence rate of hot Jupiters and metallicity of their host stars \citep{Santos2004,Valenti2005}.
This connection has been interpreted as arising from the greater ease of forming giant planets in high-metallicity disks \citep[e.g.,][]{Mordasini2012}, eventually producing hot Jupiters either through disk migration or via scattering events between planets, further enhancing this correlation \citep{Dawson2013,Buchhave2018}.

TOI-2803\,b is the only planet in our sample orbiting a host with sub-solar metallicity, with our adopted spectroscopic metallicity of $\feh = -0.11\pm0.08$~dex.
However, we note that the \texttt{SpecMatch-Emp} and \texttt{ZASPE} analyses of the PFS template spectrum returned much lower spectroscopic metallicities of $\feh = -0.43\pm0.09$ and $-0.35\pm0.07$~dex respectively (see discussion in \S\ref{ssec:toi2803_spec}).
If the low metallicity results were accurate, TOI-2803\,b would be a surprising outlier, with only a handful of other hot Jupiters reported to orbit host stars with lower metallicities.
Resolving this discrepancy in the stellar properties may be possible with future higher signal-to-noise observations or more detailed stellar modelling.


\begin{figure}
\epsscale{1.2}
\plotone{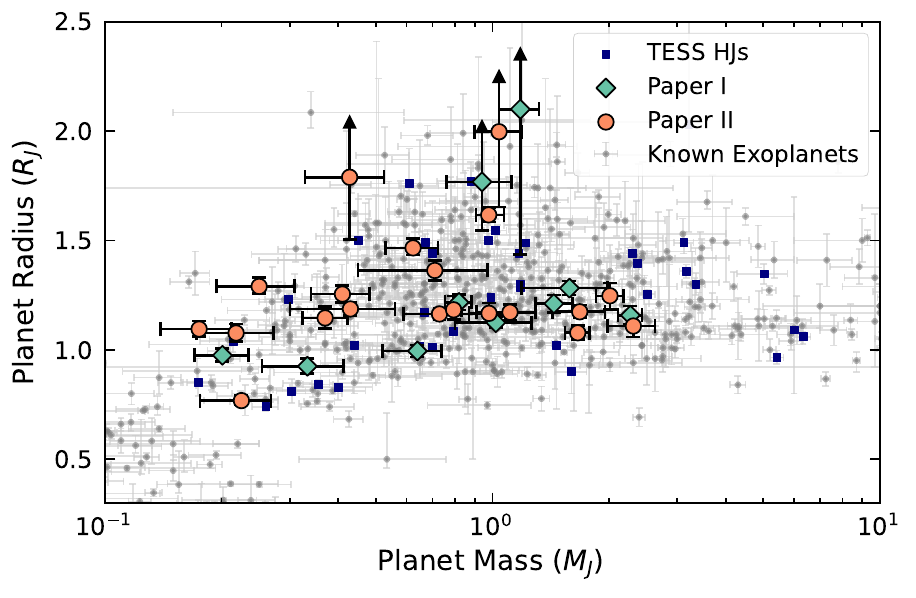}
\plotone{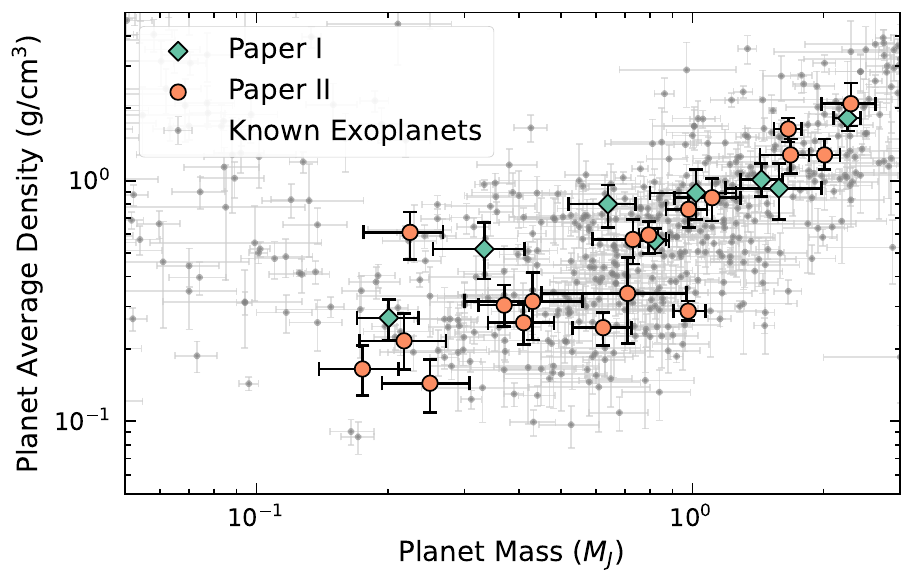}
\caption{
\textbf{Top:} Planet mass-radius distribution for the newly confirmed hot Jupiters from our survey.
For the objects with grazing transits, we plot the mode of the radius posterior distribution and the 2-$\sigma$ lower limit.
\textbf{Bottom:} Planet bulk density as a function of planet mass.
In this plot, we have excluded the planets with grazing transits due to the large uncertainty in the planet density.
\label{fig:hj_pop_mass_radius}}
\end{figure}

The observations for all twenty planets presented in this paper are consistent with the planets being on circular orbits, and as such we have adopted those fits for our fiducial set of parameters.
This is unsurprising given the short tidal circularization timescales for most of these planets.
We also performed fits for all the systems where eccentricity was allowed to float, allowing us to obtain upper limits on the eccentricity for each system.
In most cases, the limited number of RV measurements per system do not allow us to constrain the eccentricity to much better than $e \lesssim 0.1$.
However, for TOI-3364\,b, a planet on a $P = 5.88$~day orbit with the longest tidal circularization timescale in our sample, we were able to put a 1$\sigma$ upper limit of $e < 0.036$.

We also note that for TOI-3976\,b, the longest period planet in our sample ($P = 6.61$~days), the eccentric fit found $e = 0.180_{-0.056}^{+0.060}$, roughly 3$\sigma$ greater than zero.
We compared the circular and eccentric models using the Bayesian Information Criterion (BIC; \citealt{Schwarz1978}), finding that the circular model is favored by $\Delta$(BIC) = 5.
Additional future observations may help place better constraints on the eccentricity of this longer-period hot Jupiter.

We also present the distribution of planet radii and bulk densities as a function of planet mass in Figure \ref{fig:hj_pop_mass_radius}.
Four of the objects in our sample (TOI-2364\,b, TOI-2583\,b, TOI-2587\,b and TOI-3976\,b) are sub-Saturns, with masses less than $0.3$~\Mjup.
Three of them are highly inflated, with bulk densities $\rho_\mathrm{P} < 0.25$~\gcc, similar to other objects in this regime.
TOI-2364\,b does not appear to be inflated, nor would we expect it to be inflated given the comparatively low incident flux it receives; the host star (K0V, $\Teff\approx5300$~K) is one of the cooler stars in our sample.

\begin{figure}
\epsscale{1.2}
\plotone{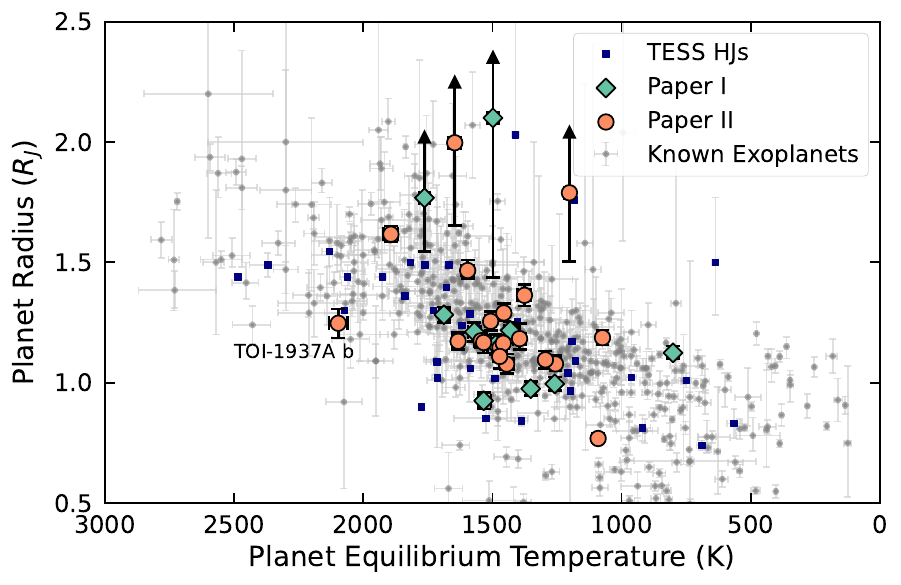}
\caption{
Distribution of planet radius as a function of planet equilibrium temperature, computed at the planets' semimajor axis and assuming no albedo and perfect heat redistribution.
TOI-1937A\,b is anomalously small for its orbit, and is labelled in the plot.
\label{fig:hj_pop_teq_radius}}
\end{figure}

Finally, we show in Figure \ref{fig:hj_pop_teq_radius} the distribution of planet radii with planet equilibrium temperature.
Our planets are consistent with the radius inflation trend seen with the previously known sample \citep[e.g.,][]{Demory2011}.
One noteworthy planet is TOI-1937A\,b, which is relatively smaller than expected given its close-in ($P < 1$~day) orbit.

\subsection{TOI-1937A b: Youngest Hot Jupiter, or Just Tidally Spun Up?}
\label{disc:toi1937}

Section~\ref{ssec:toi1937} described the evidence for, and against,
the membership of TOI-1937 in NGC\,2516 (150\,Myr).  The position and
kinematics of the star suggest a probability of $\approx 60\%$ that it
is a member of the cluster's trailing tidal tail.  The stellar
rotation period is fast relative to the field, but in detail would be
more consistent with a gyrochronal age of 300 to 400\,Myr, not
150\,Myr.  The super-solar metallicity is approximately consistent
with that of other cluster members, though a homogeneous assessment
using identical instruments and pipelines would help confirm this
suggestion.  The non-detection of lithium would be quite anomalous for
a 150\,Myr Solar-mass star, and seems to be the best evidence for
TOI-1937 being a field interloper.  If TOI-1937A were a cluster
member, the presence of the hot Jupiter would need to have both
somehow spun down the star, and also caused it to deplete its lithium
much faster than is typical for Sun-like stars.

\begin{figure}[t]
	\begin{center}
		\leavevmode
			\includegraphics[width=0.49\textwidth]{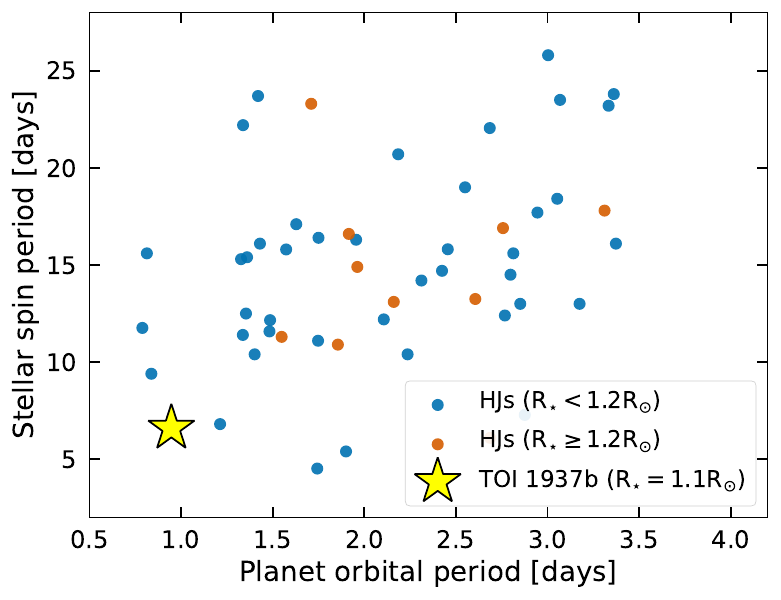}
	\end{center}
	\vspace{-0.6cm}
	\caption{
    Stellar spin periods and planetary orbital periods for hot Jupiters.
    Data were collected by \citet{penev_2018}; we only show hot Jupiter
    systems with stellar spin period S/N ratios of at least 3.  For a dwarf star,
    the $1.2R_\odot$ radius boundary corresponds to a stellar
    effective temperature of 6000$\,$K, slightly below the Kraft break.
    Stars hotter than the Kraft break experience less efficient stellar spindown
    as the outer convection zone shrinks.
    \label{fig:porb_prot}
	}
\end{figure}

The alternative is that TOI-1937A is simply a tidally spun-up field
star.  If we take the age of the star to be $\gtrsim 1\,$Gyr, as
suggested by the lack of lithium, then by gyrochronal arguments, the
rotation period should be $\gtrsim 11\,$days, based on the observed
rotation sequence of NGC-6811 \citep{curtis_2019}.  The actual 6.6-day
stellar spin period is highly discordant from this prediction.
Combined with the 22.7-hour orbital period, which is much shorter than
that of most hot Jupiters (Figure~\ref{fig:porb_prot}), TOI-1937 would
seem to be a system that could provide new constraints on tidal
spin-up, and perhaps eventually orbital decay.

\section{Conclusions} \label{sec:conclusion}

We have confirmed and characterized twenty new short-period giant planets detected by the \TESS mission, based on extensive ground-based photometric, spectroscopic, and imaging observations.
These objects join a host of other giant planets that have been discovered by \TESS over the last few years (\citealp[e.g.,][]{Rodriguez2019,Zhou2019a,Brahm2020,Davis2020,Nielsen2020,Ikwut-Ukwa2021,Rodriguez2021,Sha2021,Wong2021,Knudstrup2022,Rodriguez2022,Psaridi2022,Yee2022a}), showcasing how \TESS is rapidly transforming our knowledge of hot Jupiters by providing a uniform, all-sky sample of these planets.

\cite{Yee2021b} found that the sample of known transiting hot Jupiters was only $\sim 40\%$ complete at a magnitude-limit of $G < 12.5$.
Since that publication, many planets have been newly confirmed, closing the gap between the number of expected planets and the number of known planets.
Our survey has contributed to the confirmation of thirty of these objects, and is continuing to observe more than 100 \TESS planet candidates.
This has only been possible with the collaboration of the broader community, in particular the many members of TFOP whose observations are key for guiding target selection, host star and planet characterization, and making efficient use of follow-up resources.

The ultimate goal of our survey is to use \TESS to assemble a relatively complete and unbiased sample of hot Jupiters with which we can study the demographics of this population of planets.
We expect that \TESS will be mostly complete to detecting such planets orbiting stars down to a magnitude limit of $G < 12.5$, and the $\approx 400$ objects expected in such a sample will enable new investigations into the period, radius, and bulk density distributions of hot Jupiters, along with their correlations with host star properties like stellar metallicity.
This large sample of planets will also be useful for selecting the most promising targets for additional studies, such as transit timing and obliquity measurements, as well as atmospheric characterization by missions like the James Webb Space Telescope (JWST) and the Atmospheric Remote-sensing Infrared Exoplanet Large-survey (ARIEL; \citealt{ARIEL_Tinetti2016}).


\begin{subfigures}
\label{fig:all_multiplots}
\makeatletter\onecolumngrid@push\makeatother
\clearpage
\begin{figure*}
\centering
\includegraphics[width=380pt]{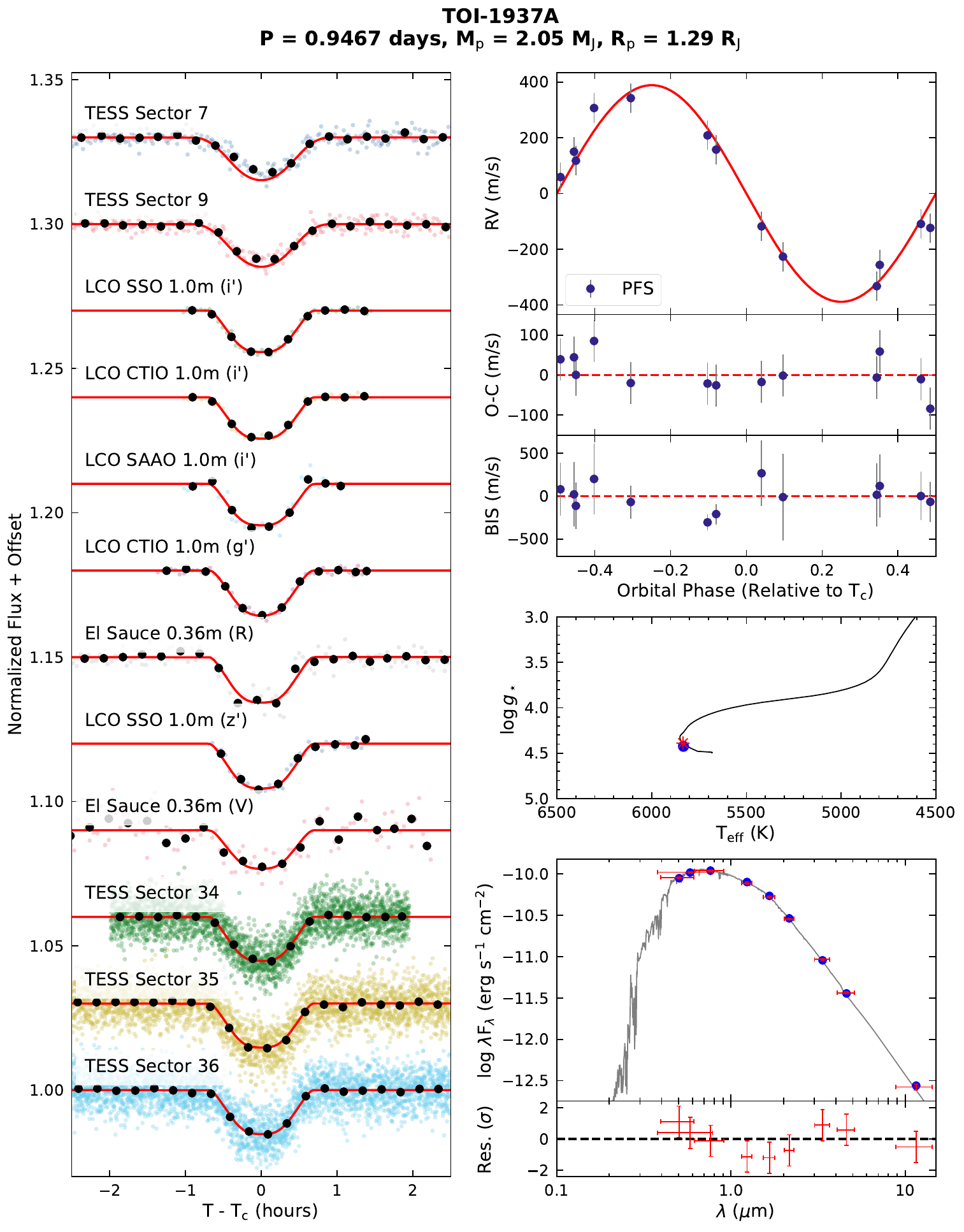}
\caption{\Exofast fit results for TOI-1937A\,b.
\textbf{Left:} \TESS and ground-based light-curves, phase-folded onto the best-fit period and time of conjunction. The faint colored points are the unbinned data, while the black dots show the time-series data binned to 10-min cadence. The best-fit transit model in each band is shown as the red line.
\textbf{Top right:} RV observations, also phased onto the best-fit orbital period. The fitted per-instrument jitter term $\sigma_\mathrm{jit}$ has been added in quadrature to the instrumental uncertainties to produce the gray error bars. The red line shows the best-fit circular model for the RVs.
We plot the residuals after subtracting the model in the middle subpanel, and the phased bisector span measurements in the lower subpanel.
\textbf{Middle right:} The best-fit stellar \Teff and \logg are plotted as the blue point.
The best-fit MIST stellar evolution track is plotted in black, with the red asterisk shows the position along the track corresponding to the best-fit stellar age.
The discrepancy between the blue point and red asterisk are well within the fitted uncertainties in each parameter, indicating no tension between the different constraints on the stellar properties.
\textbf{Bottom right:} The observed stellar fluxes from the \Gaia, Tycho, 2MASS and WISE catalogs are plotted in red, with horizontal error bars corresponding to the width of the photometric bandpass.
The blue points show the best-fit model flux derived from the stellar properties and MIST bolometric correction grid.
We plot in gray an atmospheric model from \citet{Kurucz1993} corresponding to the best-fit stellar parameters for illustrative purposes only, as the fit is performed directly to the MIST grid.
The complete figure set for all TOIs (20 images) is available in the full electronic version of the paper.
The TESS and ground-based time-series photometry, as well as the RV measurements, are available as Data behind the Figure.}
\label{fig:toi1937_multiplot}
\end{figure*}

\def\ExToiMultiplot{1937}
\foreach \toi in \input{tois_HJPaperII.txt}
{
\unless\ifnum\toi=\ExToiMultiplot{
\IfFileExists{toi\toi_multiplot.pdf}{
\begin{figure*}[p]
\includegraphics[width=500pt]{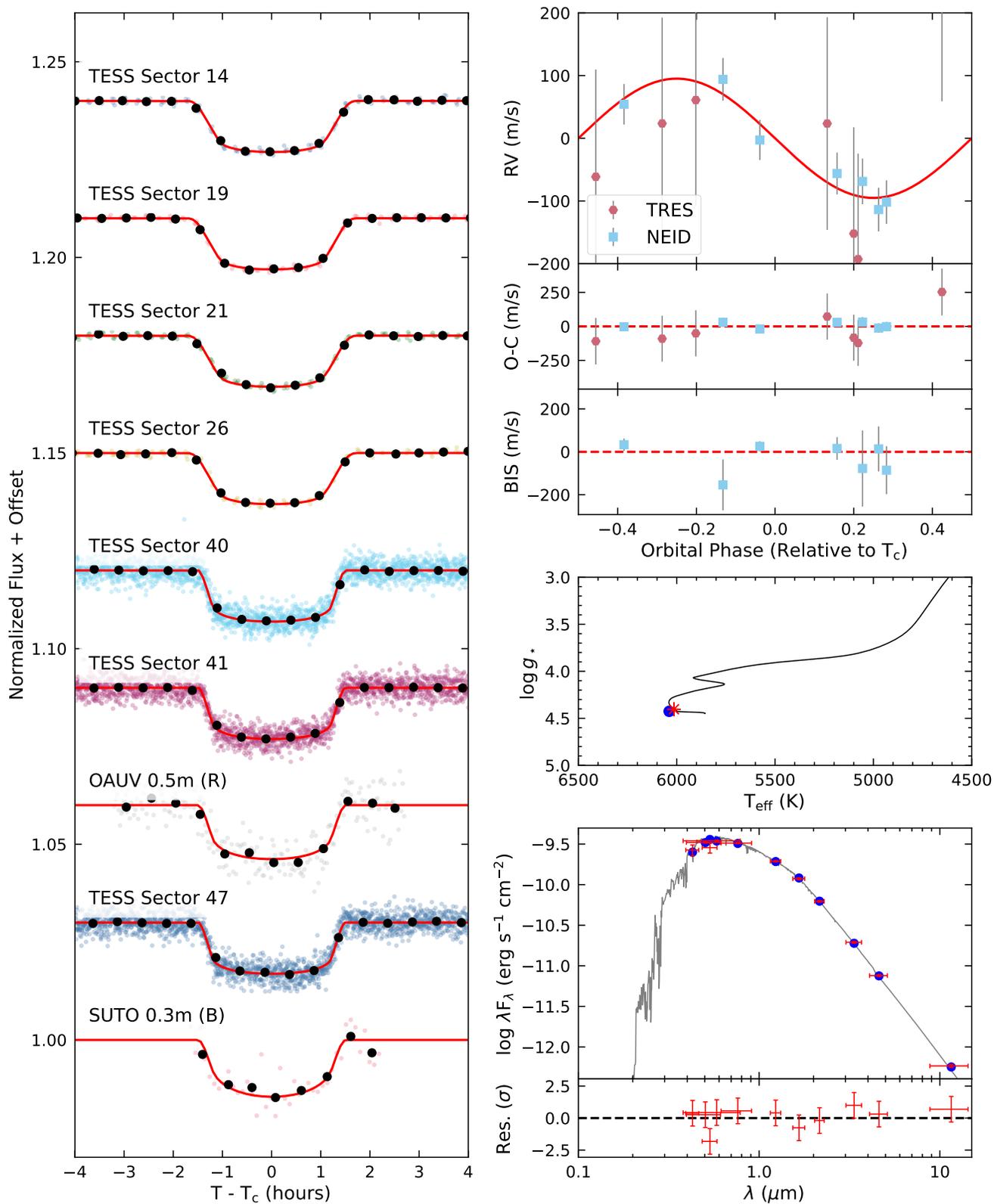}
\caption{Same as Figure \ref{fig:toi1937_multiplot}, but for TOI-\toi\,b.}
\label{fig:toi\toi_multiplot}
\end{figure*}
}{\textcolor{red}{Missing figure for TOI-\toi.}}
}
\fi
}
\end{subfigures}

\begin{rotatepage}
\movetableright=-1in
\movetabledown=2.3in
\begin{rotatetable}
\begin{deluxetable*}{l>{\centering}cccccc}
\tablecaption{Median Values and 68\% Confidence Intervals for Fitted Stellar and Planetary Parameters \label{tab:fitted_props}}
\tabletypesize{\footnotesize}
\input{planet_fit_results_0}
\end{deluxetable*}
\end{rotatetable}
\addtocounter{table}{-1}

\movetableright=-1in
\movetabledown=2.3in
\begin{rotatetable}
\begin{deluxetable*}{l>{\centering}cccccc}
\tablecaption{\textit{(Continued)}}
\tabletypesize{\footnotesize}
\input{planet_fit_results_1}
\end{deluxetable*}
\end{rotatetable}
\addtocounter{table}{-1}

\movetableright=-1in
\movetabledown=2.3in
\begin{rotatetable}
\begin{deluxetable*}{l>{\centering}cccccc}
\tablecaption{\textit{(Continued)}}
\tabletypesize{\footnotesize}
\input{planet_fit_results_2}
\end{deluxetable*}
\end{rotatetable}
\addtocounter{table}{-1}

\movetableright=-1in
\movetabledown=2.3in
\begin{rotatetable}
\begin{deluxetable*}{l>{\centering}cccccc}
\tablecaption{\textit{(Continued)}}
\tabletypesize{\footnotesize} 
\input{planet_fit_results_3}

\tablecommentsmod{
This table contains the fit results from the preferred fit for each target.\\
Table 3 in \citet{ExoFASTv2_Eastman19} provides a detailed description of all derived and fitted parameters.\\ 
$^a$\,This is 68\% upper limit on eccentricity derived from the eccentric fits. \\
$^b$\,The tidal circularization timescale is computed with Equation (3) of \citet{Adams2006}, assuming a tidal quality factor $Q_S = 10^6$. \\
$^c$\,The stellar metallicity when the star was formed, that define the grid points for the MIST stellar evolutionary tracks.\\ 
$^d$\,The equal evolutionary phase (EEP) corresponds to specific points in the stellar evolutionary tracks, as described in \citet{MIST0_Dotter2016}. \\ 
Table \ref{tab:fitted_props} is published in its entirety in the electronic
edition of the journal. This version only shows the
results from the preferred fit for each target. The full version includes
these results and fits where the eccentricity was allowed to float. Note that
the full version also includes the results from the additional fit parameters outlined in Table \ref{tab:additional_fit_params}.}
\end{deluxetable*}
\end{rotatetable}
\end{rotatepage}

\makeatletter\onecolumngrid@pop\makeatother
\clearpage
\begin{acknowledgements} \label{sec:acknowledgments}

We thank the anonymous reviewer for helpful comments that improved this manuscript.

This paper includes data collected by the \TESS mission that are publicly available from the Mikulski Archive for Space Telescopes (MAST).
The raw \TESS data can be accessed at \dataset[10.17909/t9-nmc8-f686]{http://dx.doi.org/10.17909/t9-nmc8-f686} (SPOC 2-minute light-curves), \dataset[10.17909/t9-wpz1-8s54]{https://dx.doi.org/10.17909/t9-wpz1-8s54} (TESS-SPOC full-frame image light-curves), \dataset[10.17909/t9-r086-e880]{https://dx.doi.org/10.17909/t9-r086-e880} (QLP light-curves), and \dataset[10.17909/t9-ayd0-k727]{https://dx.doi.org/10.17909/t9-ayd0-k727} (CDIPS light-curves).
The Data Validation reports are available at \dataset[10.17909/t9-2tc5-a751]{http://dx.doi.org/10.17909/t9-2tc5-a751} and \dataset[10.17909/t9-yjj5-4t42]{http://dx.doi.org/10.17909/t9-yjj5-4t42}.
Funding for the \TESS mission is provided by NASA's Science Mission Directorate. We acknowledge the use of public \TESS data from pipelines at the \TESS Science Office and at the \TESS Science Processing Operations Center.
Resources supporting this work were provided by the NASA High-End Computing (HEC) Program through the NASA Advanced Supercomputing (NAS) Division at Ames Research Center for the production of the SPOC data products.
We also acknowledge the use of data from the Exoplanet Follow-up Observation Program website, which is operated by the California Institute of Technology, under contract with the National Aeronautics and Space Administration under the Exoplanet Exploration Program \citep{ExoFoP,ExoFoPTESS}.

This research made use of Lightkurve, a Python package for \Kepler and \TESS data analysis \citep{Lightkurve18}.
We acknowledge that the work reported on in this paper was substantially performed using the Princeton Research Computing resources at Princeton University which is consortium of groups led by the Princeton Institute for Computational Science and Engineering (PICSciE) and Office of Information Technology's Research Computing.

Some of the data presented herein were obtained at the W. M. Keck Observatory, which is operated as a scientific partnership among the California Institute of Technology, the University of California and the National Aeronautics and Space Administration. The Observatory was made possible by the generous financial support of the W. M. Keck Foundation.
Keck telescope time was granted by NOIRLab (Prop. IDs 2021B-0162, 2022A-543544, PI: Yee) through the Mid-Scale Innovations Program (MSIP). MSIP is funded by NSF.
The authors wish to recognize and acknowledge the very significant cultural role and reverence that the summit of Maunakea has always had within the indigenous Hawaiian community.  We are most fortunate to have the opportunity to conduct observations from this mountain.

This paper contains data taken with the NEID instrument, which was funded by the NASA-NSF Exoplanet Observational Research (NN-EXPLORE) partnership and built by Pennsylvania State University. NEID is installed on the WIYN telescope, which is operated by the National Optical Astronomy Observatory, and the NEID archive is operated by the NASA Exoplanet Science Institute at the California Institute of Technology. NN-EXPLORE is managed by the Jet Propulsion Laboratory, California Institute of Technology under contract with the National Aeronautics and Space Administration.
Data presented herein were obtained at the WIYN Observatory from telescope time allocated to NN-EXPLORE through the scientific partnership of the National Aeronautics and Space Administration, the National Science Foundation, and NOIRLab.
This work was supported by a NASA WIYN PI Data Award, administered by the NASA Exoplanet Science Institute.
The authors are honored to be permitted to conduct astronomical research on Iolkam Du’ag (Kitt Peak), a mountain with particular significance to the Tohono O’odham.

This paper includes data gathered with the 6.5 meter Magellan Telescopes located at Las Campanas Observatory, Chile.

This research has used data from the CTIO/SMARTS 1.5m telescope, which is operated as part of the SMARTS Consortium by RECONS (\url{www.recons.org}) members Todd Henry, Hodari James, Wei-Chun Jao, and Leonardo Paredes. At the telescope, observations were carried out by Roberto Aviles and Rodrigo Hinojosa.
The CHIRON data were obtained from telescope time allocated under the NN-EXPLORE program with support from the National Aeronautics and Space Administration.

This work makes use of observations from the LCOGT network. Part of the LCOGT telescope time was granted by NOIRLab through the Mid-Scale Innovations Program (MSIP). MSIP is funded by NSF.

JH acknowledges funding from NASA grant 80NSSC21K0335. 
This work was supported by an LSSTC Catalyst Fellowship awarded by LSST Corporation to T.D. with funding from the John Templeton Foundation grant ID \#62192. 
K.K.M. acknowledges support from the New York Community Trust’s Fund for Astrophysical Research. N.L.-B. thanks the NASA Massachusetts Space Grant for support. 
A.A.B., N.A.M and B.S.S. acknowledge the support of the Ministry of Science and Higher Education of the Russian Federation under the grant 075-15-2020-780 (N13.1902.21.0039). 
Adam Popowicz acknowledges grant BK-246/RAu-11/2022. 
The research leading to these results has received funding from the ARC grant
for Concerted Research Actions, financed by the Wallonia-Brussels Federation. TRAPPIST is funded by the Belgian Fund for Scientific Research (Fond National de la Recherche Scientifique, FNRS) under the grant PDR T.0120.21. TRAPPIST-North is a project funded by the University of Liege (Belgium), in collaboration with Cadi Ayyad University of Marrakech (Morocco).
M. Gillon is F.R.S-FNRS Research Director. E. Jehin is F.R.S-FNRS Senior Research Associate.
This publication benefits from the support of the French Community of Belgium in the context of the FRIA Doctoral Grant awarded to MT. 

\end{acknowledgements}

\facilities{
TESS, Keck:I (HIRES), Magellan:Clay (PFS), WIYN (NEID, NESSI), CTIO:1.5m (CHIRON), FLWO:1.5m (TRES), FLWO:1.2m (KeplerCam), LCOGT, TRAPPIST-North, WASP, CMO:2.5m (SPP), Gemini:Gillett ('Alopeke), Gemini:South (Zorro), Hale (PHARO), SOAR (HRCam), Shane (ShARCS).
}

\software{
astropy \citep{Astropy13,Astropy18}, lightkurve \citep{Lightkurve18}, EXOFASTv2 \citep{ExoFAST_Eastman2013,ExoFASTv2_Eastman19}, SpecMatch-Emp \citep{SpecMatchEmp_Yee2017}, SpecMatch-Synth \citep{SpecMatchSynth_Petigura2015}, AstroImageJ \citep{AstroImageJ_Collins17}, TAPIR \citep{TAPIR_Jensen2013}, numpy \citep{Numpy}, scipy \citep{Scipy}, pandas \citep{Pandas20,Pandas_McKinney10}, matplotlib \citep{Matplotlib}.
}

\appendix
\section{Additional Fit Parameters}
We present in Table \ref{tab:additional_fit_params} the median and 68\% confidence intervals for additional fit parameters not listed in Table \ref{tab:fitted_props} for the adopted fits.
These are the linear and quadratic limb-darkening parameters $(u_1, u_2)$ in each band; additional flux dilution from neighboring stars in each band $(D)$; the relative RV offset for each instrument $\gamma_\mathrm{rel}$ (\ms); and the RV jitter for each instrument $\sigma_J$ (\ms).
\begin{ownpage}
\startlongtable
\begin{deluxetable*}{lcccccc} \label{tab:additional_fit_params}
\tablecaption{Additional Fit Parameters (Median and 68\% Confidence Intervals)}
\tabletypesize{\footnotesize}
\tablehead{\colhead{Parameter}}
\startdata
\input{aux_fit_results}
\enddata
\end{deluxetable*}
\end{ownpage}

\bibliography{manuscript,instruments,software,catalogs}
\bibliographystyle{aasjournal}

\end{document}

%% file: authors_alpha.tex
\author[0000-0003-0125-0039]{Owen~Alfaro}				
\affiliation{George Mason University, 4400 University Drive, Fairfax, VA 22030, USA}
\author[0000-0003-1464-9276]{Khalid~Barkaoui}			
\affiliation{Astrobiology Research Unit, Université de Liège, 19C Allée du 6 Août, 4000 Liège, Belgium}
\affiliation{Department of Earth, Atmospheric and Planetary Sciences, Massachusetts Institute of Technology, Cambridge, MA 02139, USA}
\affiliation{Instituto de Astrof\'isica de Canarias (IAC), E-38205 La Laguna, Tenerife, Spain}
\author[0000-0001-7708-2364]{Corey~Beard}				
\affiliation{Department of Physics \& Astronomy, University of California, Irvine, Irvine, CA 92697, USA}
\author[0000-0003-3469-0989]{Alexander~A.~Belinski}		
\affiliation{Sternberg Astronomical Institute, M.V. Lomonosov Moscow State University, 13, Universitetskij pr., 119234, Moscow, Russia}
\author[0000-0001-6285-9847]{Zouhair~Benkhaldoun}		
\affiliation{Oukaimeden Observatory, High Energy Physics and Astrophysics Laboratory, Faculty of sciences Semlalia, Cadi Ayyad University, Marrakech, Morocco}
\author[0000-0001-6981-8722]{Paul~Benni}				
\affiliation{Acton Sky Portal (Private Observatory), Acton, MA, USA}
\author[0000-0003-4647-7114]{Krzysztof~Bernacki}		
\affiliation{Silesian University of Technology, Department of Electronics, Electrical Engineering and Microelectronics, Akademicka 16, 44-100 Gliwice, Poland}
\author[0000-0001-6037-2971]{Andrew~W.~Boyle}			
\affiliation{Department of Astronomy, California Institute of Technology, Pasadena, CA 91125, USA}
\author[0000-0003-1305-3761]{R.~Paul~Butler}			
\affiliation{Earth and Planets Laboratory, Carnegie Institution for Science, 5241 Broad Branch Road, NW, Washington, DC 20015, USA}
\author[0000-0003-1963-9616]{Douglas~A.~Caldwell}		
\affiliation{SETI Institute}
\affiliation{NASA Ames Research Center, Moffett Field, CA 94035, USA}
\author[0000-0003-1125-2564]{Ashley~Chontos}			
\affiliation{Institute for Astronomy, University of Hawai'i, Honolulu, HI 96822, USA}
\author[0000-0002-8035-4778]{Jessie~L.~Christiansen}	
\affiliation{Caltech/IPAC-NASA Exoplanet Science Institute, 770 S. Wilson Avenue, Pasadena, CA 91106, USA}
\author[0000-0002-5741-3047]{David~R.~Ciardi}			
\affiliation{Caltech/IPAC-NASA Exoplanet Science Institute, 770 S. Wilson Avenue, Pasadena, CA 91106, USA}
\author[0000-0003-2781-3207]{Kevin~I.~Collins}			
\affiliation{George Mason University, 4400 University Drive, Fairfax, VA, 22030 USA}
\author[0000-0003-2239-0567]{Dennis~M.~Conti}
\affiliation{American Association of Variable Star Observers, 49 Bay State Road, Cambridge, MA 02138, USA}
\author[0000-0002-5226-787X]{Jeffrey~D.~Crane}			
\affiliation{Observatories of the Carnegie Institution for Science, 813 Santa Barbara Street, Pasadena, CA 91101, USA}
\author[0000-0002-6939-9211]{Tansu~Daylan}				
\affiliation{Department of Astrophysical Sciences, Princeton University, 4 Ivy Lane, Princeton, NJ 08544}
\affiliation{LSSTC Catalyst Fellow}
\author[0000-0001-8189-0233]{Courtney~D.~Dressing}		
\affiliation{Department of Astronomy,  University of California Berkeley, Berkeley, CA 94720, USA}
\author[0000-0003-3773-5142]{Jason~D.~Eastman}			
\affiliation{Center for Astrophysics \textbar \ Harvard \& Smithsonian, 60 Garden St, Cambridge, MA 02138, USA}
\author[0000-0002-2482-0180]{Zahra~Essack}				
\affiliation{Department of Earth, Atmospheric and Planetary Sciences, Massachusetts Institute of Technology, Cambridge, MA 02139, USA}
\affiliation{Department of Physics and Kavli Institute for Astrophysics and Space Research, Massachusetts Institute of Technology, Cambridge, MA 02139, USA}
\author[0000-0002-5674-2404]{Phil Evans}				
\affiliation{El Sauce Observatory, Coquimbo, 1870000, Chile}
\author[0000-0002-0885-7215]{Mark~E.~Everett}			
\affiliation{NSF’s National Optical-Infrared Astronomy Research Laboratory, 950 N. Cherry Avenue, Tucson, AZ 85719, USA}
\author[0000-0001-9309-0102]{Sergio~Fajardo-Acosta}							
\affiliation{NASA Exoplanet Science Institute, Caltech/IPAC, Mail Code 100-22, 1200 E. California Boulevard, Pasadena, CA 91125, USA}
\author[0000-0002-6482-2180]{Raquel~For\'{e}s-Toribio}	
\affiliation{Departamento de Astronom\'{\i}a y Astrof\'{\i}sica, Universidad de Valencia, E-46100 Burjassot, Valencia, Spain}
\affiliation{Observatorio Astron\'omico, Universidad de Valencia, E-46980 Paterna, Valencia, Spain} 
\author[0000-0001-9800-6248]{Elise~Furlan}				
\affiliation{NASA Exoplanet Science Institute, Caltech/IPAC, Mail Code 100-22, 1200 E. California Boulevard, Pasadena, CA 91125, USA}
\author{Mourad~Ghachoui}								
\affiliation{Oukaimeden Observatory, High Energy Physics and Astrophysics Laboratory, Faculty of sciences Semlalia, Cadi Ayyad University, Marrakech, Morocco}
\affiliation{Astrobiology Research Unit, Université de Liège, 19C Allée du 6 Août, 4000 Liège, Belgium}
\author{Micha\"{e}l~Gillon}								
\affiliation{Astrobiology Research Unit, Université de Liège, 19C Allée du 6 Août, 4000 Liège, Belgium}
\author[0000-0002-3439-1439]{Coel~Hellier}				
\affiliation{Astrophysics Group, Keele University, Staffordshire, ST5 5BG, UK}
\author[0000-0002-0473-4437]{Ian~Helm}					
\affiliation{George Mason University, 4400 University Drive, Fairfax, VA 22030, USA}
\author[0000-0001-8638-0320]{Andrew~W.~Howard}			
\affiliation{Department of Astronomy, California Institute of Technology, Pasadena, CA 91125, USA}
\author[0000-0002-2532-2853]{Steve~B.~Howell}			
\affiliation{NASA Ames Research Center, Moffett Field, CA 94035, USA}
\author[0000-0002-0531-1073]{Howard~Isaacson}			
\affiliation{Department of Astronomy,  University of California Berkeley, Berkeley, CA 94720, USA}
\affiliation{University of Southern Queensland, Centre for Astrophysics, West Street, Toowoomba, QLD 4350, Australia}
\author{Emmanuel~Jehin}									
\affiliation{Space sciences, Technologies and Astrophysics Research (STAR) Institute, Universit\'e de Li\`ege, Belgium}
\author[0000-0002-4715-9460]{Jon~M.~Jenkins}			
\affiliation{NASA Ames Research Center, Moffett Field, CA 94035, USA}
\author[0000-0002-4625-7333]{Eric~L.~N.~Jensen}
\affiliation{Department of Physics \& Astronomy, Swarthmore College, Swarthmore PA 19081, USA}
\author[0000-0003-0497-2651]{John F.\ Kielkopf}			
\affiliation{Department of Physics and Astronomy, University of Louisville, Louisville, KY 40292, USA}
\author{Didier~Laloum}									
\affiliation{American Association of Variable Star Observers, 49 Bay State Road, Cambridge, MA 02138, USA}
\author[0000-0001-5782-3719]{Naunet~Leonhardes-Barboza} 
\affiliation{Department of Astronomy, Wellesley College, Wellesley, MA 02481, USA}\author[0000-0003-0828-6368]{Pablo~Lewin}				
\affiliation{The Maury Lewin Astronomical Observatory, Glendora, CA 91741, USA}
\author[0000-0002-9632-9382]{Sarah~E.~Logsdon}			
\affiliation{NSF’s National Optical-Infrared Astronomy Research Laboratory, 950 N. Cherry Avenue, Tucson, AZ 85719, USA}
\author[0000-0001-7047-8681]{Jack~Lubin}				
\affiliation{Department of Physics \& Astronomy, University of California, Irvine, Irvine, CA 92697, USA}
\author[0000-0003-2527-1598]{Michael~B.~Lund}			
\affiliation{Caltech/IPAC-NASA Exoplanet Science Institute, 770 S. Wilson Avenue, Pasadena, CA 91106, USA}
\author[0000-0003-2562-9043]{Mason~G.~MacDougall}		
\affiliation{Department of Physics \& Astronomy, University of California Los Angeles, Los Angeles, CA 90095, USA}
\author[0000-0003-3654-1602]{Andrew~W.~Mann}			
\affiliation{Department of Physics and Astronomy, The University of North Carolina at Chapel Hill, Chapel Hill, NC 27599-3255, USA}
\author[0000-0003-4147-5195]{Natalia A.\ Maslennikova}
\affiliation{Sternberg Astronomical Institute, M.V. Lomonosov Moscow State University, 13, Universitetskij pr., 119234, Moscow, Russia}
\affiliation{Faculty of Physics, Moscow State University, 1 bldg. 2, Leninskie Gory, Moscow 119991, Russia}
\author[0000-0001-8879-7138]{Bob~Massey}				
\affiliation{Villa '39 Observatory, Landers, CA 92285, USA}
\author[0000-0001-9504-1486]{Kim~K.~McLeod}				
\affiliation{Department of Astronomy, Wellesley College, Wellesley, MA 02481, USA}
\author[0000-0001-9833-2959]{Jose~A.~Mu\~noz}			
\affiliation{Departamento de Astronom\'{\i}a y Astrof\'{\i}sica, Universidad de Valencia, E-46100 Burjassot, Valencia, Spain}
\affiliation{Observatorio Astron\'omico, Universidad de Valencia, E-46980 Paterna, Valencia, Spain}
\author[0000-0003-3848-3418]{Patrick~Newman}			
\affiliation{George Mason University, 4400 University Drive, Fairfax, VA 22030, USA}
\author[0000-0002-4650-2984]{Valeri Orlov}				
\affiliation{Instituto de Astronom\'{\i}a, Universidad Nacional Aut\'onoma de M\'exico,\\ Apdo. Postal 70-264, Cd. Universitaria, 04510\\ CDMX, M\'exico\\}
\author[0000-0002-8864-1667]{Peter~Plavchan}			
\affiliation{George Mason University, 4400 University Drive, Fairfax, VA 22030, USA}
\author[0000-0003-3184-5228]{Adam~Popowicz}				
\affiliation{Silesian University of Technology, Department of Electronics, Electrical Engineering and Microelectronics, Akademicka 16, 44-100 Gliwice, Poland}
\author[0000-0003-1572-7707]{Francisco~J.~Pozuelos}		
\affiliation{Astrobiology Research Unit, Université de Liège, 19C Allée du 6 Août, 4000 Liège, Belgium}
\affiliation{Space sciences, Technologies and Astrophysics Research (STAR) Institute, Universit\'e de Li\`ege, Belgium}
\affiliation{Instituto de Astrof\'isica de Andaluc\'ia (IAA-CSIC), Glorieta de la Astronom\'ia s/n, 18008 Granada, Spain}
\author{Tyler~A.~Pritchard}								
\affiliation{University of Maryland, College Park, MD 20742, USA}
\affiliation{NASA Goddard Space Flight Center, Greenbelt, MD 20771, USA}
\author[0000-0002-3940-2360]{Don~J.~Radford}			
\affiliation{Brierfield Observatory, New South Wales, Australia}
\author[0000-0003-4701-8497]{Michael~Reefe}				
\affiliation{George Mason University, 4400 University Drive, Fairfax, VA 22030, USA}
\affiliation{Department of Physics and Kavli Institute for Astrophysics and Space Research, Massachusetts Institute of Technology, Cambridge, MA 02139, USA}
\author[0000-0003-2058-6662]{George~R.~Ricker}			
\affiliation{Department of Physics and Kavli Institute for Astrophysics and Space Research, Massachusetts Institute of Technology, Cambridge, MA 02139, USA}
\author{Alexander~Rudat}								
\affiliation{Department of Physics and Kavli Institute for Astrophysics and Space Research, Massachusetts Institute of Technology, Cambridge, MA 02139, USA}
\author[0000-0003-1713-3208]{Boris~S.~Safonov}			
\affiliation{Sternberg Astronomical Institute, M.V. Lomonosov Moscow State University, 13, Universitetskij pr., 119234, Moscow, Russia}
\author[0000-0001-8227-1020]{Richard~P.~Schwarz}		
\affiliation{Center for Astrophysics \textbar \ Harvard \& Smithsonian, 60 Garden St, Cambridge, MA 02138, USA}
\author[0000-0001-9580-4869]{Heidi~Schweiker}			
\affiliation{NSF’s National Optical-Infrared Astronomy Research Laboratory, 950 N. Cherry Avenue, Tucson, AZ 85719, USA}
\author[0000-0003-1038-9702]{Nicholas~J.~Scott}			
\affiliation{NASA Ames Research Center, Moffett Field, CA 94035, USA}
\author[0000-0002-6892-6948]{S.~Seager}					
\affiliation{Department of Physics and Kavli Institute for Astrophysics and Space Research, Massachusetts Institute of Technology, Cambridge, MA 02139, USA}
\affiliation{Department of Earth, Atmospheric and Planetary Sciences, Massachusetts Institute of Technology, Cambridge, MA 02139, USA}
\affiliation{Department of Aeronautics and Astronautics, MIT, 77 Massachusetts Avenue, Cambridge, MA 02139, USA}
\author[0000-0002-8681-6136]{Stephen~A.~Shectman}		
\affiliation{Observatories of the Carnegie Institution for Science, 813 Santa Barbara Street, Pasadena, CA 91101, USA}
\author[0000-0003-2163-1437]{Chris Stockdale}			
\affiliation{Hazelwood Observatory, Australia}
\author[0000-0001-5603-6895]{Thiam-Guan~Tan}			
\affiliation{Perth Exoplanet Survey Telescope, Perth, Australia}
\affiliation{Curtin Institute of Radio Astronomy, Curtin University, Bentley, 6102, Australia}
\author{Johanna~K.~Teske}								
\affiliation{Earth and Planets Laboratory, Carnegie Institution for Science, 5241 Broad Branch Road, NW, Washington, DC 20015, USA}
\author{Neil~B.~Thomas} 								
\affiliation{Department of Astronautics, United States Air Force Academy, CO 80840, USA}
\author{Mathilde~Timmermans}							
\affiliation{Astrobiology Research Unit, Université de Liège, 19C Allée du 6 Août, 4000 Liège, Belgium}
\author[0000-0001-6763-6562]{Roland~Vanderspek}			
\affiliation{Department of Physics and Kavli Institute for Astrophysics and Space Research, Massachusetts Institute of Technology, Cambridge, MA 02139, USA}
\author[0000-0002-4501-564X]{David~Vermilion}			
\affiliation{George Mason University, 4400 University Drive, Fairfax, VA 22030, USA}
\author{David~Watanabe}									
\affiliation{Planetary Discoveries, Fredericksburg, VA 22405, USA}
\author[0000-0002-3725-3058]{Lauren~M.~Weiss}			
\affiliation{Department of Physics and Astronomy, University of Notre Dame, Notre Dame, IN 46556, USA}
\author[0000-0001-6604-5533]{Richard~G.~West}			
\affiliation{Centre for Exoplanets and Habitability, University of Warwick, Coventry, CV4 7AL, UK}
\affiliation{Department of Physics, University of Warwick, Coventry, CV4 7AL, UK}
\author[0000-0002-4290-6826]{Judah~Van~Zandt}			
\affiliation{Department of Physics \& Astronomy, University of California Los Angeles, Los Angeles, CA 90095, USA}
\author{Michal~Zejmo}									
\affiliation{Janusz Gil Institute of Astronomy, University of Zielona Gora, Prof. Szafrana 2, PL-65-516 Zielona Gora, Poland}
\author[0000-0002-0619-7639]{Carl~Ziegler}				
\affiliation{Department of Physics, Engineering and Astronomy, Stephen F. Austin State University, 1936 North Street, Nacogdoches, TX 75962, USA}

%% file: target_summary.tex
TOI-1937A\,b & 268301217 & 13.02 & $5814^{+91}_{-93}$ & $1.080^{+0.025}_{-0.024}$ & 0.947 & $1.247^{+0.059}_{-0.062}$ & $2.01^{+0.17}_{-0.16}$ \\
TOI-2364\,b & 39414571 & 12.09 & $5306^{+76}_{-68}$ & $0.886^{+0.021}_{-0.017}$ & 4.020 & $0.768^{+0.023}_{-0.018}$ & $0.225^{+0.043}_{-0.049}$ \\
TOI-2583A\,b & 7548817 & 12.46 & $5936^{+65}_{-68}$ & $1.477^{+0.036}_{-0.032}$ & 4.521 & $1.290^{+0.040}_{-0.033}$ & $0.250^{+0.058}_{-0.056}$ \\
TOI-2587A\,b & 68007716 & 11.41 & $5760^{+80}_{-79}$ & $1.726^{+0.049}_{-0.047}$ & 5.457 & $1.077^{+0.042}_{-0.040}$ & $0.218^{+0.054}_{-0.046}$ \\
TOI-2796\,b & 220076110 & 12.36 & $5764^{+81}_{-78}$ & $1.069 \pm 0.024$ & 4.808 & $1.59\,(> 1.54)$ & $0.44^{+0.10}_{-0.11}$ \\
TOI-2803A\,b & 124379043 & 12.43 & $6280^{+99}_{-96}$ & $1.245^{+0.022}_{-0.021}$ & 1.962 & $1.616^{+0.034}_{-0.032}$ & $0.975^{+0.083}_{-0.070}$ \\
TOI-2818\,b & 151483286 & 11.84 & $5721^{+88}_{-83}$ & $1.229^{+0.032}_{-0.031}$ & 4.040 & $1.363^{+0.046}_{-0.045}$ & $0.71 \pm 0.26$ \\
TOI-2842\,b & 178162579 & 12.46 & $5910 \pm 100$ & $1.265^{+0.040}_{-0.037}$ & 3.551 & $1.146^{+0.051}_{-0.048}$ & $0.370^{+0.052}_{-0.047}$ \\
TOI-2977\,b & 361343239 & 12.44 & $5691^{+94}_{-93}$ & $1.073^{+0.024}_{-0.020}$ & 2.351 & $1.174^{+0.031}_{-0.027}$ & $1.68^{+0.26}_{-0.25}$ \\
TOI-3023\,b & 454248975 & 12.03 & $5760^{+85}_{-88}$ & $1.668^{+0.046}_{-0.033}$ & 3.901 & $1.466^{+0.043}_{-0.032}$ & $0.62^{+0.10}_{-0.09}$ \\
TOI-3364\,b & 280655495 & 11.29 & $5706^{+95}_{-91}$ & $1.419^{+0.036}_{-0.030}$ & 5.877 & $1.091^{+0.038}_{-0.032}$ & $1.67^{+0.12}_{-0.13}$ \\
TOI-3688A\,b & 245509452 & 12.37 & $5950 \pm 100$ & $1.302^{+0.038}_{-0.035}$ & 3.246 & $1.167^{+0.048}_{-0.044}$ & $0.98^{+0.10}_{-0.11}$ \\
TOI-3807\,b & 289661991 & 12.03 & $5772^{+84}_{-80}$ & $1.468 \pm 0.037$ & 2.899 & $2.00\,(> 1.65)$ & $1.04^{+0.15}_{-0.14}$ \\
TOI-3819\,b & 95660472 & 12.41 & $5859^{+72}_{-71}$ & $1.538 \pm 0.037$ & 3.244 & $1.172^{+0.036}_{-0.035}$ & $1.11^{+0.18}_{-0.20}$ \\
TOI-3912\,b & 156648452 & 12.31 & $5725^{+69}_{-68}$ & $1.392^{+0.035}_{-0.034}$ & 3.494 & $1.274^{+0.041}_{-0.040}$ & $0.406^{+0.071}_{-0.068}$ \\
TOI-3976A\,b & 154293917 & 12.22 & $5975^{+70}_{-69}$ & $1.501^{+0.039}_{-0.038}$ & 6.608 & $1.095^{+0.036}_{-0.035}$ & $0.175^{+0.037}_{-0.036}$ \\
TOI-4087\,b & 310002617 & 11.71 & $6060^{+74}_{-67}$ & $1.112^{+0.021}_{-0.020}$ & 3.177 & $1.164^{+0.025}_{-0.024}$ & $0.73 \pm 0.14$ \\
TOI-4145A\,b & 279947414 & 12.06 & $5281^{+86}_{-76}$ & $0.859^{+0.018}_{-0.017}$ & 4.066 & $1.187^{+0.032}_{-0.031}$ & $0.43 \pm 0.13$ \\
TOI-4463A\,b & 8599009 & 10.95 & $5640^{+89}_{-82}$ & $1.062^{+0.027}_{-0.024}$ & 2.881 & $1.183^{+0.064}_{-0.045}$ & $0.794^{+0.039}_{-0.040}$ \\
TOI-4791\,b & 100389539 & 11.32 & $6058^{+99}_{-94}$ & $1.409^{+0.039}_{-0.038}$ & 4.281 & $1.110 \pm 0.050$ & $2.31^{+0.32}_{-0.33}$ \\

%% file: tess_summary.tex
TOI-1937A & 7,9 & CDIPS & 1800 \\
$\cdots$ & 34--36 & SPOC & 120 \\
TOI-2364 & 6 & SPOC & 1800 \\
$\cdots$ & 33 & SPOC & 600 \\
TOI-2583A & 14,25,26 & SPOC & 1800 \\
$\cdots$ & 40 & SPOC & 120 \\
TOI-2587A & 7 & SPOC & 1800 \\
$\cdots$ & 34 & SPOC & 600 \\
TOI-2796 & 6 & SPOC & 1800 \\
$\cdots$ & 32 & SPOC & 600 \\
TOI-2803A & 6 & QLP & 1800 \\
$\cdots$ & 33 & QLP & 600 \\
TOI-2818 & 7,8 & QLP & 1800 \\
$\cdots$ & 34 & QLP & 600 \\
TOI-2842 & 7 & QLP & 1800 \\
$\cdots$ & 33 & QLP & 600 \\
TOI-2977 & 9 & QLP & 1800 \\
$\cdots$ & 36,37 & QLP & 600 \\
TOI-3023 & 10,12 & SPOC & 1800 \\
$\cdots$ & 37 & QLP & 600 \\
$\cdots$ & 38 & SPOC & 600 \\
TOI-3364 & 9 & SPOC & 1800 \\
$\cdots$ & 35,36 & SPOC & 600 \\
TOI-3688A & 18 & QLP & 1800 \\
TOI-3807 & 14,20 & SPOC & 1800 \\
$\cdots$ & 47 & SPOC & 120 \\
TOI-3819 & 20 & SPOC & 1800 \\
$\cdots$ & 44--47 & SPOC & 120 \\
TOI-3912 & 23 & SPOC & 1800 \\
$\cdots$ & 50 & SPOC & 120 \\
TOI-3976A & 16,23,24 & SPOC & 1800 \\
TOI-4087 & 14,19,21,26 & SPOC & 1800 \\
$\cdots$ & 40,41,47 & SPOC & 120 \\
TOI-4145A & 18,19,25,26 & SPOC & 1800 \\
TOI-4463A & 26 & SPOC & 1800 \\
TOI-4791 & 33,34 & SPOC & 600 \\

%% file: sg1_summary.tex
TOI-1937A & LCO SSO/Sinistro & 1.0 & $i^\prime$ & 2020 Jan 22 & 207 & Y & 0.8 & BJD$_\mathrm{TDB}$, (BJD$_\mathrm{TDB}$)$^2$, Shape (S) \\
$\cdots$ & LCO CTIO/Sinistro & 1.0 & $i^\prime$ & 2020 Jan 29 & 207 & Y & 1.0 & BJD$_\mathrm{TDB}$, (BJD$_\mathrm{TDB}$)$^2$, Shape (S) \\
$\cdots$ & LCO SAAO/Sinistro & 1.0 & $i^\prime$ & 2020 Jan 31 & 207 & Y & 2.7 & BJD$_\mathrm{TDB}$, (BJD$_\mathrm{TDB}$)$^2$, Shape (S) \\
$\cdots$ & LCO CTIO/Sinistro & 1.0 & $g^\prime$ & 2020 Feb 13 & 207 & Y & 1.2 & BJD$_\mathrm{TDB}$, (BJD$_\mathrm{TDB}$)$^2$, Shape (S) \\
$\cdots$ & El Sauce & 0.36 & $R$ & 2020 Feb 14 & 180 & Y & 4.0 & Airmass \\
$\cdots$ & LCO SSO/Sinistro & 1.0 & $z^\prime$ & 2020 Apr 22 & 207 & Y & 1.5 & BJD$_\mathrm{TDB}$, (BJD$_\mathrm{TDB}$)$^2$, Shape (S) \\
$\cdots$ & El Sauce & 0.36 & $V$ & 2020 Dec 29 & 180 & Y & 5.7 & Airmass \\
TOI-2364 & TRAPPIST-North & 0.6 & $z^\prime$ & 2020 Nov 28 & 25 & N & -- & -- \\
$\cdots$ & Hazelwood & 0.318 & $R$ & 2021 Jan 08 & 180 & Y & 2.2 & Total Counts \\
TOI-2583A & WASP & -- & WASP & 2004 May 14 & 40 & N & -- & -- \\
$\cdots$ & TRAPPIST-North & 0.6 & $I+z$ & 2021 May 07 & 20 & Y & 2.7 & Meridian Flip \\
$\cdots$ & OAUV/T50 & 0.5 & $R$ & 2021 May 16 & 150 & N & -- & -- \\
$\cdots$ & FLWO/KeplerCam & 1.2 & $B$ & 2021 Jun 13 & 180 & Y & 2.3 & Airmass \\
$\cdots$ & FLWO/KeplerCam & 1.2 & $z^\prime$ & 2021 Jun 13 & 180 & Y & 2.2 & Airmass \\
$\cdots$ & Mt. Lemmon/ULMT & 0.6 & $g^\prime$ & 2021 Jun 22 & 128 & Y & 2.1 & -- \\
TOI-2587A & FLWO/KeplerCam & 1.2 & $i^\prime$ & 2021 Mar 31 & 28 & Y & 7.2 & Airmass \\
$\cdots$ & Mt. Lemmon/ULMT & 0.6 & $r^\prime$ & 2021 Mar 31 & 64 & Y & 1.9 & Total Counts, J.D.-2400000 \\
TOI-2796 & LCO SSO/SBIG-6303 & 0.4 & $i^\prime$ & 2021 Sep 01 & 120 & Y & 3.4 & Airmass, BJD$_\mathrm{TDB}$ \\
$\cdots$ & El Sauce & 0.36 & $B$ & 2021 Nov 03 & 180 & Y & 7.1 & Airmass, Sky/Pixel \\
$\cdots$ & CMO/RC600 & 0.6 & $g^\prime$ & 2021 Dec 06 & 50 & Y & 2.3 & Airmass \\
TOI-2803A & WASP & -- & WASP & 2006 Oct 20 & 40 & N & -- & -- \\
$\cdots$ & Brierfield & 0.36 & $R$ & 2021 Oct 06 & 180 & Y & 2.7 & Airmass \\
$\cdots$ & LCO CTIO/SBIG-6303 & 0.4 & $i^\prime$ & 2021 Oct 25 & 140 & Y & 3.9 & Airmass, Total Counts, FWHM \\
$\cdots$ & El Sauce & 0.36 & $B$ & 2021 Oct 28 & 180 & Y & 4.1 & Airmass \\
$\cdots$ & LCO SSO/SBIG-6303 & 0.4 & $g^\prime$ & 2021 Dec 02 & 140 & Y & 6.7 & -- \\
$\cdots$ & Brierfield & 0.36 & $B$ & 2021 Dec 02 & 240 & Y & 7.0 & Airmass \\
TOI-2818 & El Sauce & 0.51 & $R$ & 2021 Dec 08 & 30 & Y & 3.0 & Airmass, Total Counts \\
$\cdots$ & LCO CTIO/SBIG-6303 & 0.4 & $g^\prime$ & 2021 Dec 12 & 60 & Y & 9.6 & -- \\
TOI-2842 & El Sauce & 0.36 & $R$ & 2021 Nov 05 & 180 & Y & 3.3 & Airmass \\
$\cdots$ & El Sauce & 0.51 & $B$ & 2021 Dec 07 & 180 & Y & 2.8 & Airmass, Sky/Pixel \\
TOI-2977 & El Sauce & 0.51 & $R$ & 2022 Jan 08 & 60 & Y & 2.3 & Y position \\
$\cdots$ & PEST & 0.3 & $g^\prime$ & 2022 Feb 26 & 250 & Y & 3.3 & FWHM \\
$\cdots$ & PEST & 0.3 & $i^\prime$ & 2022 Feb 26 & 250 & Y & 4.7 & FWHM \\
TOI-3023 & Brierfield & 0.36 & $i^\prime$ & 2021 Jun 25 & 240 & Y & 2.2 & -- \\
$\cdots$ & El Sauce & 0.51 & $R$ & 2022 Jan 26 & 60 & Y & 3.0 & Airmass, FWHM \\
TOI-3364 & Hazelwood & 0.318 & $R$ & 2021 Dec 12 & 120 & Y & 2.6 & FWHM \\
TOI-3688A & MLO & 0.356 & $i^\prime$ & 2021 Oct 09 & 90 & N & -- & -- \\
$\cdots$ & FLWO/KeplerCam & 1.2 & $i^\prime$ & 2021 Dec 01 & 34 & Y & 2.7 & Airmass \\
TOI-3807 & GMU/SBIG-16803 & 0.8 & $R$ & 2021 Nov 23 & 60 & Y & 3.8 & Airmass, FWHM, Sky/Pixel \\
$\cdots$ & Acton Sky Portal & 0.36 & $r^\prime$ & 2021 Nov 24 & 35 & Y & 5.1 & Airmass \\
$\cdots$ & FLWO/KeplerCam & 1.2 & $i^\prime$ & 2022 Feb 16 & 96 & Y & 4.5 & Airmass \\
$\cdots$ & FLWO/KeplerCam & 1.2 & $B$ & 2022 Feb 16 & 96 & Y & 4.7 & Airmass \\
$\cdots$ & FLWO/KeplerCam & 1.2 & $B$ & 2022 Mar 23 & 120 & Y & 2.4 & FWHM \\
$\cdots$ & FLWO/KeplerCam & 1.2 & $z$ & 2022 Mar 23 & 120 & Y & 2.9 & FWHM \\
TOI-3819 & GMU/SBIG-16803 & 0.8 & $R$ & 2021 Nov 19 & 70 & N & -- & -- \\
$\cdots$ & GMU/SBIG-16803 & 0.8 & $R$ & 2022 Feb 11 & 70 & Y & 4.4 & Airmass, Sky/Pixel, X position \\
$\cdots$ & FLWO/KeplerCam & 1.2 & $i^\prime$ & 2022 Apr 18 & 40 & Y & 1.9 & Airmass \\
TOI-3912 & OAUV/T50 & 0.5 & $R$ & 2021 Jul 15 & 120 & Y & 3.1 & BJD$_\mathrm{TDB}$, Airmass \\
$\cdots$ & SUTO/OTIVAR & 0.3 & $B$ & 2022 Mar 20 & 300 & Y & 6.2 & Airmass \\
$\cdots$ & FLWO/KeplerCam & 1.2 & $i^\prime$ & 2022 Mar 24 & 40 & Y & 2.1 & Airmass, Total Counts \\
$\cdots$ & Villa '39 & 0.355 & $B$ & 2022 Mar 31 & 360 & Y & 2.8 & Airmass \\
$\cdots$ & Villa '39 & 0.355 & $i^\prime$ & 2022 Mar 31 & 240 & Y & 2.5 & Airmass, Meridian Flip \\
TOI-3976A & FLWO/KeplerCam & 1.2 & $i^\prime$ & 2022 Apr 22 & 36 & Y & 2.8 & Airmass \\
TOI-4087 & OAUV/T50 & 0.5 & $R$ & 2021 Oct 04 & 75 & Y & 4.0 & Total Counts, Y position \\
$\cdots$ & SUTO/OTIVAR & 0.3 & $B$ & 2022 Apr 19 & 300 & Y & 3.8 & Airmass \\
TOI-4145A & OPM/RC8 & 0.2 & $I$ & 2021 Aug 23 & 120 & Y & 8.8 & Airmass \\
$\cdots$ & SUTO/OTIVAR & 0.3 & $B$ & 2021 Dec 28 & 180 & Y & 6.3 & Airmass \\
TOI-4463A & Mt. Lemmon/ULMT & 0.6 & $r^\prime$ & 2022 Apr 25 & 48 & Y & 1.5 & Total Counts \\
$\cdots$ & Brierfield & 0.36 & $R$ & 2022 Jun 07 & 120 & Y & 6.8 & Airmass \\
$\cdots$ & Whitin/CDK700 & 0.7 & $z^\prime$ & 2022 Jun 16 & 30 & Y & 2.5 & Airmass \\
$\cdots$ & Whitin/CDK700 & 0.7 & $g^\prime$ & 2022 Jun 16 & 30 & Y & 1.8 & Airmass \\
TOI-4791 & PEST & 0.3 & $R$ & 2022 Jan 17 & 120 & Y & 3.0 & FWHM, Sky/Pixel\\

%% file: sg3_summary.tex
TOI-1937A & Gemini-S (8 m) & Zorro & 562 nm & 2020 Mar 13 & Speckle & $\Delta$mag = 5.01 at $0\farcs5$ \\
$\cdots$ & Gemini-S (8 m) & Zorro & 832 nm & 2020 Mar 13 & Speckle & $\Delta$mag = 5.97 at $0\farcs5$ \\
$\cdots$ & SOAR (4.1 m) & HRCam & $I_c$ & 2020 Dec 03 & Speckle & $\Delta$mag = 4.6 at $1\farcs0$ \\
TOI-2364 & SOAR (4.1 m) & HRCam & $I_c$ & 2020 Dec 03 & Speckle & $\Delta$mag = 6.2 at $1\farcs0$ \\
$\cdots$ & Palomar (5 m) & PHARO & Br$\gamma$ & 2021 Feb 24 & AO & $\Delta$mag = 6.791 at $0\farcs5$ \\
$\cdots$ & Gemini-S (8 m) & Zorro & 562 nm & 2021 Feb 27 & Speckle & $\Delta$mag = 5.01 at $0\farcs5$ \\
$\cdots$ & Gemini-S (8 m) & Zorro & 832 nm & 2021 Feb 27 & Speckle & $\Delta$mag = 6.2 at $0\farcs5$ \\
TOI-2583A & Shane (3 m) & ShARCS & $J$ & 2021 Jun 01 & AO & -- \\
$\cdots$ & Shane (3 m) & ShARCS & $K_s$ & 2021 Jun 01 & AO & -- \\
TOI-2587A & WIYN (3.5 m) & NESSI & 562 nm & 2021 Apr 24 & Speckle & $\Delta$mag = 4.0 at $1\farcs0$ \\
$\cdots$ & WIYN (3.5 m) & NESSI & 832 nm & 2021 Apr 24 & Speckle & $\Delta$mag = 5.1 at $1\farcs0$ \\
$\cdots$ & SOAR (4.1 m) & HRCam & $I_c$ & 2021 Nov 20 & Speckle & $\Delta$mag = 6.6 at $1\farcs0$ \\
TOI-2796 & SOAR (4.1 m) & HRCam & $I_c$ & 2021 Oct 18 & Speckle & $\Delta$mag = 6.8 at $1\farcs0$ \\
TOI-2803A & SOAR (4.1 m) & HRCam & $I_c$ & 2021 Oct 01 & Speckle & $\Delta$mag = 5.6 at $1\farcs0$ \\
TOI-2818 & SOAR (4.1 m) & HRCam & $I_c$ & 2021 Oct 01 & Speckle & $\Delta$mag = 6.8 at $1\farcs0$ \\
TOI-2842 & SOAR (4.1 m) & HRCam & $I_c$ & 2021 Nov 20 & Speckle & $\Delta$mag = 6.4 at $1\farcs0$ \\
TOI-2977 & SOAR (4.1 m) & HRCam & $I_c$ & 2022 Mar 20 & Speckle & $\Delta$mag = 5.7 at $1\farcs0$ \\
TOI-3023 & SOAR (4.1 m) & HRCam & $I_c$ & 2022 Apr 15 & Speckle & $\Delta$mag = 5.6 at $1\farcs0$ \\
TOI-3364 & SOAR (4.1 m) & HRCam & $I_c$ & 2021 Nov 20 & Speckle & $\Delta$mag = 7.1 at $1\farcs0$ \\
TOI-3688A & SAI-2.5m (2.5 m) & Speckle Polarimeter & $I_c$ & 2021 Sep 09 & Speckle & $\Delta$mag = 4.6 at $1\farcs0$ \\
TOI-3807 & WIYN (3.5 m) & NESSI & 832 nm & 2022 Apr 17 & Speckle & $\Delta$mag = 4.9 at $1\farcs0$ \\
TOI-3912 & WIYN (3.5 m) & NESSI & 832 nm & 2022 Apr 18 & Speckle & $\Delta$mag = 5.0 at $1\farcs0$ \\
TOI-3976A & WIYN (3.5 m) & NESSI & 832 nm & 2022 Apr 21 & Speckle & $\Delta$mag = 5.2 at $1\farcs0$ \\
TOI-4087 & WIYN (3.5 m) & NESSI & 832 nm & 2022 Apr 20 & Speckle & $\Delta$mag = 4.6 at $1\farcs0$ \\
$\cdots$ & SAI-2.5m (2.5 m) & Speckle Polarimeter & $I_c$ & 2022 May 13 & Speckle & $\Delta$mag = 6.6 at $1\farcs0$ \\
TOI-4145A & SAI-2.5m (2.5 m) & Speckle Polarimeter & $I_c$ & 2021 Oct 30 & Speckle & $\Delta$mag = 5.1 at $1\farcs0$ \\
$\cdots$ & Gemini-N (8 m) & 'Alopeke & 562 nm & 2022 Feb 15 & Speckle & $\Delta$mag = 3.64 at $0\farcs5$ \\
$\cdots$ & Gemini-N (8 m) & 'Alopeke & 832 nm & 2022 Feb 15 & Speckle & $\Delta$mag = 5.25 at $0\farcs5$ \\
TOI-4463A & SAI-2.5m (2.5 m) & Speckle Polarimeter & $I_c$ & 2021 Oct 22 & Speckle & $\Delta$mag = 6.3 at $1\farcs0$ \\
$\cdots$ & SOAR (4.1 m) & HRCam & $I_c$ & 2022 Apr 15 & Speckle & $\Delta$mag = 6.9 at $1\farcs0$ \\
$\cdots$ & WIYN (3.5 m) & NESSI & 832 nm & 2022 May 05 & Speckle & $\Delta$mag = 5.5 at $1\farcs0$ \\
$\cdots$ & Palomar (5 m) & PHARO & Br$\gamma$ & 2022 May 21 & AO & $\Delta$mag = 6.755 at $0\farcs5$ \\
$\cdots$ & Palomar (5 m) & PHARO & $H$cont & 2022 May 21 & AO & $\Delta$mag = 7.385 at $0\farcs5$ \\

%% file: rv_summary.tex
TOI-1937A & Magellan-Clay/PFS & 13 & 8.5 & 2020 Feb 04 & 2020 Nov 04 \\
TOI-2364 & Magellan-Clay/PFS & 6 & 3.2 & 2021 Oct 19 & 2022 Jan 23 \\
$\cdots$ & FLWO/TRES & 4 & 38.1 & 2017 Dec 22 & 2019 Oct 31 \\
TOI-2583A & Keck-I/HIRES & 6 & 6.5 & 2021 Oct 10 & 2022 Jun 11 \\
$\cdots$ & FLWO/TRES & 2 & 44.5 & 2021 Mar 22 & 2021 Mar 29 \\
TOI-2587A & WIYN/NEID & 10 & 6.7 & 2021 Nov 10 & 2022 Apr 22 \\
$\cdots$ & FLWO/TRES & 2 & 35.9 & 2021 Apr 23 & 2021 May 07 \\
TOI-2796 & Keck-I/HIRES & 3 & 5.5 & 2021 Nov 24 & 2022 Jan 08 \\
$\cdots$ & WIYN/NEID & 7 & 8.8 & 2021 Dec 06 & 2022 Jan 04 \\
$\cdots$ & FLWO/TRES & 2 & 48.9 & 2021 Oct 01 & 2021 Oct 13 \\
TOI-2803A & Magellan-Clay/PFS & 6 & 6.0 & 2022 Jan 13 & 2022 Mar 24 \\
$\cdots$ & FLWO/TRES & 3 & 87.8 & 2021 Nov 11 & 2021 Nov 14 \\
TOI-2818 & CTIO-1.5m/CHIRON & 7 & 27.0 & 2021 Dec 21 & 2022 Mar 15 \\
TOI-2842 & Magellan-Clay/PFS & 6 & 6.8 & 2022 Jan 13 & 2022 Mar 25 \\
$\cdots$ & FLWO/TRES & 2 & 51.4 & 2021 Nov 02 & 2021 Nov 11 \\
TOI-2977 & CTIO-1.5m/CHIRON & 6 & 33.5 & 2022 Mar 24 & 2022 May 22 \\
TOI-3023 & Magellan-Clay/PFS & 6 & 5.0 & 2022 Jan 21 & 2022 Jun 19 \\
TOI-3364 & Magellan-Clay/PFS & 6 & 2.3 & 2022 Jan 13 & 2022 Mar 22 \\
TOI-3688A & WIYN/NEID & 7 & 11.3 & 2021 Nov 28 & 2022 Jan 10 \\
$\cdots$ & FLWO/TRES & 2 & 47.9 & 2021 Sep 18 & 2021 Oct 09 \\
TOI-3807 & WIYN/NEID & 5 & 9.4 & 2022 Mar 07 & 2022 Jun 07 \\
$\cdots$ & FLWO/TRES & 2 & 31.2 & 2021 Nov 29 & 2021 Dec 09 \\
TOI-3819 & WIYN/NEID & 6 & 11.4 & 2021 Dec 13 & 2022 Jan 07 \\
$\cdots$ & FLWO/TRES & 2 & 39.0 & 2021 Nov 11 & 2021 Nov 19 \\
TOI-3912 & Keck-I/HIRES & 7 & 5.7 & 2022 Apr 21 & 2022 May 30 \\
$\cdots$ & FLWO/TRES & 2 & 33.8 & 2022 Feb 11 & 2022 Feb 16 \\
TOI-3976A & Keck-I/HIRES & 12 & 6.0 & 2022 Apr 21 & 2022 Aug 01 \\
$\cdots$ & FLWO/TRES & 2 & 32.0 & 2022 Jan 31 & 2022 Feb 16 \\
TOI-4087 & WIYN/NEID & 7 & 14.2 & 2022 May 09 & 2022 Jun 07 \\
$\cdots$ & FLWO/TRES & 8 & 34.7 & 2022 Feb 06 & 2022 Jun 14 \\
TOI-4145A & WIYN/NEID & 8 & 7.2 & 2021 Nov 28 & 2022 Mar 18 \\
$\cdots$ & FLWO/TRES & 2 & 34.4 & 2021 Sep 20 & 2021 Sep 22 \\
TOI-4463A & WIYN/NEID & 6 & 3.9 & 2022 Mar 07 & 2022 Jun 01 \\
$\cdots$ & FLWO/TRES & 2 & 27.5 & 2021 Sep 16 & 2021 Sep 29 \\
TOI-4791 & CTIO-1.5m/CHIRON & 6 & 52.0 & 2022 Mar 16 & 2022 Mar 26 \\
$\cdots$ & FLWO/TRES & 2 & 47.5 & 2022 Jan 10 & 2022 Feb 07 \\

%% file: target_props_0.tex
\tablecolumns{7}
\tablehead{
\colhead{\textbf{Target}} & \colhead{TOI-1937} & \colhead{TOI-2364} & \colhead{TOI-2583} & \colhead{TOI-2587} & \colhead{TOI-2796} & \colhead{Source} 
}
\startdata
\multicolumn{7}{l}{\textbf{Identifiers}}\\
TIC & 268301217 & 39414571 & 7548817 & 68007716 & 220076110 & \\
Gaia DR3 & 5489726768531119616 & 3022644120017418496 & 2115776451371641984 & 3085216850017388672 & 3222453935725951488 & \\
2MASS & 07452898-5222599 & 05563119-0500430 & 18090407+4520127 & 07434358-0104235 & 05363665+0053466 & \\
Tycho-2 & -- & -- & -- & 4831-01170-1 & -- & \\
WISE & J074528.97-522259.6 & J055631.19-050043.2 & J180904.08+452012.7 & J074343.60-010423.9 & J053636.65+005346.5 & \\
\multicolumn{7}{l}{\textbf{Astrometric Measurements}}\\
R.A. (J2000) & 07:45:28.973 & 05:56:31.207 & 18:09:04.087 & 07:43:43.611 & 05:36:36.653 & 1\\
Decl. (J2000) & -52:22:59.73 & -05:00:43.31 & +45:20:12.84 & -01:04:24.01 & +00:53:46.61 & 1\\
$\mu_{{\alpha}}$ (mas yr$^{{-1}}$) & -5.627 $\pm$ 0.013 & 12.225 $\pm$ 0.019 & 7.831 $\pm$ 0.011 & 22.004 $\pm$ 0.023 & 2.537 $\pm$ 0.018 & 1\\
$\mu_{{\delta}}$ (mas yr$^{{-1}}$) & 11.309 $\pm$ 0.013 & -15.462 $\pm$ 0.015 & 4.991 $\pm$ 0.012 & -25.296 $\pm$ 0.019 & 2.635 $\pm$ 0.012 & 1\\
Parallax (mas) & 2.389 $\pm$ 0.011 & 4.574 $\pm$ 0.019 & 1.781 $\pm$ 0.009 & 2.626 $\pm$ 0.022 & 2.838 $\pm$ 0.015 & 1\\
$b$ ($^{{\circ}}$) & -13.549 & -14.493 & 26.060 & 11.112 & -16.157 & 1\\
$l$ ($^{{\circ}}$) & 265.308 & 211.153 & 72.654 & 219.944 & 203.334 & 1\\
\multicolumn{7}{l}{\textbf{Photometric Measurements}}\\
$T$ (mag) & 12.493 $\pm$ 0.006 & 11.552 $\pm$ 0.006 & 12.045 $\pm$ 0.007 & 10.966 $\pm$ 0.006 & 11.916 $\pm$ 0.022 & 2\\
$G$ (mag) & 13.005 $\pm$ 0.003 & 12.082 $\pm$ 0.003 & 12.453 $\pm$ 0.003 & 11.405 $\pm$ 0.003 & 12.353 $\pm$ 0.003 & 1\\
$G_\mathrm{{BP}}$ (mag) & 13.417 $\pm$ 0.003 & 12.511 $\pm$ 0.003 & 12.759 $\pm$ 0.003 & 11.738 $\pm$ 0.003 & 12.695 $\pm$ 0.003 & 1\\
$G_\mathrm{{RP}}$ (mag) & 12.421 $\pm$ 0.004 & 11.488 $\pm$ 0.004 & 11.983 $\pm$ 0.004 & 10.909 $\pm$ 0.004 & 11.851 $\pm$ 0.004 & 1\\
$B_T$ (mag) & -- & -- & -- & 12.263 $\pm$ 0.226 & -- & 3\\
$V_T$ (mag) & -- & -- & -- & 11.612 $\pm$ 0.173 & -- & 3\\
$J$ (mag) & 11.717 $\pm$ 0.029\tablenotemark{a} & 10.909 $\pm$ 0.024 & -- & 10.354 $\pm$ 0.023 & 11.286 $\pm$ 0.022 & 4\\
$H$ (mag) & 11.324 $\pm$ 0.026\tablenotemark{a} & 10.436 $\pm$ 0.022 & 11.192 $\pm$ 0.020 & 10.079 $\pm$ 0.023 & 10.976 $\pm$ 0.022 & 4\\
$K_s$ (mag) & 11.226 $\pm$ 0.021\tablenotemark{a} & 10.363 $\pm$ 0.021 & -- & 9.991 $\pm$ 0.023 & 10.936 $\pm$ 0.023 & 4\\
$W1$ (mag) & 11.135 $\pm$ 0.023\tablenotemark{a} & 10.290 $\pm$ 0.022 & 11.082 $\pm$ 0.023\tablenotemark{b} & 9.954 $\pm$ 0.022 & 10.885 $\pm$ 0.023 & 5\\
$W2$ (mag) & 11.155 $\pm$ 0.020\tablenotemark{a} & 10.362 $\pm$ 0.020 & 11.112 $\pm$ 0.021\tablenotemark{b} & 9.998 $\pm$ 0.019 & 10.942 $\pm$ 0.021 & 5\\
$W3$ (mag) & 11.160 $\pm$ 0.086\tablenotemark{a} & 10.657 $\pm$ 0.105 & 11.015 $\pm$ 0.088\tablenotemark{b} & 9.940 $\pm$ 0.060 & 10.840 $\pm$ 0.127 & 5\\
\multicolumn{\totalcolumns}{c}{\vrule height 24pt width 0pt \small\it\fnum@table\ continued on next page}
\enddata

%% file: target_props_1.tex
\tablecolumns{7}
\tablehead{
\colhead{\textbf{Target}} & \colhead{TOI-2803} & \colhead{TOI-2818} & \colhead{TOI-2842} & \colhead{TOI-2977}\tablenotemark{c} & \colhead{TOI-3023} & \colhead{Source} 
}
\startdata
\multicolumn{7}{l}{\textbf{Identifiers}}\\
TIC & 124379043 & 151483286 & 178162579 & 361343239 & 454248975 & \\
Gaia DR3 & 2913482170369046656 & 5545555570947605376 & 3046327742219111808 & 5305241494923209856 & 5227928170880093568 & \\
2MASS & 06122753-2329331 & 07561435-3507000 & 07121060-1038217 & 09354239-6021271 & 11011602-7221246 & \\
Tycho-2 & -- & -- & -- & -- & -- & \\
WISE & J061227.53-232933.0 & J075614.35-350700.1 & J071210.60-103821.6 & J093542.37-602126.9 & J110115.86-722124.2 & \\
\multicolumn{7}{l}{\textbf{Astrometric Measurements}}\\
R.A. (J2000) & 06:12:27.538 & 07:56:14.356 & 07:12:10.610 & 09:35:42.363 & 11:01:15.790 & 1\\
Decl. (J2000) & -23:29:32.98 & -35:07:00.16 & -10:38:21.62 & -60:21:26.90 & -72:21:24.12 & 1\\
$\mu_{{\alpha}}$ (mas yr$^{{-1}}$) & 0.749 $\pm$ 0.009 & -6.034 $\pm$ 0.010 & 3.268 $\pm$ 0.013 & -21.720 $\pm$ 0.118 & -66.732 $\pm$ 0.011 & 1\\
$\mu_{{\delta}}$ (mas yr$^{{-1}}$) & 11.908 $\pm$ 0.011 & -8.990 $\pm$ 0.012 & 4.195 $\pm$ 0.014 & 20.892 $\pm$ 0.108 & 25.773 $\pm$ 0.011 & 1\\
Parallax (mas) & 1.997 $\pm$ 0.012 & 3.170 $\pm$ 0.010 & 2.174 $\pm$ 0.013 & 2.919 $\pm$ 0.094 & 2.546 $\pm$ 0.009 & 1\\
$b$ ($^{{\circ}}$) & -18.658 & -3.394 & -0.295 & -6.150 & -11.266 & 1\\
$l$ ($^{{\circ}}$) & 230.374 & 251.037 & 224.751 & 280.905 & 294.716 & 1\\
\multicolumn{7}{l}{\textbf{Photometric Measurements}}\\
$T$ (mag) & 12.075 $\pm$ 0.006 & 11.394 $\pm$ 0.006 & 12.033 $\pm$ 0.006 & 11.938 $\pm$ 0.007 & 11.485 $\pm$ 0.006 & 2\\
$G$ (mag) & 12.417 $\pm$ 0.003 & 11.830 $\pm$ 0.003 & 12.455 $\pm$ 0.003 & 12.492 $\pm$ 0.003 & 12.028 $\pm$ 0.003 & 1\\
$G_\mathrm{{BP}}$ (mag) & 12.669 $\pm$ 0.003 & 12.166 $\pm$ 0.003 & 12.778 $\pm$ 0.003 & 12.748 $\pm$ 0.003 & 12.464 $\pm$ 0.003 & 1\\
$G_\mathrm{{RP}}$ (mag) & 12.006 $\pm$ 0.004 & 11.329 $\pm$ 0.004 & 11.970 $\pm$ 0.004 & 11.791 $\pm$ 0.005 & 11.420 $\pm$ 0.004 & 1\\
$B_T$ (mag) & -- & -- & -- & -- & -- & 3\\
$V_T$ (mag) & -- & -- & -- & -- & -- & 3\\
$J$ (mag) & 11.607 $\pm$ 0.024 & 10.743 $\pm$ 0.027 & 11.410 $\pm$ 0.024 & 11.128 $\pm$ 0.026 & 10.680 $\pm$ 0.024 & 4\\
$H$ (mag) & 11.355 $\pm$ 0.026 & 10.454 $\pm$ 0.026 & 11.152 $\pm$ 0.023 & 10.776 $\pm$ 0.026 & 10.312 $\pm$ 0.023 & 4\\
$K_s$ (mag) & 11.282 $\pm$ 0.025 & 10.403 $\pm$ 0.025 & 11.062 $\pm$ 0.019 & 10.666 $\pm$ 0.026 & 10.243 $\pm$ 0.023 & 4\\
$W1$ (mag) & 11.201 $\pm$ 0.023 & 10.329 $\pm$ 0.022 & 11.050 $\pm$ 0.023 & 10.610 $\pm$ 0.023 & 10.159 $\pm$ 0.022 & 5\\
$W2$ (mag) & 11.243 $\pm$ 0.021 & 10.374 $\pm$ 0.020 & 11.101 $\pm$ 0.022 & 10.651 $\pm$ 0.021 & 10.189 $\pm$ 0.020 & 5\\
$W3$ (mag) & 11.255 $\pm$ 0.129 & 10.199 $\pm$ 0.053 & 11.247 $\pm$ 0.146 & 10.641 $\pm$ 0.104 & 9.986 $\pm$ 0.033 & 5\\
\multicolumn{\totalcolumns}{c}{\vrule height 24pt width 0pt \small\it\fnum@table\ continued on next page}
\enddata

%% file: target_props_2.tex
\tablecolumns{7}
\tablehead{
\colhead{\textbf{Target}} & \colhead{TOI-3364} & \colhead{TOI-3688} & \colhead{TOI-3807} & \colhead{TOI-3819} & \colhead{TOI-3912} & \colhead{Source} 
}
\startdata
\multicolumn{7}{l}{\textbf{Identifiers}}\\
TIC & 280655495 & 245509452 & 289661991 & 95660472 & 156648452 & \\
Gaia DR3 & 5409268527708808320 & 454187981188647168 & 1117620107545810944 & 876608292608416512 & 1255520718461480960 & \\
2MASS & 09380910-4920069 & 02370772+5451046 & 09162700+6904269 & 08072719+2923194 & 14231236+2404357 & \\
Tycho-2 & 8176-00923-1 & 3691-01688-1 & 4376-00887-1 & -- & -- & \\
WISE & J093809.09-492006.7 & J023707.74+545104.7 & J091627.00+690426.8 & J080727.18+292319.2 & J142312.36+240435.5 & \\
\multicolumn{7}{l}{\textbf{Astrometric Measurements}}\\
R.A. (J2000) & 09:38:09.064 & 02:37:07.762 & 09:16:27.016 & 08:07:27.181 & 14:23:12.359 & 1\\
Decl. (J2000) & -49:20:06.79 & +54:51:04.37 & +69:04:26.86 & +29:23:19.11 & +24:04:35.63 & 1\\
$\mu_{{\alpha}}$ (mas yr$^{{-1}}$) & -26.417 $\pm$ 0.014 & 18.777 $\pm$ 0.009 & -0.654 $\pm$ 0.009 & -6.144 $\pm$ 0.015 & 7.746 $\pm$ 0.011 & 1\\
$\mu_{{\delta}}$ (mas yr$^{{-1}}$) & 14.037 $\pm$ 0.013 & -11.398 $\pm$ 0.013 & -1.238 $\pm$ 0.010 & -22.673 $\pm$ 0.012 & -0.364 $\pm$ 0.012 & 1\\
Parallax (mas) & 3.628 $\pm$ 0.013 & 2.494 $\pm$ 0.011 & 2.327 $\pm$ 0.011 & 1.779 $\pm$ 0.016 & 2.128 $\pm$ 0.013 & 1\\
$b$ ($^{{\circ}}$) & 2.278 & -4.934 & 37.871 & 28.443 & 69.117 & 1\\
$l$ ($^{{\circ}}$) & 273.783 & 137.832 & 144.331 & 192.471 & 29.666 & 1\\
\multicolumn{7}{l}{\textbf{Photometric Measurements}}\\
$T$ (mag) & 10.814 $\pm$ 0.006 & 11.843 $\pm$ 0.013 & 11.582 $\pm$ 0.007 & 11.982 $\pm$ 0.009 & 11.875 $\pm$ 0.007 & 2\\
$G$ (mag) & 11.293 $\pm$ 0.003 & 12.357 $\pm$ 0.003 & 12.025 $\pm$ 0.003 & 12.404 $\pm$ 0.003 & 12.298 $\pm$ 0.003 & 1\\
$G_\mathrm{{BP}}$ (mag) & 11.664 $\pm$ 0.003 & 12.772 $\pm$ 0.003 & 12.368 $\pm$ 0.003 & 12.727 $\pm$ 0.003 & 12.624 $\pm$ 0.003 & 1\\
$G_\mathrm{{RP}}$ (mag) & 10.760 $\pm$ 0.004 & 11.776 $\pm$ 0.004 & 11.520 $\pm$ 0.004 & 11.921 $\pm$ 0.004 & 11.812 $\pm$ 0.004 & 1\\
$B_T$ (mag) & 12.422 $\pm$ 0.193 & 13.075 $\pm$ 0.272 & 12.986 $\pm$ 0.207 & -- & -- & 3\\
$V_T$ (mag) & 11.929 $\pm$ 0.171 & 12.318 $\pm$ 0.188 & 11.956 $\pm$ 0.133 & -- & -- & 3\\
$J$ (mag) & 10.156 $\pm$ 0.022 & 11.129 $\pm$ 0.026 & 10.956 $\pm$ 0.027 & 11.398 $\pm$ 0.020 & 11.270 $\pm$ 0.021 & 4\\
$H$ (mag) & 9.876 $\pm$ 0.022 & 10.819 $\pm$ 0.024 & 10.634 $\pm$ 0.030 & 11.143 $\pm$ 0.019 & 11.016 $\pm$ 0.019 & 4\\
$K_s$ (mag) & 9.783 $\pm$ 0.019 & 10.732 $\pm$ 0.020 & 10.617 $\pm$ 0.019 & 11.077 $\pm$ 0.017 & 10.922 $\pm$ 0.018 & 4\\
$W1$ (mag) & 9.691 $\pm$ 0.023 & 10.125 $\pm$ 0.122 & 10.591 $\pm$ 0.022 & 11.033 $\pm$ 0.022 & 10.898 $\pm$ 0.023 & 5\\
$W2$ (mag) & 9.729 $\pm$ 0.020 & 10.662 $\pm$ 0.021 & 10.644 $\pm$ 0.021 & 11.078 $\pm$ 0.022 & 10.931 $\pm$ 0.019 & 5\\
$W3$ (mag) & 9.593 $\pm$ 0.037 & 10.656 $\pm$ 0.101 & 10.591 $\pm$ 0.073 & 10.827 $\pm$ 0.120 & 10.858 $\pm$ 0.081 & 5\\
\multicolumn{\totalcolumns}{c}{\vrule height 24pt width 0pt \small\it\fnum@table\ continued on next page}
\enddata

%% file: target_props_3.tex
\tablecolumns{7}
\tablehead{
\colhead{\textbf{Target}} & \colhead{TOI-3976} & \colhead{TOI-4087} & \colhead{TOI-4145} & \colhead{TOI-4463} & \colhead{TOI-4791} & \colhead{Source} 
}
\startdata
\multicolumn{7}{l}{\textbf{Identifiers}}\\
TIC & 154293917 & 310002617 & 279947414 & 8599009 & 100389539 & \\
Gaia DR3 & 1585783740516554624 & 1725490333641502976 & 568619413331898240 & 4524517290642206336 & 2930141657016297216 & \\
2MASS & 14572548+4416274 & 14314234+8322215 & 02374306+8016028 & 18372595+1843470 & 07173322-1949443 & \\
Tycho-2 & -- & 4634-01080-1 & 4503-00293-1 & 1574-00343-1 & 5973-02018-1 & \\
WISE & J145725.44+441627.4 & J143142.22+832221.7 & J023743.14+801602.6 & J183725.95+184347.5 & J071733.21-194944.2 & \\
\multicolumn{7}{l}{\textbf{Astrometric Measurements}}\\
R.A. (J2000) & 14:57:25.441 & 14:31:42.182 & 02:37:43.141 & 18:37:25.951 & 07:17:33.211 & 1\\
Decl. (J2000) & +44:16:27.57 & +83:22:21.75 & +80:16:02.57 & +18:43:47.85 & -19:49:44.18 & 1\\
$\mu_{{\alpha}}$ (mas yr$^{{-1}}$) & -19.831 $\pm$ 0.008 & -16.514 $\pm$ 0.014 & 11.694 $\pm$ 0.010 & -5.157 $\pm$ 0.017 & -14.195 $\pm$ 0.018 & 1\\
$\mu_{{\delta}}$ (mas yr$^{{-1}}$) & 10.118 $\pm$ 0.011 & 10.456 $\pm$ 0.014 & -13.534 $\pm$ 0.011 & 50.619 $\pm$ 0.017 & 6.098 $\pm$ 0.018 & 1\\
Parallax (mas) & 1.914 $\pm$ 0.009 & 3.220 $\pm$ 0.011 & 4.879 $\pm$ 0.010 & 5.628 $\pm$ 0.018 & 3.104 $\pm$ 0.018 & 1\\
$b$ ($^{{\circ}}$) & 59.505 & 33.091 & 18.352 & 11.414 & -3.410 & 1\\
$l$ ($^{{\circ}}$) & 75.443 & 119.586 & 127.501 & 48.278 & 233.498 & 1\\
\multicolumn{7}{l}{\textbf{Photometric Measurements}}\\
$T$ (mag) & 11.814 $\pm$ 0.007 & 11.296 $\pm$ 0.027 & 11.524 $\pm$ 0.006 & 10.467 $\pm$ 0.006 & 10.896 $\pm$ 0.006 & 2\\
$G$ (mag) & 12.211 $\pm$ 0.003 & 11.701 $\pm$ 0.003 & 12.049 $\pm$ 0.003 & 10.951 $\pm$ 0.003 & 11.313 $\pm$ 0.003 & 1\\
$G_\mathrm{{BP}}$ (mag) & 12.511 $\pm$ 0.003 & 12.007 $\pm$ 0.003 & 12.472 $\pm$ 0.003 & 11.324 $\pm$ 0.003 & 11.624 $\pm$ 0.003 & 1\\
$G_\mathrm{{RP}}$ (mag) & 11.751 $\pm$ 0.004 & 11.235 $\pm$ 0.004 & 11.463 $\pm$ 0.004 & 10.411 $\pm$ 0.004 & 10.836 $\pm$ 0.004 & 1\\
$B_T$ (mag) & -- & 12.509 $\pm$ 0.167 & 13.152 $\pm$ 0.263 & 11.778 $\pm$ 0.066 & 11.908 $\pm$ 0.086 & 3\\
$V_T$ (mag) & -- & 12.148 $\pm$ 0.156 & 12.179 $\pm$ 0.160 & 11.106 $\pm$ 0.066 & 11.579 $\pm$ 0.102 & 3\\
$J$ (mag) & 11.238 $\pm$ 0.021 & 10.754 $\pm$ 0.022 & 10.775 $\pm$ 0.023 & 9.802 $\pm$ 0.021 & 10.319 $\pm$ 0.026 & 4\\
$H$ (mag) & 10.960 $\pm$ 0.018 & 10.493 $\pm$ 0.029 & 10.380 $\pm$ 0.028 & 9.454 $\pm$ 0.018 & 10.108 $\pm$ 0.023 & 4\\
$K_s$ (mag) & -- & 10.436 $\pm$ 0.025 & 10.286 $\pm$ 0.021 & 9.402 $\pm$ 0.015 & 10.019 $\pm$ 0.023 & 4\\
$W1$ (mag) & 10.883 $\pm$ 0.023 & 10.390 $\pm$ 0.023 & 10.198 $\pm$ 0.023 & 9.343 $\pm$ 0.023 & 9.959 $\pm$ 0.023 & 5\\
$W2$ (mag) & 10.904 $\pm$ 0.021 & 10.432 $\pm$ 0.019 & 10.251 $\pm$ 0.020 & 9.393 $\pm$ 0.021 & 9.990 $\pm$ 0.020 & 5\\
$W3$ (mag) & 10.869 $\pm$ 0.068 & 10.381 $\pm$ 0.047 & 10.216 $\pm$ 0.059 & 9.397 $\pm$ 0.036 & 10.030 $\pm$ 0.050 & 5\\
\enddata

%% file: gaia_comp_props.tex
\\[-\normalbaselineskip]\multicolumn{\totalcolumns}{l}{\textbf{TOI-1937}} \\
\hline \\
Gaia DR3 ID & 5489726768531119616 & 5489726768531118848 \\
TIC ID & 268301217 & 766593811 \\
Ang. Sep. ($"$) & -- & 2.48 \\
Proj. Sep. (AU) & -- & 1030 \\
Parallax (mas) & $2.411 \pm 0.011$ & $2.351 \pm 0.089$ \\
$\mu_{{\alpha}}$ (mas/yr) & $-5.627 \pm 0.013$ & $-5.39 \pm 0.10$ \\
$\mu_{{\delta}}$ (mas/yr) & $11.309 \pm 0.013$ & $11.349 \pm 0.096$ \\
RV (km/s) & $24.8 \pm 2.5$ & -- \\
$G$ (mag) & $13.005 \pm 0.003$ & $17.649 \pm 0.003$ \\
$G_\mathrm{BP}$ (mag) & $13.417 \pm 0.003$ & $17.9 \pm 0.1$ \\
$G_\mathrm{RP}$ (mag) & $12.421 \pm 0.004$ & $16.25 \pm 0.02$ \\
$T$ (mag) & $12.493 \pm 0.006$ & $16.86 \pm 0.08$ \\
\hline \\
\multicolumn{\totalcolumns}{l}{\textbf{TOI-2583}} \\
\hline
Gaia DR3 ID & 2115776451371641984 & 2115776451371641472 \\
TIC ID & 7548817 & 7548819 \\
Ang. Sep. ($"$) & -- & 5.26 \\
Proj. Sep. (AU) & -- & 2927 \\
Parallax (mas) & $1.7970 \pm 0.0090$ & $1.66 \pm 0.13$ \\
$\mu_{{\alpha}}$ (mas/yr) & $7.831 \pm 0.011$ & $7.59 \pm 0.16$ \\
$\mu_{{\delta}}$ (mas/yr) & $4.991 \pm 0.012$ & $4.80 \pm 0.19$ \\
RV (km/s) & $-38.3 \pm 1.1$ & -- \\
$G$ (mag) & $12.453 \pm 0.003$ & $18.707 \pm 0.004$ \\
$G_\mathrm{BP}$ (mag) & $12.759 \pm 0.003$ & $19.34 \pm 0.09$ \\
$G_\mathrm{RP}$ (mag) & $11.983 \pm 0.004$ & $17.39 \pm 0.03$ \\
$T$ (mag) & $12.045 \pm 0.007$ & $17.87 \pm 0.03$ \\
\hline \\
\multicolumn{\totalcolumns}{l}{\textbf{TOI-2587}} \\
\hline
Gaia DR3 ID & 3085216850017388672 & 3085216850017388032 \\
TIC ID & 68007716 & 68007714 \\
Ang. Sep. ($"$) & -- & 8.88 \\
Proj. Sep. (AU) & -- & 3334 \\
Parallax (mas) & $2.664 \pm 0.022$ & $2.625 \pm 0.019$ \\
$\mu_{{\alpha}}$ (mas/yr) & $22.004 \pm 0.023$ & $21.595 \pm 0.019$ \\
$\mu_{{\delta}}$ (mas/yr) & $-25.296 \pm 0.019$ & $-25.542 \pm 0.016$ \\
RV (km/s) & $-24.71 \pm 0.35$ & $-24.4 \pm 1.8$ \\
$G$ (mag) & $11.405 \pm 0.003$ & $13.702 \pm 0.003$ \\
$G_\mathrm{BP}$ (mag) & $11.738 \pm 0.003$ & $14.162 \pm 0.003$ \\
$G_\mathrm{RP}$ (mag) & $10.909 \pm 0.004$ & $13.077 \pm 0.004$ \\
$T$ (mag) & $10.966 \pm 0.006$ & $13.147 \pm 0.006$ \\
\hline \\
\multicolumn{\totalcolumns}{l}{\textbf{TOI-2803}} \\
\hline
Gaia DR3 ID & 2913482170369046656 & 2913482170369045120 \\
TIC ID & 124379043 & 124379044 \\
Ang. Sep. ($"$) & -- & 19.54 \\
Proj. Sep. (AU) & -- & 9656 \\
Parallax (mas) & $2.023 \pm 0.012$ & $2.024 \pm 0.012$ \\
$\mu_{{\alpha}}$ (mas/yr) & $0.7490 \pm 0.0090$ & $0.7950 \pm 0.0090$ \\
$\mu_{{\delta}}$ (mas/yr) & $11.908 \pm 0.011$ & $11.972 \pm 0.011$ \\
RV (km/s) & $24.7 \pm 1.2$ & $24.9 \pm 1.4$ \\
$G$ (mag) & $12.417 \pm 0.003$ & $12.753 \pm 0.003$ \\
$G_\mathrm{BP}$ (mag) & $12.669 \pm 0.003$ & $13.025 \pm 0.003$ \\
$G_\mathrm{RP}$ (mag) & $12.006 \pm 0.004$ & $12.326 \pm 0.004$ \\
$T$ (mag) & $12.075 \pm 0.006$ & $12.395 \pm 0.006$ \\
\hline \\
\multicolumn{\totalcolumns}{l}{\textbf{TOI-3688}} \\
\hline
Gaia DR3 ID & 454187981188647168 & 454187981192925568 \\
TIC ID & 245509452 & 245509454 \\
Ang. Sep. ($"$) & -- & 4.96 \\
Proj. Sep. (AU) & -- & 1976 \\
Parallax (mas) & $2.512 \pm 0.011$ & $2.51 \pm 0.18$ \\
$\mu_{{\alpha}}$ (mas/yr) & $18.7770 \pm 0.0090$ & $18.86 \pm 0.17$ \\
$\mu_{{\delta}}$ (mas/yr) & $-11.398 \pm 0.013$ & $-11.13 \pm 0.25$ \\
RV (km/s) & $0.69 \pm 0.61$ & -- \\
$G$ (mag) & $12.357 \pm 0.003$ & $18.434 \pm 0.004$ \\
$G_\mathrm{BP}$ (mag) & $12.772 \pm 0.003$ & $19.51 \pm 0.06$ \\
$G_\mathrm{RP}$ (mag) & $11.776 \pm 0.004$ & $17.08 \pm 0.01$ \\
$T$ (mag) & $11.84 \pm 0.01$ & $17.03 \pm 0.01$ \\
\hline \\
\multicolumn{\totalcolumns}{l}{\textbf{TOI-3976}} \\
\hline
Gaia DR3 ID & 1585783740516554624 & 1585783706157262720 \\
TIC ID & 154293917 & 1102055625 \\
Ang. Sep. ($"$) & -- & 6.81 \\
Proj. Sep. (AU) & -- & 3526 \\
Parallax (mas) & $1.9310 \pm 0.0090$ & $1.98 \pm 0.21$ \\
$\mu_{{\alpha}}$ (mas/yr) & $-19.8310 \pm 0.0080$ & $-19.81 \pm 0.19$ \\
$\mu_{{\delta}}$ (mas/yr) & $10.118 \pm 0.011$ & $9.45 \pm 0.27$ \\
RV (km/s) & $-44.68 \pm 0.69$ & -- \\
$G$ (mag) & $12.211 \pm 0.003$ & $19.158 \pm 0.004$ \\
$G_\mathrm{BP}$ (mag) & $12.511 \pm 0.003$ & $20.1 \pm 0.1$ \\
$G_\mathrm{RP}$ (mag) & $11.751 \pm 0.004$ & $17.89 \pm 0.03$ \\
$T$ (mag) & $11.814 \pm 0.007$ & $18.29 \pm 0.04$ \\
\hline \\
\multicolumn{\totalcolumns}{l}{\textbf{TOI-4145}} \\
\hline
Gaia DR3 ID & 568619413331898240 & 568619417628568704 \\
TIC ID & 279947414 & 629870229 \\
Ang. Sep. ($"$) & -- & 1.74 \\
Proj. Sep. (AU) & -- & 356 \\
Parallax (mas) & $4.9017 \pm 0.0096$ & $5.71 \pm 0.16$ \\
$\mu_{{\alpha}}$ (mas/yr) & $11.690 \pm 0.010$ & $11.93 \pm 0.20$ \\
$\mu_{{\delta}}$ (mas/yr) & $-13.534 \pm 0.011$ & $-16.83 \pm 0.59$ \\
RV (km/s) & $1.16 \pm 0.54$ & -- \\
$G$ (mag) & $12.049 \pm 0.003$ & $17.367 \pm 0.005$ \\
$G_\mathrm{BP}$ (mag) & $12.472 \pm 0.003$ & -- \\
$G_\mathrm{RP}$ (mag) & $11.463 \pm 0.004$ & -- \\
$T$ (mag) & $11.524 \pm 0.006$ & $16.6 \pm 0.6$ \\
\hline \\
\multicolumn{\totalcolumns}{l}{\textbf{TOI-4463}} \\
\hline
Gaia DR3 ID & 4524517290642206336 & 4524517290642204928 \\
TIC ID & 8599009 & 8599017 \\
Ang. Sep. ($"$) & -- & 11.37 \\
Proj. Sep. (AU) & -- & 2009 \\
Parallax (mas) & $5.659 \pm 0.018$ & $5.644 \pm 0.067$ \\
$\mu_{{\alpha}}$ (mas/yr) & $-5.157 \pm 0.017$ & $-5.258 \pm 0.068$ \\
$\mu_{{\delta}}$ (mas/yr) & $50.619 \pm 0.017$ & $50.977 \pm 0.072$ \\
RV (km/s) & $-90.38 \pm 0.32$ & -- \\
$G$ (mag) & $10.951 \pm 0.003$ & $16.768 \pm 0.003$ \\
$G_\mathrm{BP}$ (mag) & $11.324 \pm 0.003$ & $18.28 \pm 0.02$ \\
$G_\mathrm{RP}$ (mag) & $10.411 \pm 0.004$ & $15.539 \pm 0.005$ \\
$T$ (mag) & $10.467 \pm 0.006$ & $15.482 \pm 0.008$ \\

%% file: comp_props.tex
\\[-\normalbaselineskip]\multicolumn{3}{l}{\textbf{Stellar Properties}}\\
\Teff (K) & $5737.0^{+90.0}_{-92.0}$ & $4614.0^{+83.0}_{-81.0}$\\
\feh (dex) & $0.017^{+0.076}_{-0.078}$ & $0.044^{+0.072}_{-0.073}$\\
Age (Gyr) & $5.5^{+3.0}_{-2.8}$ & $5.5^{+3.0}_{-2.8}$\\
\Mstar (\Msun) & $0.978 \pm 0.046$ & $0.734^{+0.023}_{-0.024}$\\
\Rstar (\Rsun) & $1.004^{+0.037}_{-0.034}$ & $0.698 \pm 0.016$\\
$\log{{g}}$ (cgs) & $4.424^{+0.040}_{-0.041}$ & $4.615^{+0.013}_{-0.014}$\\
\multicolumn{3}{l}{\textbf{Synthetic Photometry}}\\
$G$ (mag) & $12.57 \pm 0.02$ & $14.50^{+0.10}_{-0.09}$\\
$G_\mathrm{{BP}}$ (mag) & $12.94 \pm 0.02$ & $15.1 \pm 0.1$\\
$G_\mathrm{{RP}}$ (mag) & $12.03 \pm 0.02$ & $13.76^{+0.09}_{-0.08}$\\
$T$ (mag) & $12.02 \pm 0.02$ & $13.74^{+0.09}_{-0.08}$\\
$B_T$ (mag) & $13.58 \pm 0.03$ & $16.3 \pm 0.2$\\
$V_T$ (mag) & $12.80 \pm 0.02$ & $15.0 \pm 0.1$\\
$J$ (mag) & $11.42 \pm 0.02$ & $12.81^{+0.07}_{-0.06}$\\
$H$ (mag) & $11.06 \pm 0.02$ & $12.18 \pm 0.05$\\
$K$ (mag) & $11.01 \pm 0.02$ & $12.08 \pm 0.05$\\
$W1$ (mag) & $10.98 \pm 0.02$ & $12.04 \pm 0.05$\\
$W2$ (mag) & $11.00 \pm 0.02$ & $12.10 \pm 0.05$\\
$W3$ (mag) & $10.96 \pm 0.02$ & $12.00 \pm 0.05$\\

%% file: spec_props.tex
TOI-1937 & SpecMatch & PFS & 5798 $\pm$ 110 & 1.16 $\pm$ 0.21 & -- & 0.22 $\pm$ 0.09 & 8.0 $\pm$ 1.0 & 3.8 $\pm$ 0.2 & Y \\
TOI-2364 & SpecMatch & PFS & 5271 $\pm$ 110 & 0.96 $\pm$ 0.10 & -- & 0.36 $\pm$ 0.09 & 1.3 $\pm$ 1.0 & 4.5 $\pm$ 0.2 & Y \\
TOI-2583 & SpecMatch & HIRES & 5867 $\pm$ 110 & 1.39 $\pm$ 0.25 & -- & 0.15 $\pm$ 0.09 & 2.4 $\pm$ 1.0 & 3.7 $\pm$ 0.2 & Y \\
$\cdots$ & SPC & TRES & 6059 $\pm$ 50 & -- & 4.35 $\pm$ 0.10 & 0.29 $\pm$ 0.08 & 5.2 $\pm$ 0.5 & -- & N \\
TOI-2587 & SpecMatch & NEID & 5766 $\pm$ 110 & 1.92 $\pm$ 0.34 & -- & 0.17 $\pm$ 0.09 & 2.8 $\pm$ 1.0 & 4.2 $\pm$ 0.2 & Y \\
$\cdots$ & SPC & TRES & 5792 $\pm$ 50 & -- & 4.22 $\pm$ 0.10 & 0.15 $\pm$ 0.08 & 4.2 $\pm$ 0.5 & -- & N \\
TOI-2796 & SpecMatch & NEID & 5721 $\pm$ 110 & 1.27 $\pm$ 0.18 & -- & 0.22 $\pm$ 0.09 & 2.7 $\pm$ 1.0 & 4.3 $\pm$ 0.2 & Y \\
$\cdots$ & SPC & TRES & 5742 $\pm$ 58 & -- & 4.44 $\pm$ 0.10 & 0.32 $\pm$ 0.08 & 6.0 $\pm$ 0.5 & -- & N \\
TOI-2803 & SpecMatch & PFS & 6151 $\pm$ 110 & 1.80 $\pm$ 0.32 & -- & -0.43 $\pm$ 0.09 & 2.2 $\pm$ 1.0 & 3.6 $\pm$ 0.2 & N \\
$\cdots$ & SPC & TRES & 6058 $\pm$ 108 & -- & 4.26 $\pm$ 0.18 & -0.11 $\pm$ 0.10 & 5.4 $\pm$ 0.8 & -- & N \\
$\cdots$ & SPC & TRES & 6277 $\pm$ 70  & -- & 4.30 (fixed)    & -0.11 $\pm$ 0.08 & 5.4 $\pm$ 0.8 & -- & Y \\
$\cdots$ & ZASPE & PFS & 6436 $\pm$ 80 & -- & 4.30 (fixed)    & -0.35 $\pm$ 0.07 & 4.8 $\pm$ 0.3 & -- & N \\
TOI-2818 & SpecMatch & CHIRON & 5715 $\pm$ 110 & 1.24 $\pm$ 0.22 & -- & 0.02 $\pm$ 0.09 & 2.9 $\pm$ 1.0 & 4.0 $\pm$ 0.2 & Y \\
$\cdots$ & SPC & CHIRON & 5767 $\pm$ 50 & -- & 4.27 $\pm$ 0.10 & -0.06 $\pm$ 0.10 & 5.2 $\pm$ 0.5 & -- & N \\
TOI-2842 & SpecMatch & PFS & 5942 $\pm$ 110 & 1.59 $\pm$ 0.29 & -- & 0.26 $\pm$ 0.09 & 5.2 $\pm$ 1.0 & 3.6 $\pm$ 0.2 & Y \\
$\cdots$ & SPC & TRES & 6004 $\pm$ 75 & -- & 4.30 $\pm$ 0.13 & 0.35 $\pm$ 0.08 & 7.6 $\pm$ 0.5 & -- & N \\
TOI-2977 & SpecMatch & CHIRON & 5674 $\pm$ 110 & 1.15 $\pm$ 0.21 & -- & 0.04 $\pm$ 0.09 & 4.7 $\pm$ 1.0 & 4.0 $\pm$ 0.2 & Y \\
$\cdots$ & SPC & CHIRON & 5790 $\pm$ 74 & -- & 4.25 $\pm$ 0.21 & 0.05 $\pm$ 0.10 & 6.2 $\pm$ 0.5 & -- & N \\
TOI-3023 & SpecMatch & PFS & 5651 $\pm$ 110 & 1.85 $\pm$ 0.33 & -- & 0.09 $\pm$ 0.09 & 3.0 $\pm$ 1.0 & 4.0 $\pm$ 0.2 & Y \\
TOI-3364 & SpecMatch & PFS & 5626 $\pm$ 110 & 1.49 $\pm$ 0.27 & -- & 0.39 $\pm$ 0.09 & 2.0 $\pm$ 1.0 & 4.3 $\pm$ 0.2 & Y \\
TOI-3688 & SpecMatch & NEID & 5909 $\pm$ 110 & 1.69 $\pm$ 0.31 & -- & 0.26 $\pm$ 0.09 & 4.3 $\pm$ 1.0 & 4.2 $\pm$ 0.2 & Y \\
$\cdots$ & SPC & TRES & 6110 $\pm$ 50 & -- & 4.33 $\pm$ 0.10 & 0.47 $\pm$ 0.08 & 6.3 $\pm$ 0.5 & -- & N \\
TOI-3807 & SpecMatch & NEID & 5755 $\pm$ 110 & 1.88 $\pm$ 0.34 & -- & 0.31 $\pm$ 0.09 & 3.1 $\pm$ 1.0 & 3.9 $\pm$ 0.2 & Y \\
$\cdots$ & SPC & TRES & 5942 $\pm$ 50 & -- & 4.34 $\pm$ 0.10 & 0.41 $\pm$ 0.08 & 5.5 $\pm$ 0.5 & -- & N \\
TOI-3819 & SpecMatch & NEID & 5847 $\pm$ 110 & 1.95 $\pm$ 0.35 & -- & 0.29 $\pm$ 0.09 & 5.8 $\pm$ 1.0 & 4.0 $\pm$ 0.2 & Y \\
$\cdots$ & SPC & TRES & 5964 $\pm$ 50 & -- & 4.29 $\pm$ 0.10 & 0.41 $\pm$ 0.08 & 6.8 $\pm$ 0.5 & -- & N \\
TOI-3912 & SpecMatch & HIRES & 5730 $\pm$ 110 & 1.48 $\pm$ 0.27 & -- & 0.19 $\pm$ 0.09 & 2.3 $\pm$ 1.0 & 3.7 $\pm$ 0.2 & Y \\
$\cdots$ & SPC & TRES & 5829 $\pm$ 50 & -- & 4.29 $\pm$ 0.10 & 0.22 $\pm$ 0.08 & 4.4 $\pm$ 0.5 & -- & N \\
TOI-3976 & SpecMatch & HIRES & 5942 $\pm$ 110 & 1.33 $\pm$ 0.24 & -- & 0.17 $\pm$ 0.09 & 2.0 $\pm$ 1.0 & 3.6 $\pm$ 0.2 & Y \\
$\cdots$ & SPC & TRES & 6032 $\pm$ 64 & -- & 4.30 $\pm$ 0.11 & 0.36 $\pm$ 0.08 & 5.3 $\pm$ 0.5 & -- & N \\
TOI-4087 & SpecMatch & NEID & 5888 $\pm$ 110 & 1.18 $\pm$ 0.21 & -- & 0.27 $\pm$ 0.09 & 4.6 $\pm$ 1.0 & 4.0 $\pm$ 0.2 & Y \\
$\cdots$ & SPC & TRES & 5957 $\pm$ 50 & -- & 4.43 $\pm$ 0.10 & 0.27 $\pm$ 0.08 & 6.5 $\pm$ 0.5 & -- & N \\
TOI-4145 & SpecMatch & NEID & 5266 $\pm$ 110 & 0.87 $\pm$ 0.09 & -- & 0.20 $\pm$ 0.09 & 2.2 $\pm$ 1.0 & 4.8 $\pm$ 0.2 & Y \\
$\cdots$ & SPC & TRES & 5380 $\pm$ 50 & -- & 4.63 $\pm$ 0.10 & 0.24 $\pm$ 0.08 & 3.8 $\pm$ 0.5 & -- & N \\
TOI-4463 & SpecMatch & NEID & 5566 $\pm$ 110 & 1.14 $\pm$ 0.21 & -- & 0.28 $\pm$ 0.09 & 1.6 $\pm$ 1.0 & 4.2 $\pm$ 0.2 & Y \\
$\cdots$ & SPC & TRES & 5646 $\pm$ 50 & -- & 4.42 $\pm$ 0.10 & 0.17 $\pm$ 0.08 & 3.4 $\pm$ 0.5 & -- & N \\
TOI-4791 & SpecMatch & CHIRON & 6011 $\pm$ 110 & 1.77 $\pm$ 0.32 & -- & 0.23 $\pm$ 0.09 & 10.9 $\pm$ 1.0 & 3.3 $\pm$ 0.2 & Y \\
$\cdots$ & SPC & CHIRON & 6128 $\pm$ 50 & -- & 4.27 $\pm$ 0.10 & 0.12 $\pm$ 0.10 & 11.2 $\pm$ 0.5 & -- & N \\
$\cdots$ & SPC & TRES & 6231 $\pm$ 50 & -- & 4.31 $\pm$ 0.10 & 0.42 $\pm$ 0.08 & 11.9 $\pm$ 0.5 & -- & N \\

%% file: tois_HJPaperII.txt
1937
2364
2583
2587
2796
2803
2818
2842
2977
3023
3364
3688
3807
3819
3912
3976
4087
4145
4463
4791%

%% file: planet_fit_results_0.tex
\tablecolumns{7}
\tablehead{
 & & \colhead{TOI-1937 b} & \colhead{TOI-2364 b} & \colhead{TOI-2583 b} & \colhead{TOI-2587 b} & \colhead{TOI-2796 b} 
}
\startdata
\multicolumn{7}{l}{\textbf{Planet Parameters}}\\
$P$ (days) & Period & $0.94667944 \pm 0.00000047$ & $4.0197517 \pm 0.0000043$ & $4.5207265 \pm 0.0000049$ & $5.456640 \pm 0.000011$ & $4.8084983^{+0.0000057}_{-0.0000056}$\\
$T_c$ (BJD$_\mathrm{TDB}$) & Time of conjunction & $2459085.91023 \pm 0.00012$ & $2459058.21396 \pm 0.00033$ & $2459216.01039 \pm 0.00036$ & $2458950.09242 \pm 0.00076$ & $2459026.48882 \pm 0.00045$\\
$T_{14}$ (days) & Transit duration & $0.05737 \pm 0.00057$ & $0.11563^{+0.00097}_{-0.00093}$ & $0.1836^{+0.0013}_{-0.0012}$ & $0.1734 \pm 0.0023$ & $0.0818 \pm 0.0017$\\
$\tau$ (days) & Ingress/egress duration & $0.0220^{+0.0014}_{-0.0012}$ & $0.00988^{+0.00066}_{-0.00040}$ & $0.0163^{+0.0013}_{-0.0010}$ & $0.0204^{+0.0020}_{-0.0021}$ & $0.04089^{+0.00087}_{-0.00083}$\\
$a/R_\star$ & Planet-star separation & $3.85^{+0.09}_{-0.10}$ & $11.84^{+0.22}_{-0.33}$ & $8.31^{+0.23}_{-0.27}$ & $7.92^{+0.38}_{-0.30}$ & $11.43^{+0.33}_{-0.30}$\\
$\left(R_P / R_\star\right)^2$ & Transit depth & $0.0141 \pm 0.0016$ & $0.00794^{+0.00016}_{-0.00015}$ & $0.00806^{+0.00015}_{-0.00014}$ & $0.00411 \pm 0.00013$ & $0.0234\,(> 0.0224)$\\
$i$ (deg) & Inclination & $77.00^{+0.44}_{-0.49}$ & $88.98^{+0.62}_{-0.54}$ & $88.16^{+0.98}_{-0.76}$ & $84.81^{+0.49}_{-0.40}$ & $84.90\,(< 85.22)$\\
$K$ (m/s) & RV semi-amplitude & $386 \pm 28$ & $29.7^{+5.6}_{-6.4}$ & $27.1 \pm 6.1$ & $22.7^{+5.4}_{-4.7}$ & $50^{+11}_{-12}$\\
$a$ (AU) & Semimajor axis & $0.01932^{+0.00035}_{-0.00039}$ & $0.04871^{+0.00069}_{-0.00079}$ & $0.0571^{+0.0010}_{-0.0013}$ & $0.0635^{+0.0025}_{-0.0013}$ & $0.0569^{+0.0010}_{-0.0011}$\\
$R_P$ ($R_\mathrm{J}$) & Planet radius & $1.247^{+0.059}_{-0.062}$ & $0.768^{+0.023}_{-0.018}$ & $1.290^{+0.040}_{-0.033}$ & $1.077^{+0.042}_{-0.040}$ & $1.59\,(> 1.54)$\\
$M_P$ ($M_\mathrm{J}$) & Planet mass & $2.01^{+0.17}_{-0.16}$ & $0.225^{+0.043}_{-0.049}$ & $0.250^{+0.058}_{-0.056}$ & $0.218^{+0.054}_{-0.046}$ & $0.44^{+0.10}_{-0.11}$\\
$\rho_P$ (g cm$^{-3}$) & Planet density & $1.28^{+0.21}_{-0.17}$ & $0.61^{+0.13}_{-0.14}$ & $0.144^{+0.037}_{-0.035}$ & $0.216^{+0.064}_{-0.052}$ & $0.03\,(< 0.15)$\\
$\log{g_P}$ (cgs) & Planet surface gravity & $3.506^{+0.047}_{-0.048}$ & $2.97^{+0.08}_{-0.11}$ & $2.57^{+0.09}_{-0.11}$ & $2.67 \pm 0.11$ & $2.39\,(< 2.67)$\\
$b \equiv a\cos{i}/R_\star$ & Transit impact parameter & $0.8653^{+0.0086}_{-0.0084}$ & $0.21^{+0.10}_{-0.13}$ & $0.27^{+0.10}_{-0.14}$ & $0.716^{+0.027}_{-0.036}$ & $1.00\,(> 0.97)$\\
$e$ & Eccentricity & 0.0 (fixed) & 0.0 (fixed) & 0.0 (fixed) & 0.0 (fixed) & 0.0 (fixed)\\
$e_\mathrm{lim}$\tablenotemark{a} & 1-$\sigma$ upper limit on eccentricity & $< 0.034$ & $< 0.067$ & $< 0.089$ & $< 0.171$ & $< 0.172$\\
$\tau_\mathrm{circ}$ (Gyr)\tablenotemark{b} & Tidal circularization timescale & $0.00168^{+0.00042}_{-0.00033}$ & $1.02 \pm 0.26$ & $0.166^{+0.051}_{-0.046}$ & $0.78^{+0.31}_{-0.23}$ & $0.010\,(< 0.144)$\\
$\langle F \rangle$ (Gerg~s$^{-1}$~cm$^{-2}$) & Incident flux & $4.39^{+0.30}_{-0.29}$ & $0.322^{+0.020}_{-0.016}$ & $1.022^{+0.054}_{-0.044}$ & $0.991^{+0.066}_{-0.069}$ & $0.479^{+0.029}_{-0.027}$\\
$T_\mathrm{eq}$ (K) & Planet equilibirum temperature & $2097^{+35}_{-36}$ & $1091^{+17}_{-14}$ & $1456^{+19}_{-16}$ & $1445^{+23}_{-26}$ & $1205^{+18}_{-17}$\\
\multicolumn{7}{l}{\textbf{Stellar Parameters}}\\
$M_\star$ ($M_\odot$) & Stellar mass & $1.072^{+0.059}_{-0.064}$ & $0.954^{+0.041}_{-0.046}$ & $1.215^{+0.063}_{-0.082}$ & $1.15^{+0.14}_{-0.07}$ & $1.063^{+0.057}_{-0.062}$\\
$R_\star$ ($R_\odot$) & Stellar radius & $1.080^{+0.025}_{-0.024}$ & $0.886^{+0.021}_{-0.017}$ & $1.477^{+0.036}_{-0.032}$ & $1.726^{+0.049}_{-0.047}$ & $1.069 \pm 0.024$\\
$\log{g_\star}$ (cgs) & Stellar surface gravity & $4.401^{+0.026}_{-0.029}$ & $4.524^{+0.018}_{-0.029}$ & $4.184^{+0.028}_{-0.035}$ & $4.024^{+0.057}_{-0.039}$ & $4.400 \pm 0.030$\\
$\rho_\star$ (g cm$^{-3}$) & Stellar density & $1.198^{+0.090}_{-0.092}$ & $1.94^{+0.11}_{-0.16}$ & $0.531^{+0.045}_{-0.050}$ & $0.316^{+0.048}_{-0.034}$ & $1.22^{+0.11}_{-0.09}$\\
$L_\star$ ($L_\odot$) & Stellar luminosity & $1.202^{+0.088}_{-0.086}$ & $0.560^{+0.037}_{-0.027}$ & $2.445^{+0.084}_{-0.085}$ & $2.96 \pm 0.14$ & $1.136^{+0.069}_{-0.063}$\\
$T_\mathrm{eff}$ (K) & Stellar effective temperature & $5814^{+91}_{-93}$ & $5306^{+76}_{-68}$ & $5936^{+65}_{-68}$ & $5760^{+80}_{-79}$ & $5764^{+81}_{-78}$\\
$[\mathrm{Fe/H}]$ (dex) & Metallicity & $0.196 \pm 0.084$ & $0.344 \pm 0.085$ & $0.158^{+0.084}_{-0.085}$ & $0.13^{+0.09}_{-0.10}$ & $0.239^{+0.076}_{-0.079}$\\
$[\mathrm{Fe/H}]_0$ (dex)\tablenotemark{c} & Initial metallicity & $0.192^{+0.077}_{-0.078}$ & $0.309^{+0.079}_{-0.080}$ & $0.212^{+0.073}_{-0.074}$ & $0.167^{+0.084}_{-0.098}$ & $0.228^{+0.071}_{-0.073}$\\
Age (Gyr) & Stellar age & $3.6^{+3.1}_{-2.3}$ & $3.5^{+4.2}_{-2.4}$ & $4.4^{+1.9}_{-1.2}$ & $6.7^{+1.7}_{-2.6}$ & $4.0^{+3.3}_{-2.5}$\\
EEP\tablenotemark{d} & Equal evolutionary phase & $351^{+39}_{-27}$ & $334^{+25}_{-36}$ & $404 \pm 28$ & $451.4^{+5.9}_{-41.0}$ & $354^{+39}_{-28}$\\
$A_V$ (mag) & Visual extinction & $0.540^{+0.091}_{-0.098}$ & $0.102^{+0.099}_{-0.069}$ & $0.061^{+0.031}_{-0.038}$ & $0.093^{+0.050}_{-0.058}$ & $0.145^{+0.073}_{-0.072}$\\
d (pc) & Distance & $418.7 \pm 1.9$ & $218.63^{+0.89}_{-0.88}$ & $561.2^{+2.9}_{-2.8}$ & $380.9 \pm 3.2$ & $352.4 \pm 1.8$\\
\multicolumn{\totalcolumns}{c}{\vrule height 24pt width 0pt \small\it\fnum@table\ continued on next page}
\enddata

%% file: planet_fit_results_1.tex
\tablecolumns{7}
\tablehead{
 & & \colhead{TOI-2803 b} & \colhead{TOI-2818 b} & \colhead{TOI-2842 b} & \colhead{TOI-2977 b} & \colhead{TOI-3023 b} 
}
\startdata
\multicolumn{7}{l}{\textbf{Planet Parameters}}\\
$P$ (days) & Period & $1.96229325 \pm 0.00000082$ & $4.0397090^{+0.0000024}_{-0.0000023}$ & $3.5514058^{+0.0000077}_{-0.0000078}$ & $2.3505614 \pm 0.0000025$ & $3.9014971 \pm 0.0000031$\\
$T_c$ (BJD$_\mathrm{TDB}$) & Time of conjunction & $2459207.68640^{+0.00022}_{-0.00023}$ & $2459023.44641 \pm 0.00025$ & $2459172.17909 \pm 0.00077$ & $2459373.83677 \pm 0.00029$ & $2459071.19220 \pm 0.00030$\\
$T_{14}$ (days) & Transit duration & $0.12879^{+0.00080}_{-0.00074}$ & $0.1585^{+0.0013}_{-0.0012}$ & $0.1073 \pm 0.0022$ & $0.12225^{+0.00084}_{-0.00081}$ & $0.2067^{+0.0010}_{-0.0009}$\\
$\tau$ (days) & Ingress/egress duration & $0.01545^{+0.00047}_{-0.00024}$ & $0.0182 \pm 0.0013$ & $0.0216^{+0.0022}_{-0.0020}$ & $0.01258^{+0.00039}_{-0.00024}$ & $0.01750^{+0.00082}_{-0.00031}$\\
$a/R_\star$ & Planet-star separation & $5.512^{+0.043}_{-0.070}$ & $8.63^{+0.28}_{-0.25}$ & $8.07 \pm 0.30$ & $6.806^{+0.059}_{-0.090}$ & $6.53^{+0.06}_{-0.13}$\\
$\left(R_P / R_\star\right)^2$ & Transit depth & $0.01781 \pm 0.00034$ & $0.01300 \pm 0.00030$ & $0.00867^{+0.00037}_{-0.00036}$ & $0.01265^{+0.00034}_{-0.00033}$ & $0.00815 \pm 0.00012$\\
$i$ (deg) & Inclination & $89.0^{+0.7}_{-1.0}$ & $87.73^{+0.75}_{-0.55}$ & $84.42^{+0.34}_{-0.35}$ & $89.19^{+0.57}_{-0.82}$ & $88.9^{+0.8}_{-1.0}$\\
$K$ (m/s) & RV semi-amplitude & $146.7^{+11.0}_{-8.9}$ & $91 \pm 33$ & $45.2^{+5.8}_{-5.4}$ & $266^{+39}_{-38}$ & $74 \pm 11$\\
$a$ (AU) & Semimajor axis & $0.03185 \pm 0.00052$ & $0.0493^{+0.0010}_{-0.0008}$ & $0.0475^{+0.0010}_{-0.0011}$ & $0.03386^{+0.00067}_{-0.00050}$ & $0.0505^{+0.0015}_{-0.0009}$\\
$R_P$ ($R_\mathrm{J}$) & Planet radius & $1.616^{+0.034}_{-0.032}$ & $1.363^{+0.046}_{-0.045}$ & $1.146^{+0.051}_{-0.048}$ & $1.174^{+0.031}_{-0.027}$ & $1.466^{+0.043}_{-0.032}$\\
$M_P$ ($M_\mathrm{J}$) & Planet mass & $0.975^{+0.083}_{-0.070}$ & $0.71 \pm 0.26$ & $0.370^{+0.052}_{-0.047}$ & $1.68^{+0.26}_{-0.25}$ & $0.62^{+0.10}_{-0.09}$\\
$\rho_P$ (g cm$^{-3}$) & Planet density & $0.287^{+0.026}_{-0.023}$ & $0.34^{+0.14}_{-0.13}$ & $0.304^{+0.065}_{-0.056}$ & $1.28^{+0.21}_{-0.20}$ & $0.245^{+0.039}_{-0.038}$\\
$\log{g_P}$ (cgs) & Planet surface gravity & $2.966^{+0.035}_{-0.033}$ & $2.98^{+0.14}_{-0.18}$ & $2.843^{+0.071}_{-0.074}$ & $3.479^{+0.062}_{-0.069}$ & $2.856^{+0.062}_{-0.070}$\\
$b \equiv a\cos{i}/R_\star$ & Transit impact parameter & $0.097^{+0.093}_{-0.067}$ & $0.34^{+0.07}_{-0.11}$ & $0.785^{+0.020}_{-0.022}$ & $0.097^{+0.095}_{-0.067}$ & $0.13^{+0.11}_{-0.09}$\\
$e$ & Eccentricity & 0.0 (fixed) & 0.0 (fixed) & 0.0 (fixed) & 0.0 (fixed) & 0.0 (fixed)\\
$e_\mathrm{lim}$\tablenotemark{a} & 1-$\sigma$ upper limit on eccentricity & $< 0.097$ & $< 0.147$ & $< 0.090$ & $< 0.126$ & $< 0.097$\\
$\tau_\mathrm{circ}$ (Gyr)\tablenotemark{b} & Tidal circularization timescale & $0.00542^{+0.00059}_{-0.00057}$ & $0.190^{+0.086}_{-0.074}$ & $0.149^{+0.049}_{-0.038}$ & $0.089^{+0.017}_{-0.016}$ & $0.111^{+0.019}_{-0.020}$\\
$\langle F \rangle$ (Gerg~s$^{-1}$~cm$^{-2}$) & Incident flux & $2.92^{+0.18}_{-0.17}$ & $0.814^{+0.064}_{-0.056}$ & $1.064^{+0.087}_{-0.077}$ & $1.290^{+0.091}_{-0.084}$ & $1.475^{+0.092}_{-0.089}$\\
$T_\mathrm{eq}$ (K) & Planet equilibirum temperature & $1893^{+29}_{-28}$ & $1376^{+26}_{-24}$ & $1471^{+29}_{-27}$ & $1544 \pm 26$ & $1596^{+24}_{-25}$\\
\multicolumn{7}{l}{\textbf{Stellar Parameters}}\\
$M_\star$ ($M_\odot$) & Stellar mass & $1.118^{+0.056}_{-0.054}$ & $0.977^{+0.063}_{-0.049}$ & $1.135^{+0.077}_{-0.079}$ & $0.936^{+0.057}_{-0.041}$ & $1.12^{+0.11}_{-0.06}$\\
$R_\star$ ($R_\odot$) & Stellar radius & $1.245^{+0.022}_{-0.021}$ & $1.229^{+0.032}_{-0.031}$ & $1.265^{+0.040}_{-0.037}$ & $1.073^{+0.024}_{-0.020}$ & $1.668^{+0.046}_{-0.033}$\\
$\log{g_\star}$ (cgs) & Stellar surface gravity & $4.298^{+0.011}_{-0.015}$ & $4.250^{+0.034}_{-0.030}$ & $4.288^{+0.038}_{-0.040}$ & $4.350^{+0.012}_{-0.013}$ & $4.048^{+0.017}_{-0.021}$\\
$\rho_\star$ (g cm$^{-3}$) & Stellar density & $0.822^{+0.019}_{-0.031}$ & $0.745^{+0.074}_{-0.063}$ & $0.789^{+0.093}_{-0.084}$ & $1.078^{+0.028}_{-0.042}$ & $0.346^{+0.009}_{-0.020}$\\
$L_\star$ ($L_\odot$) & Stellar luminosity & $2.17^{+0.14}_{-0.12}$ & $1.46^{+0.11}_{-0.10}$ & $1.76^{+0.15}_{-0.13}$ & $1.089^{+0.095}_{-0.083}$ & $2.78^{+0.22}_{-0.21}$\\
$T_\mathrm{eff}$ (K) & Stellar effective temperature & $6280^{+99}_{-96}$ & $5721^{+88}_{-83}$ & $5910 \pm 100$ & $5691^{+94}_{-93}$ & $5760^{+85}_{-88}$\\
$[\mathrm{Fe/H}]$ (dex) & Metallicity & $-0.105^{+0.068}_{-0.072}$ & $-0.02^{+0.11}_{-0.09}$ & $0.233^{+0.087}_{-0.090}$ & $0.013^{+0.087}_{-0.085}$ & $0.085^{+0.084}_{-0.083}$\\
$[\mathrm{Fe/H}]_0$ (dex)\tablenotemark{c} & Initial metallicity & $-0.025^{+0.058}_{-0.056}$ & $0.040^{+0.093}_{-0.081}$ & $0.250 \pm 0.076$ & $0.058^{+0.080}_{-0.078}$ & $0.132^{+0.073}_{-0.075}$\\
Age (Gyr) & Stellar age & $3.7^{+1.5}_{-1.3}$ & $9.5^{+2.6}_{-2.8}$ & $4.7^{+3.0}_{-2.3}$ & $9.8^{+2.6}_{-3.0}$ & $7.0^{+1.7}_{-2.2}$\\
EEP\tablenotemark{d} & Equal evolutionary phase & $378^{+23}_{-29}$ & $426.7^{+9.5}_{-16.0}$ & $396^{+27}_{-47}$ & $408^{+10}_{-18}$ & $449.7^{+5.4}_{-28.0}$\\
$A_V$ (mag) & Visual extinction & $0.062^{+0.046}_{-0.043}$ & $0.13^{+0.11}_{-0.08}$ & $0.21 \pm 0.11$ & $0.30 \pm 0.12$ & $0.61^{+0.10}_{-0.11}$\\
d (pc) & Distance & $501.1 \pm 2.9$ & $315.0 \pm 1.0$ & $460.2 \pm 2.6$ & $355.7^{+9.5}_{-9.0}$ & $392.8 \pm 1.4$\\
\multicolumn{\totalcolumns}{c}{\vrule height 24pt width 0pt \small\it\fnum@table\ continued on next page}
\enddata

%% file: planet_fit_results_2.tex
\tablecolumns{7}
\tablehead{
 & & \colhead{TOI-3364 b} & \colhead{TOI-3688 b} & \colhead{TOI-3807 b} & \colhead{TOI-3819 b} & \colhead{TOI-3912 b} 
}
\startdata
\multicolumn{7}{l}{\textbf{Planet Parameters}}\\
$P$ (days) & Period & $5.8768918 \pm 0.0000069$ & $3.246075 \pm 0.000012$ & $2.8989727^{+0.0000038}_{-0.0000039}$ & $3.2443141 \pm 0.0000055$ & $3.4936264 \pm 0.0000038$\\
$T_c$ (BJD$_\mathrm{TDB}$) & Time of conjunction & $2459090.99481^{+0.00040}_{-0.00041}$ & $2459108.0507 \pm 0.0014$ & $2459218.13454 \pm 0.00055$ & $2459502.74370 \pm 0.00038$ & $2459442.81871^{+0.00037}_{-0.00036}$\\
$T_{14}$ (days) & Transit duration & $0.1936 \pm 0.0013$ & $0.1451^{+0.0038}_{-0.0037}$ & $0.0742 \pm 0.0019$ & $0.1158^{+0.0015}_{-0.0014}$ & $0.1482 \pm 0.0014$\\
$\tau$ (days) & Ingress/egress duration & $0.0149^{+0.0011}_{-0.0007}$ & $0.0132^{+0.0012}_{-0.0008}$ & $0.03710^{+0.00095}_{-0.00094}$ & $0.0220^{+0.0015}_{-0.0014}$ & $0.0176^{+0.0015}_{-0.0014}$\\
$a/R_\star$ & Planet-star separation & $10.24^{+0.20}_{-0.32}$ & $7.53^{+0.23}_{-0.25}$ & $6.14^{+0.19}_{-0.18}$ & $6.43 \pm 0.17$ & $7.16^{+0.25}_{-0.24}$\\
$\left(R_P / R_\star\right)^2$ & Transit depth & $0.00624^{+0.00024}_{-0.00022}$ & $0.00850^{+0.00043}_{-0.00042}$ & $0.0150\,(> 0.0136)$ & $0.00613 \pm 0.00013$ & $0.00885 \pm 0.00019$\\
$i$ (deg) & Inclination & $88.79^{+0.75}_{-0.68}$ & $87.9^{+1.2}_{-1.0}$ & $79.96\,(< 80.81)$ & $82.79^{+0.30}_{-0.31}$ & $85.64^{+0.55}_{-0.52}$\\
$K$ (m/s) & RV semi-amplitude & $168 \pm 10$ & $119^{+11}_{-13}$ & $130^{+18}_{-16}$ & $131^{+20}_{-23}$ & $51.2^{+8.5}_{-8.4}$\\
$a$ (AU) & Semimajor axis & $0.0675^{+0.0011}_{-0.0016}$ & $0.04560^{+0.00089}_{-0.00098}$ & $0.0421^{+0.0007}_{-0.0011}$ & $0.04611^{+0.00069}_{-0.00096}$ & $0.0463^{+0.0012}_{-0.0010}$\\
$R_P$ ($R_\mathrm{J}$) & Planet radius & $1.091^{+0.038}_{-0.032}$ & $1.167^{+0.048}_{-0.044}$ & $2.00\,(> 1.65)$ & $1.172^{+0.036}_{-0.035}$ & $1.274^{+0.041}_{-0.040}$\\
$M_P$ ($M_\mathrm{J}$) & Planet mass & $1.67^{+0.12}_{-0.13}$ & $0.98^{+0.10}_{-0.11}$ & $1.04^{+0.15}_{-0.14}$ & $1.11^{+0.18}_{-0.20}$ & $0.406^{+0.071}_{-0.068}$\\
$\rho_P$ (g cm$^{-3}$) & Planet density & $1.60^{+0.19}_{-0.20}$ & $0.76^{+0.13}_{-0.12}$ & $0.065\,(< 0.289)$ & $0.85 \pm 0.17$ & $0.243^{+0.053}_{-0.047}$\\
$\log{g_P}$ (cgs) & Planet surface gravity & $3.542^{+0.039}_{-0.046}$ & $3.251^{+0.056}_{-0.064}$ & $1.72\,(< 2.98)$ & $3.302^{+0.072}_{-0.087}$ & $2.792^{+0.078}_{-0.087}$\\
$b \equiv a\cos{i}/R_\star$ & Transit impact parameter & $0.22^{+0.11}_{-0.13}$ & $0.27^{+0.13}_{-0.16}$ & $1.03\,(> 1.00)$ & $0.808^{+0.012}_{-0.013}$ & $0.544^{+0.045}_{-0.052}$\\
$e$ & Eccentricity & 0.0 (fixed) & 0.0 (fixed) & 0.0 (fixed) & 0.0 (fixed) & 0.0 (fixed)\\
$e_\mathrm{lim}$\tablenotemark{a} & 1-$\sigma$ upper limit on eccentricity & $< 0.036$ & $< 0.099$ & $< 0.098$ & $< 0.093$ & $< 0.102$\\
$\tau_\mathrm{circ}$ (Gyr)\tablenotemark{b} & Tidal circularization timescale & $8.0^{+1.4}_{-1.5}$ & $0.255^{+0.065}_{-0.056}$ & $0.0032\,(< 0.0293)$ & $0.287^{+0.075}_{-0.067}$ & $0.088^{+0.026}_{-0.021}$\\
$\langle F \rangle$ (Gerg~s$^{-1}$~cm$^{-2}$) & Incident flux & $0.579^{+0.042}_{-0.037}$ & $1.258^{+0.098}_{-0.090}$ & $1.67^{+0.10}_{-0.10}$ & $1.614^{+0.083}_{-0.075}$ & $1.187^{+0.065}_{-0.063}$\\
$T_\mathrm{eq}$ (K) & Planet equilibirum temperature & $1264^{+22}_{-21}$ & $1534^{+29}_{-28}$ & $1646^{+25}_{-24}$ & $1633^{+21}_{-19}$ & $1512 \pm 20$\\
\multicolumn{7}{l}{\textbf{Stellar Parameters}}\\
$M_\star$ ($M_\odot$) & Stellar mass & $1.186^{+0.059}_{-0.083}$ & $1.199^{+0.072}_{-0.076}$ & $1.181^{+0.064}_{-0.092}$ & $1.242^{+0.057}_{-0.076}$ & $1.088^{+0.086}_{-0.070}$\\
$R_\star$ ($R_\odot$) & Stellar radius & $1.419^{+0.036}_{-0.030}$ & $1.302^{+0.038}_{-0.035}$ & $1.468 \pm 0.037$ & $1.538 \pm 0.037$ & $1.392^{+0.035}_{-0.034}$\\
$\log{g_\star}$ (cgs) & Stellar surface gravity & $4.209^{+0.020}_{-0.033}$ & $4.288^{+0.031}_{-0.035}$ & $4.174^{+0.031}_{-0.035}$ & $4.156^{+0.027}_{-0.029}$ & $4.188^{+0.037}_{-0.038}$\\
$\rho_\star$ (g cm$^{-3}$) & Stellar density & $0.587^{+0.035}_{-0.053}$ & $0.767^{+0.073}_{-0.074}$ & $0.521^{+0.050}_{-0.045}$ & $0.478^{+0.040}_{-0.038}$ & $0.569^{+0.061}_{-0.056}$\\
$L_\star$ ($L_\odot$) & Stellar luminosity & $1.93^{+0.16}_{-0.14}$ & $1.92^{+0.16}_{-0.15}$ & $2.15^{+0.14}_{-0.11}$ & $2.51 \pm 0.11$ & $1.876^{+0.060}_{-0.059}$\\
$T_\mathrm{eff}$ (K) & Stellar effective temperature & $5706^{+95}_{-91}$ & $5950 \pm 100$ & $5772^{+84}_{-80}$ & $5859^{+72}_{-71}$ & $5725^{+69}_{-68}$\\
$[\mathrm{Fe/H}]$ (dex) & Metallicity & $0.387^{+0.072}_{-0.081}$ & $0.262^{+0.085}_{-0.086}$ & $0.278^{+0.084}_{-0.081}$ & $0.273^{+0.084}_{-0.082}$ & $0.187^{+0.087}_{-0.090}$\\
$[\mathrm{Fe/H}]_0$ (dex)\tablenotemark{c} & Initial metallicity & $0.392^{+0.062}_{-0.071}$ & $0.277^{+0.073}_{-0.070}$ & $0.301^{+0.074}_{-0.070}$ & $0.297^{+0.077}_{-0.072}$ & $0.227^{+0.076}_{-0.080}$\\
Age (Gyr) & Stellar age & $5.4^{+2.4}_{-1.5}$ & $3.3^{+2.2}_{-1.7}$ & $5.6^{+2.7}_{-1.5}$ & $4.5^{+1.6}_{-1.1}$ & $7.5^{+2.8}_{-2.5}$\\
EEP\tablenotemark{d} & Equal evolutionary phase & $414^{+21}_{-25}$ & $374^{+38}_{-37}$ & $420^{+22}_{-25}$ & $408^{+22}_{-27}$ & $432^{+11}_{-22}$\\
$A_V$ (mag) & Visual extinction & $0.22 \pm 0.11$ & $0.52 \pm 0.11$ & $0.119^{+0.084}_{-0.075}$ & $0.077^{+0.046}_{-0.049}$ & $0.034 \pm 0.023$\\
d (pc) & Distance & $275.60^{+0.99}_{-0.98}$ & $400.9 \pm 1.8$ & $429.0 \pm 2.0$ & $562.1^{+5.0}_{-4.9}$ & $470.0 \pm 2.8$\\
\multicolumn{\totalcolumns}{c}{\vrule height 24pt width 0pt \small\it\fnum@table\ continued on next page}
\enddata

%% file: planet_fit_results_3.tex
\tablecolumns{7}
\tablehead{
 & & \colhead{TOI-3976 b} & \colhead{TOI-4087 b} & \colhead{TOI-4145 b} & \colhead{TOI-4463 b} & \colhead{TOI-4791 b} 
}
\startdata
\multicolumn{7}{l}{\textbf{Planet Parameters}}\\
$P$ (days) & Period & $6.607662^{+0.000016}_{-0.000015}$ & $3.17748350 \pm 0.00000094$ & $4.0664428 \pm 0.0000058$ & $2.8807198^{+0.0000028}_{-0.0000027}$ & $4.280880^{+0.000022}_{-0.000023}$\\
$T_c$ (BJD$_\mathrm{TDB}$) & Time of conjunction & $2459011.15791 \pm 0.00077$ & $2459244.566414^{+0.000093}_{-0.000092}$ & $2458925.88211^{+0.00017}_{-0.00016}$ & $2459291.64004 \pm 0.00032$ & $2459237.59508 \pm 0.00036$\\
$T_{14}$ (days) & Transit duration & $0.1882 \pm 0.0022$ & $0.12434^{+0.00049}_{-0.00048}$ & $0.08773^{+0.00098}_{-0.00095}$ & $0.0772^{+0.0015}_{-0.0014}$ & $0.1389^{+0.0015}_{-0.0014}$\\
$\tau$ (days) & Ingress/egress duration & $0.0173^{+0.0014}_{-0.0013}$ & $0.01344^{+0.00052}_{-0.00049}$ & $0.0271^{+0.0017}_{-0.0016}$ & $0.0330^{+0.0061}_{-0.0041}$ & $0.0175^{+0.0014}_{-0.0013}$\\
$a/R_\star$ & Planet-star separation & $10.63^{+0.36}_{-0.35}$ & $8.63 \pm 0.14$ & $12.07^{+0.24}_{-0.23}$ & $8.17^{+0.21}_{-0.24}$ & $8.46^{+0.27}_{-0.26}$\\
$\left(R_P / R_\star\right)^2$ & Transit depth & $0.00562 \pm 0.00013$ & $0.011565^{+0.000099}_{-0.000097}$ & $0.02015^{+0.00038}_{-0.00036}$ & $0.01311^{+0.00091}_{-0.00054}$ & $0.00655 \pm 0.00045$\\
$i$ (deg) & Inclination & $87.28^{+0.39}_{-0.36}$ & $87.82^{+0.37}_{-0.33}$ & $86.20 \pm 0.12$ & $83.82^{+0.21}_{-0.28}$ & $85.55^{+0.34}_{-0.33}$\\
$K$ (m/s) & RV semi-amplitude & $16.3^{+3.4}_{-3.3}$ & $90^{+17}_{-18}$ & $57^{+17}_{-18}$ & $108.7^{+3.3}_{-2.9}$ & $250^{+32}_{-34}$\\
$a$ (AU) & Semimajor axis & $0.0743^{+0.0013}_{-0.0014}$ & $0.04469^{+0.00048}_{-0.00054}$ & $0.04823^{+0.00075}_{-0.00079}$ & $0.04036^{+0.00074}_{-0.00082}$ & $0.0555^{+0.0011}_{-0.0012}$\\
$R_P$ ($R_\mathrm{J}$) & Planet radius & $1.095^{+0.036}_{-0.035}$ & $1.164^{+0.025}_{-0.024}$ & $1.187^{+0.032}_{-0.031}$ & $1.183^{+0.064}_{-0.045}$ & $1.110 \pm 0.050$\\
$M_P$ ($M_\mathrm{J}$) & Planet mass & $0.175^{+0.037}_{-0.036}$ & $0.73 \pm 0.14$ & $0.43 \pm 0.13$ & $0.794^{+0.039}_{-0.040}$ & $2.31^{+0.32}_{-0.33}$\\
$\rho_P$ (g cm$^{-3}$) & Planet density & $0.165^{+0.041}_{-0.037}$ & $0.57 \pm 0.12$ & $0.32^{+0.10}_{-0.10}$ & $0.595^{+0.080}_{-0.096}$ & $2.09^{+0.46}_{-0.40}$\\
$\log{g_P}$ (cgs) & Planet surface gravity & $2.56^{+0.09}_{-0.11}$ & $3.125^{+0.077}_{-0.096}$ & $2.88^{+0.12}_{-0.15}$ & $3.148^{+0.040}_{-0.056}$ & $3.668^{+0.071}_{-0.078}$\\
$b \equiv a\cos{i}/R_\star$ & Transit impact parameter & $0.505^{+0.049}_{-0.059}$ & $0.328^{+0.044}_{-0.051}$ & $0.800^{+0.010}_{-0.011}$ & $0.880^{+0.013}_{-0.009}$ & $0.657^{+0.027}_{-0.030}$\\
$e$ & Eccentricity & 0.0 (fixed) & 0.0 (fixed) & 0.0 (fixed) & 0.0 (fixed) & 0.0 (fixed)\\
$e_\mathrm{lim}$\tablenotemark{a} & 1-$\sigma$ upper limit on eccentricity & $< 0.207$ & $< 0.124$ & $< 0.135$ & $< 0.040$ & $< 0.152$\\
$\tau_\mathrm{circ}$ (Gyr)\tablenotemark{b} & Tidal circularization timescale & $1.39^{+0.44}_{-0.37}$ & $0.173^{+0.039}_{-0.037}$ & $0.223^{+0.079}_{-0.071}$ & $0.106^{+0.024}_{-0.027}$ & $2.62^{+0.87}_{-0.66}$\\
$\langle F \rangle$ (Gerg~s$^{-1}$~cm$^{-2}$) & Incident flux & $0.638^{+0.031}_{-0.027}$ & $1.026^{+0.056}_{-0.046}$ & $0.303^{+0.022}_{-0.018}$ & $0.861^{+0.062}_{-0.052}$ & $1.066^{+0.077}_{-0.068}$\\
$T_\mathrm{eq}$ (K) & Planet equilibirum temperature & $1295^{+16}_{-14}$ & $1458^{+20}_{-17}$ & $1074^{+19}_{-16}$ & $1395^{+25}_{-22}$ & $1472^{+26}_{-24}$\\
\multicolumn{7}{l}{\textbf{Stellar Parameters}}\\
$M_\star$ ($M_\odot$) & Stellar mass & $1.254^{+0.067}_{-0.072}$ & $1.178^{+0.039}_{-0.042}$ & $0.905^{+0.043}_{-0.044}$ & $1.056^{+0.059}_{-0.063}$ & $1.242^{+0.072}_{-0.077}$\\
$R_\star$ ($R_\odot$) & Stellar radius & $1.501^{+0.039}_{-0.038}$ & $1.112^{+0.021}_{-0.020}$ & $0.859^{+0.018}_{-0.017}$ & $1.062^{+0.027}_{-0.024}$ & $1.409^{+0.039}_{-0.038}$\\
$\log{g_\star}$ (cgs) & Stellar surface gravity & $4.183^{+0.034}_{-0.035}$ & $4.416^{+0.015}_{-0.016}$ & $4.526^{+0.020}_{-0.021}$ & $4.410^{+0.027}_{-0.032}$ & $4.234^{+0.032}_{-0.033}$\\
$\rho_\star$ (g cm$^{-3}$) & Stellar density & $0.521^{+0.055}_{-0.050}$ & $1.205^{+0.060}_{-0.058}$ & $2.01^{+0.12}_{-0.11}$ & $1.24^{+0.10}_{-0.11}$ & $0.624^{+0.062}_{-0.057}$\\
$L_\star$ ($L_\odot$) & Stellar luminosity & $2.588 \pm 0.069$ & $1.500^{+0.089}_{-0.069}$ & $0.516^{+0.040}_{-0.031}$ & $1.026^{+0.080}_{-0.062}$ & $2.40^{+0.19}_{-0.16}$\\
$T_\mathrm{eff}$ (K) & Stellar effective temperature & $5975^{+70}_{-69}$ & $6060^{+74}_{-67}$ & $5281^{+86}_{-76}$ & $5640^{+89}_{-82}$ & $6058^{+99}_{-94}$\\
$[\mathrm{Fe/H}]$ (dex) & Metallicity & $0.182^{+0.078}_{-0.079}$ & $0.237 \pm 0.079$ & $0.168^{+0.084}_{-0.087}$ & $0.250 \pm 0.085$ & $0.207^{+0.085}_{-0.089}$\\
$[\mathrm{Fe/H}]_0$ (dex)\tablenotemark{c} & Initial metallicity & $0.237^{+0.071}_{-0.072}$ & $0.205^{+0.073}_{-0.074}$ & $0.155^{+0.081}_{-0.083}$ & $0.242^{+0.077}_{-0.078}$ & $0.252^{+0.072}_{-0.073}$\\
Age (Gyr) & Stellar age & $3.8^{+1.5}_{-1.3}$ & $0.8^{+1.2}_{-0.6}$ & $4.7^{+4.2}_{-3.1}$ & $4.1^{+3.5}_{-2.6}$ & $3.3^{+1.8}_{-1.4}$\\
EEP\tablenotemark{d} & Equal evolutionary phase & $394^{+25}_{-42}$ & $313^{+26}_{-41}$ & $340^{+22}_{-30}$ & $354^{+40}_{-27}$ & $378 \pm 35$\\
$A_V$ (mag) & Visual extinction & $0.038^{+0.015}_{-0.022}$ & $0.089^{+0.072}_{-0.058}$ & $0.16^{+0.12}_{-0.10}$ & $0.15^{+0.11}_{-0.09}$ & $0.16^{+0.10}_{-0.09}$\\
d (pc) & Distance & $522.2^{+2.5}_{-2.4}$ & $310.4 \pm 1.1$ & $204.96 \pm 0.40$ & $177.64^{+0.56}_{-0.54}$ & $322.1^{+1.9}_{-1.8}$\\
\enddata

%% file: aux_fit_results.tex
\\[-\normalbaselineskip]\multicolumn{2}{l}{\textbf{TOI-1937}}\\
 &R&g'&i'&z'&TESS&V\\
~~~~$u_{1}$ &$0.369^{+0.053}_{-0.052}$&$0.598^{+0.055}_{-0.056}$&$0.309\pm0.032$&$0.270\pm0.052$&$0.305\pm0.027$&$0.462\pm0.054$\\
~~~~$u_{2}$ &$0.276^{+0.051}_{-0.050}$&$0.211\pm0.052$&$0.278\pm0.029$&$0.298\pm0.050$&$0.277\pm0.023$&$0.246\pm0.052$\\
~~~~$A_D$ &$-0.35^{+0.17}_{-0.21}$&$-0.36^{+0.17}_{-0.21}$&$-0.16^{+0.13}_{-0.16}$&$-0.23^{+0.15}_{-0.19}$&$-0.20^{+0.13}_{-0.16}$&--\\
 &PFS\\
~~~~$\gamma_{\rm rel}$ &$-59\pm15$\\
~~~~$\sigma_J$ &$52^{+16}_{-11}$\\
\hline\\[-\normalbaselineskip]\multicolumn{2}{l}{\textbf{TOI-2364}}\\
 &R&TESS\\
~~~~$u_{1}$ &$0.479^{+0.049}_{-0.050}$&$0.385^{+0.034}_{-0.033}$\\
~~~~$u_{2}$ &$0.208\pm0.050$&$0.224^{+0.036}_{-0.035}$\\
~~~~$A_D$ &--&$-0.0002\pm0.0029$\\
 &PFS&TRES\\
~~~~$\gamma_{\rm rel}$ &$-4.9^{+3.9}_{-3.8}$&$-30^{+24}_{-19}$\\
~~~~$\sigma_J$ &$7.8^{+8.2}_{-3.6}$&$19^{+44}_{-20}$\\
\hline\\[-\normalbaselineskip]\multicolumn{2}{l}{\textbf{TOI-2583}}\\
 &B&I&g'&z'&TESS\\
~~~~$u_{1}$ &$0.611\pm0.051$&$0.260\pm0.048$&$0.554\pm0.046$&$0.197^{+0.048}_{-0.049}$&$0.291\pm0.025$\\
~~~~$u_{2}$ &$0.168\pm0.051$&$0.278\pm0.049$&$0.239\pm0.049$&$0.269\pm0.050$&$0.296\pm0.025$\\
~~~~$A_D$ &--&--&--&--&$0.0020\pm0.0069$\\
 &HIRES\\
~~~~$\gamma_{\rm rel}$ &$0.2^{+4.2}_{-4.4}$\\
~~~~$\sigma_J$ &$7.3^{+11}_{-7.3}$\\
\hline\\[-\normalbaselineskip]\multicolumn{2}{l}{\textbf{TOI-2587}}\\
 &i'&r'&TESS\\
~~~~$u_{1}$ &$0.309\pm0.052$&$0.391\pm0.052$&$0.311^{+0.037}_{-0.036}$\\
~~~~$u_{2}$ &$0.282\pm0.050$&$0.272^{+0.051}_{-0.050}$&$0.284\pm0.035$\\
~~~~$A_D$ &--&--&$-0.000\pm0.014$\\
 &NEID\\
~~~~$\gamma_{\rm rel}$ &$-24502.3^{+4.1}_{-3.7}$\\
~~~~$\sigma_J$ &$8.2^{+6.4}_{-5.1}$\\
\hline\\[-\normalbaselineskip]\multicolumn{2}{l}{\textbf{TOI-2796}}\\
 &B&g'&i'&TESS\\
~~~~$u_{1}$ &$0.658\pm0.055$&$0.611\pm0.051$&$0.319^{+0.051}_{-0.052}$&$0.322\pm0.037$\\
~~~~$u_{2}$ &$0.115\pm0.053$&$0.197\pm0.051$&$0.275^{+0.051}_{-0.050}$&$0.280\pm0.035$\\
~~~~$A_D$ &--&--&--&$0.00000\pm0.00092$\\
 &HIRES&NEID\\
~~~~$\gamma_{\rm rel}$ &$-4\pm27$&$20592.2^{+8.5}_{-9.0}$\\
~~~~$\sigma_J$ &$47^{+32}_{-24}$&$19.9^{+15}_{-8.7}$\\
\hline\\[-\normalbaselineskip]\multicolumn{2}{l}{\textbf{TOI-2803}}\\
 &B&R&g'&i'&TESS\\
~~~~$u_{1}$ &$0.518\pm0.040$&$0.311\pm0.048$&$0.471\pm0.043$&$0.195\pm0.048$&$0.239\pm0.031$\\
~~~~$u_{2}$ &$0.239\pm0.038$&$0.335\pm0.049$&$0.285\pm0.048$&$0.286\pm0.049$&$0.301^{+0.033}_{-0.034}$\\
~~~~$A_D$ &--&--&--&--&$0.046\pm0.019$\\
 &PFS&TRES\\
~~~~$\gamma_{\rm rel}$ &$121.0^{+10.}_{-9.3}$&$243\pm59$\\
~~~~$\sigma_J$ &$20.3^{+19}_{-8.8}$&$61^{+28}_{-61}$\\
\hline\\[-\normalbaselineskip]\multicolumn{2}{l}{\textbf{TOI-2818}}\\
 &R&g'&TESS\\
~~~~$u_{1}$ &$0.428\pm0.042$&$0.570^{+0.055}_{-0.054}$&$0.308\pm0.028$\\
~~~~$u_{2}$ &$0.315\pm0.047$&$0.195\pm0.053$&$0.276\pm0.029$\\
~~~~$A_D$ &--&--&$0.022\pm0.014$\\
 &CHIRON\\
~~~~$\gamma_{\rm rel}$ &$58996\pm28$\\
~~~~$\sigma_J$ &$64^{+50}_{-25}$\\
\hline\\[-\normalbaselineskip]\multicolumn{2}{l}{\textbf{TOI-2842}}\\
 &B&R&TESS\\
~~~~$u_{1}$ &$0.648\pm0.056$&$0.371\pm0.053$&$0.287\pm0.038$\\
~~~~$u_{2}$ &$0.165\pm0.054$&$0.301\pm0.051$&$0.284\pm0.036$\\
~~~~$A_D$ &--&--&$-0.005\pm0.017$\\
 &PFS\\
~~~~$\gamma_{\rm rel}$ &$-1.6^{+4.2}_{-3.7}$\\
~~~~$\sigma_J$ &$5.3^{+9.9}_{-5.3}$\\
\hline\\[-\normalbaselineskip]\multicolumn{2}{l}{\textbf{TOI-2977}}\\
 &R&g'&i'&TESS\\
~~~~$u_{1}$ &$0.315\pm0.045$&$0.602^{+0.055}_{-0.054}$&$0.318^{+0.052}_{-0.051}$&$0.315\pm0.030$\\
~~~~$u_{2}$ &$0.237\pm0.049$&$0.206\pm0.053$&$0.269\pm0.050$&$0.272\pm0.029$\\
~~~~$A_D$ &$0.150\pm0.014$&$0.0966\pm0.0096$&$0.162\pm0.016$&$0.146^{+0.024}_{-0.025}$\\
 &CHIRON\\
~~~~$\gamma_{\rm rel}$ &$7839^{+34}_{-33}$\\
~~~~$\sigma_J$ &$66^{+79}_{-37}$\\
\hline\\[-\normalbaselineskip]\multicolumn{2}{l}{\textbf{TOI-3023}}\\
 &R&i'&TESS\\
~~~~$u_{1}$ &$0.342\pm0.045$&$0.291\pm0.050$&$0.288\pm0.024$\\
~~~~$u_{2}$ &$0.275\pm0.048$&$0.269\pm0.050$&$0.272\pm0.025$\\
~~~~$A_D$ &--&--&$0.003\pm0.010$\\
 &PFS\\
~~~~$\gamma_{\rm rel}$ &$-43.1^{+8.1}_{-8.3}$\\
~~~~$\sigma_J$ &$17.7^{+18}_{-7.7}$\\
\hline\\[-\normalbaselineskip]\multicolumn{2}{l}{\textbf{TOI-3364}}\\
 &R&TESS\\
~~~~$u_{1}$ &$0.376^{+0.049}_{-0.050}$&$0.335\pm0.029$\\
~~~~$u_{2}$ &$0.248\pm0.050$&$0.275\pm0.029$\\
~~~~$A_D$ &--&$-0.024^{+0.034}_{-0.035}$\\
 &PFS\\
~~~~$\gamma_{\rm rel}$ &$-90.5^{+6.4}_{-6.3}$\\
~~~~$\sigma_J$ &$13.3^{+13}_{-5.4}$\\
\hline\\[-\normalbaselineskip]\multicolumn{2}{l}{\textbf{TOI-3688}}\\
 &i'&TESS\\
~~~~$u_{1}$ &$0.280\pm0.052$&$0.279^{+0.052}_{-0.051}$\\
~~~~$u_{2}$ &$0.286^{+0.050}_{-0.051}$&$0.285\pm0.051$\\
~~~~$A_D$ &--&$0.010\pm0.031$\\
 &NEID\\
~~~~$\gamma_{\rm rel}$ &$-146.3^{+9.6}_{-8.8}$\\
~~~~$\sigma_J$ &$18^{+17}_{-10.}$\\
\hline\\[-\normalbaselineskip]\multicolumn{2}{l}{\textbf{TOI-3807}}\\
 &B&R&i'&r'&z'&TESS\\
~~~~$u_{1}$ &$0.681\pm0.042$&$0.382\pm0.051$&$0.320\pm0.051$&$0.413^{+0.050}_{-0.051}$&$0.249\pm0.050$&$0.315\pm0.032$\\
~~~~$u_{2}$ &$0.125\pm0.039$&$0.276\pm0.050$&$0.282^{+0.050}_{-0.049}$&$0.277\pm0.049$&$0.273^{+0.050}_{-0.049}$&$0.278\pm0.030$\\
~~~~$A_D$ &--&--&--&--&--&$0.178^{+0.066}_{-0.075}$\\
 &NEID\\
~~~~$\gamma_{\rm rel}$ &$-28660\pm13$\\
~~~~$\sigma_J$ &$25^{+30}_{-13}$\\
\hline\\[-\normalbaselineskip]\multicolumn{2}{l}{\textbf{TOI-3819}}\\
 &R&i'&TESS\\
~~~~$u_{1}$ &$0.384^{+0.053}_{-0.052}$&$0.314\pm0.049$&$0.294\pm0.025$\\
~~~~$u_{2}$ &$0.299^{+0.050}_{-0.051}$&$0.298^{+0.049}_{-0.048}$&$0.279\pm0.023$\\
~~~~$A_D$ &--&--&$-0.00000^{+0.00042}_{-0.00043}$\\
 &NEID\\
~~~~$\gamma_{\rm rel}$ &$30839^{+16}_{-17}$\\
~~~~$\sigma_J$ &$33^{+36}_{-16}$\\
\hline\\[-\normalbaselineskip]\multicolumn{2}{l}{\textbf{TOI-3912}}\\
 &B&R&i'&TESS\\
~~~~$u_{1}$ &$0.683^{+0.041}_{-0.042}$&$0.387\pm0.052$&$0.358^{+0.034}_{-0.035}$&$0.311\pm0.034$\\
~~~~$u_{2}$ &$0.126\pm0.039$&$0.278\pm0.050$&$0.301\pm0.034$&$0.266^{+0.035}_{-0.034}$\\
~~~~$A_D$ &--&--&--&$0.000000^{+0.000090}_{-0.000089}$\\
 &HIRES\\
~~~~$\gamma_{\rm rel}$ &$12.2\pm6.2$\\
~~~~$\sigma_J$ &$13.4^{+10.}_{-5.4}$\\
\hline\\[-\normalbaselineskip]\multicolumn{2}{l}{\textbf{TOI-3976}}\\
 &i'&TESS\\
~~~~$u_{1}$ &$0.226\pm0.050$&$0.299\pm0.030$\\
~~~~$u_{2}$ &$0.244\pm0.049$&$0.310^{+0.029}_{-0.030}$\\
~~~~$A_D$ &--&$-0.00005\pm0.00060$\\
 &HIRES\\
~~~~$\gamma_{\rm rel}$ &$2.0\pm2.6$\\
~~~~$\sigma_J$ &$6.6^{+3.7}_{-3.0}$\\
\hline\\[-\normalbaselineskip]\multicolumn{2}{l}{\textbf{TOI-4087}}\\
 &B&R&TESS\\
~~~~$u_{1}$ &$0.581\pm0.052$&$0.348\pm0.048$&$0.250^{+0.016}_{-0.017}$\\
~~~~$u_{2}$ &$0.175\pm0.051$&$0.314\pm0.049$&$0.286\pm0.019$\\
~~~~$A_D$ &--&--&$0.00000\pm0.00019$\\
 &NEID&TRES\\
~~~~$\gamma_{\rm rel}$ &$-14263^{+15}_{-14}$&$152^{+65}_{-64}$\\
~~~~$\sigma_J$ &$31^{+23}_{-12}$&$166^{+71}_{-48}$\\
\hline\\[-\normalbaselineskip]\multicolumn{2}{l}{\textbf{TOI-4145}}\\
 &B&I&TESS\\
~~~~$u_{1}$ &$0.802^{+0.054}_{-0.056}$&$0.351\pm0.051$&$0.396\pm0.029$\\
~~~~$u_{2}$ &$0.029\pm0.054$&$0.219\pm0.050$&$0.233\pm0.026$\\
~~~~$A_D$ &--&--&$0.0009\pm0.0064$\\
 &NEID\\
~~~~$\gamma_{\rm rel}$ &$1836^{+12}_{-13}$\\
~~~~$\sigma_J$ &$32^{+19}_{-10.}$\\
\hline\\[-\normalbaselineskip]\multicolumn{2}{l}{\textbf{TOI-4463}}\\
 &R&g'&r'&z'&TESS\\
~~~~$u_{1}$ &$0.411\pm0.053$&$0.626\pm0.051$&$0.427^{+0.049}_{-0.050}$&$0.267\pm0.049$&$0.342\pm0.049$\\
~~~~$u_{2}$ &$0.262\pm0.051$&$0.150^{+0.052}_{-0.051}$&$0.258\pm0.050$&$0.261\pm0.049$&$0.267\pm0.048$\\
~~~~$A_D$ &--&--&--&--&$0.010\pm0.025$\\
 &NEID\\
~~~~$\gamma_{\rm rel}$ &$-90370.8^{+2.2}_{-2.0}$\\
~~~~$\sigma_J$ &$2.5^{+5.5}_{-2.5}$\\
\hline\\[-\normalbaselineskip]\multicolumn{2}{l}{\textbf{TOI-4791}}\\
 &R&TESS\\
~~~~$u_{1}$ &$0.318\pm0.051$&$0.275\pm0.035$\\
~~~~$u_{2}$ &$0.295\pm0.050$&$0.300\pm0.035$\\
~~~~$A_D$ &--&$-0.071^{+0.069}_{-0.080}$\\
 &CHIRON\\
~~~~$\gamma_{\rm rel}$ &$33780^{+29}_{-25}$\\
~~~~$\sigma_J$ &$38^{+69}_{-38}$\\
\hline